\documentclass{article}

\usepackage{geometry}
\usepackage{amsmath,amsthm,amssymb,amsfonts}
\usepackage{graphicx}

\usepackage{tikz}
\usetikzlibrary{shapes.multipart,positioning}

\usepackage{multirow}
\usepackage{xcolor,svgcolor}
\usepackage{enumitem}
\usepackage{mathrsfs}

\usepackage{xspace}
\usepackage{booktabs}

\newcommand{\D}{\ensuremath{\mathbb{D}}\xspace}
\newcommand{\F}{\ensuremath{\mathbb{F}}\xspace}
\newcommand{\Z}{\ensuremath{\mathbb{Z}}\xspace}
\newcommand{\M}{\ensuremath{\mathfrak{M}}}
\newcommand{\MM}{\ensuremath{\mathfrak{MM}}}
\newcommand{\bnd}[2]{\ensuremath{#1\mathopen{}(#2)\mathclose{}}}
\newcommand{\bnddisplay}[2]{\ensuremath{#1\mathopen{}\left(#2\right)\mathclose{}}}
\newcommand{\softoh}[1]{\bnd{\tilde{\mathcal{O}}}{#1}}
\newcommand{\littleo}[1]{\bnd{{o}}{#1}}
\newcommand{\bigO}[1]{\bnd{\mathcal{O}}{#1}}
\newcommand{\bigOdisplay}[1]{\bnddisplay{\mathcal{O}}{#1}}
\newlength{\sfracUpperRaise}%
\newlength{\sfracLowerRaise}%
\newcommand{\sfrac}[3][0pt]{%
\setlength{\sfracUpperRaise}{#1}%
\setlength{\sfracLowerRaise}{\sfracUpperRaise}
\addtolength{\sfracLowerRaise}{5pt}%
\genfrac{}{}{}{}{\raisebox{\sfracUpperRaise}{$#2$}}{\raisebox{-\sfracLowerRaise}{$#3$}}}
\newcommand\ddfrac[2]{\sfrac[1pt]{\displaystyle #1}{\displaystyle #2}}

\newcommand{\threshold}{\ensuremath{\text{Threshold}}}
\DeclareMathOperator{\me}{\,\ensuremath{\mathrel{\textrm{--}}=}\,}
\DeclareMathOperator{\pe}{\,\ensuremath{\mathrel{+}=}\,}
\DeclareMathOperator{\fe}{\,\ensuremath{\mathrel{{\ast}}=}\,}
\DeclareMathOperator{\de}{\,\ensuremath{\mathrel{/}=}\,}

\newcommand{\Toeplitz}[1]{\ensuremath{\operatorname{\mathcal{T}_{#1}}}}
\newcommand{\Circulant}[1]{\ensuremath{\operatorname{\mathscr{C}_{#1}}}}

\DeclareMathOperator{\revdiv}{\,\overleftarrow{\operatorname{div}}}
\DeclareMathOperator{\revmul}{\,\overleftarrow{\operatorname{mul}}}
\DeclareMathOperator{\bquo}{\,\operatorname{div}}
\DeclareMathOperator{\bdiv}{\,\operatorname{div}}
\DeclareMathOperator{\bmul}{\,\operatorname{mul}}

\newcommand\DFT{\mathsf{DFT}}
\newcommand\brDFT{\mathsf{brDFT}}
\newcommand\brTFT{\mathsf{brTFT}}
\newcommand\partTFT{\mathsf{partTFT}}

\makeatletter
\DeclareRobustCommand{\cev}[1]{%
  {\mathpalette\do@cev{#1}}%
}
\newcommand{\do@cev}[2]{%
  \vbox{\offinterlineskip
    \sbox\z@{$\m@th#1 x$}%
    \ialign{##\cr
      \hidewidth\reflectbox{$\m@th#1\vec{}\mkern4mu$}\hidewidth\cr
      \noalign{\kern-\ht\z@}
      $\m@th#1#2$\cr
    }%
  }%
}
\makeatother
\newcommand{\mat}[1]{\textbf{\ensuremath{#1}}}
\newcommand{\Transpose}[1]{{{\mat{#1}}^{\intercal}}\xspace}
\newcommand{\gTranspose}[2]{{{\mat{\begingroup\setlength\arraycolsep{#1}#2\endgroup}}^{\intercal}}\xspace}

\usepackage{fancyvrb}

\usepackage{nicematrix}
\NiceMatrixOptions{%
    renew-dots,%
    xdots/shorten=.5em,%
    xdots/inter=.65em,%
}
\newenvironment{smatrix}[1][]{\left\lbrack\begin{NiceMatrix}[small,#1]}{\end{NiceMatrix}\right\rbrack}

\def\matrixsize#1#2{{{#1}\times{#2}}}
\def\MatrixProduct#1#2{{\mat{#1}\cdot\mat{#2}}}
\def\triadone{brown}
\def\triadtwo{red}
\def\triadthree{blue}
\def\triadfour{green}
\def\triadfive{magenta}
\def\triadsix{darkgray}
\def\triadseven{purple}

\usepackage{bbding}
\newcommand{\xyes}{{\color{teal}\CheckmarkBold}}
\newcommand{\xno}{{\color{purple}\XSolidBrush}}

\newcommand{\LLTr}{\ensuremath{{\text{Low}}}}
\newcommand{\Low}[1]{\ensuremath{{\LLTr\left({\mat{#1}}\right)}}}
\newcommand{\URTr}{\ensuremath{{\text{Up}}}}
\newcommand{\Up}[1]{\ensuremath{{\URTr\left({\mat{#1}}\right)}}}

\newcommand{\MUL}{\normalfont{\textbf{MUL}}\xspace}
\newcommand{\ADD}{\normalfont{\textbf{ADD}}\xspace}
\newcommand{\SCA}{\normalfont{\textbf{SCA}}\xspace}

\usepackage{threeparttable}

\usepackage{algorithm}
\usepackage{algcompatible}

\newcommand{\To}{\textbf{to}\xspace}

\newcommand{\DownTo}{\textbf{down-to}\xspace}

\algnewcommand{\IfThen}[2]{%
  \State \algorithmicif\ #1\ \algorithmicthen\ #2}
\algnewcommand{\IfThenEnd}[2]{%
  \State \algorithmicif\ #1\ \algorithmicthen\ #2\ \algorithmicend\ \algorithmicif}
\algnewcommand{\IfThenElse}[3]{%
  \State \algorithmicif\ #1\ \algorithmicthen\ #2\ \algorithmicelse\ #3}
\algnewcommand{\ElseIf}[2]{
  \State \algorithmicelse\ \algorithmicif\ #1\ \algorithmicthen\ #2}
\algnewcommand{\ElseEnd}[1]{
  \State \algorithmicelse\ #1\ \algorithmicend\ \algorithmicif}
\algnewcommand{\ForDoEnd}[3][]{
  \ifthenelse{\equal{#1}{}}
  {\State \algorithmicfor\ #2\ \algorithmicdo\ #3\ \algorithmicend\ \algorithmicfor}
  {\State\label{#1}\algorithmicfor\ #2\ \algorithmicdo\ #3\ \algorithmicend\ \algorithmicfor}
}

\renewcommand{\algorithmicensure}{{\textbf{Result:}}}
\algnewcommand\algorithmicreadonly{\textbf{Read-only:}}
\algnewcommand\READONLY{\item[\algorithmicreadonly]}%

\newcommand{\Algname}{Algorithm}
\newcommand{\Algsname}{{\Algname}s}
\newcommand{\algname}{\Algname}
\newcommand{\algsname}{\Algsname}

\usepackage{hyperref}
\makeatletter
\hypersetup{%
    breaklinks=true,
    colorlinks=true,
    linkcolor=darkred,
    citecolor=blue,
    urlcolor=darkgreen,
hypertexnames=false,
naturalnames=true,
}
\gdef\urlauthor#1#2{\g@addto@macro\@elsuads{\let\corref\@gobble%
     \def\@@tmp{#1}\raggedright\eadsep
     {\ttfamily\url{\expandafter\strip@prefix\meaning\@@tmp}}\space(#2)%
     \def\eadsep{\unskip,\space}}%
}

\gdef\emailauthor#1#2{\stepcounter{ead}%
     \g@addto@macro\@elseads{\raggedright%
      \let\corref\@gobble\def\@@tmp{#1}%
      \eadsep{\ttfamily\href{mailto:\expandafter\strip@prefix\meaning\@@tmp}{\expandafter\strip@prefix\meaning\@@tmp}}
      (#2)\def\eadsep{\unskip,\space}}%
}
\makeatother

\usepackage{biblatex}
\renewbibmacro{in:}{}

\usepackage{aliascnt}	
			
\makeatletter
\AddToHook{env/algorithmic/begin}{\def\@currentcounter{ALG@line}}
\makeatother
\usepackage[nameinlink,capitalize,noabbrev]{cleveref}
\crefalias{ALG@line}{line}
\crefformat{appendix}{#2#1#3}

\crefname{algorithm}{\algname}{\algsname}
\Crefname{algorithm}{\Algname}{\Algsname}
\Crefname{proposition}{Proposition}{Propositions}
\crefname{proposition}{Proposition}{Propositions}
\newcommand{\plinopt}{\href{https://github.com/jgdumas/plinopt}{\textsc{PLinOpt}}}

\newaliascnt{theorem}{algorithm}
\newaliascnt{definition}{algorithm}
\newaliascnt{example}{algorithm}
\newaliascnt{fact}{algorithm}
\newaliascnt{lemma}{algorithm}
\newaliascnt{notation}{algorithm}
\newaliascnt{proposition}{algorithm}
\newaliascnt{corollary}{algorithm}
\newaliascnt{remark}{algorithm}
\newaliascnt{strategy}{algorithm}
\newaliascnt{problem}{algorithm}
\newtheorem{theorem}[theorem]{Theorem}
\newtheorem{definition}[definition]{Definition}
\newtheorem{example}[example]{Example}

\newtheorem{lemma}[lemma]{Lemma}

\newtheorem{proposition}[proposition]{Proposition}
\newtheorem{corollary}[corollary]{Corollary}
\newtheorem{remark}[remark]{Remark}

\aliascntresetthe{theorem}
\aliascntresetthe{definition}
\aliascntresetthe{example}
\aliascntresetthe{fact}
\aliascntresetthe{lemma}
\aliascntresetthe{notation}
\aliascntresetthe{proposition}
\aliascntresetthe{corollary}
\aliascntresetthe{remark}
\aliascntresetthe{strategy}
\aliascntresetthe{problem}

\title{Fast in-place accumulation}
\author{Jean-Guillaume Dumas\footnote{
  {Universit\'e Grenoble Alpes}.
  {Laboratoire Jean Kuntzmann, CNRS, UMR 5224}.
  {150 place du Torrent, IMAG - CS 40700},
  {38058 Grenoble, cedex 9}
  {France}.
\href{mailto:Jean-Guillaume.Dumas@univ-grenoble-alpes.fr,Bruno.Grenet@univ-grenoble-alpes.fr}{\{firstname.lastname\}@univ-grenoble-alpes.fr}}
\and{Bruno Grenet}\footnotemark[1]}

\begin{document}
\maketitle

\begin{abstract}
This paper deals with simultaneously fast and in-place algorithms
for formulae where the result has to be linearly accumulated:
some output variables are also input variables,
linked by a linear dependency.
Fundamental examples include the in-place accumulated multiplication
of polynomials or matrices, $C\pe{AB}$ (that is with only
$\bigO{1}$ extra space).
The difficulty is to combine in-place computations with fast
algorithms:
those usually come at the expense of (potentially large) extra
temporary space, but with accumulation the output variables are not
even available to store intermediate values.
We first propose a novel automatic design of fast and in-place
accumulating algorithms for any bilinear formulae (and thus for
polynomial and matrix multiplication) and then extend it to any linear
accumulation of a collection of functions.
For this, we relax the in-place model to any
algorithm allowed to modify its inputs, provided that those are
restored to their initial state afterwards.
This allows us to ultimately derive unprecedented in-place accumulating
algorithms for fast polynomial multiplications and for Strassen-like
matrix multiplications.

We then consider the simultaneously fast and in-place computation of the
Euclidean polynomial modular remainder $R(X)\equiv{A(X)}\mod{B(X)}$.
Fast algorithms for this usually also come at the expense of
a linear amount of extra temporary space.
In particular, they require one to first compute and store the whole
quotient $Q(X)$ such that $A = BQ+R$.
We here propose an \emph{in-place} algorithm to compute the
remainder only.
If $A$ and $B$ have respective degree $m+n$ and $n$, and
$\M(k)$ denotes the complexity of a (not-in-place) algorithm to
multiply two degree-$k$ polynomials, our algorithm uses at most
$\bigOdisplay{\frac{m}{n}\M(n)\log(n)}$ arithmetic operations.
In this particular case this is a factor $\log(n)$ more than the
not-in-place algorithm. But if $\M(n) = \Theta(n^{1+\epsilon})$ for
some $\epsilon>0$, then our algorithms do match the not-in-place
complexity bound of $\bigOdisplay{\frac{m}{n}\M(n)}$.
We also propose variants that compute
-- still in-place and with the same kind of complexity bounds --
the over-place remainder $A(X)\equiv{A(X)}\mod{B(X)}$,
the accumulated remainder $R(X)\pe{A(X)}\mod{B(X)}$ and
the accumulated modular multiplication $R(X)\pe{A(X)C(X)}\mod{B(X)}$,
that is multiplication in a polynomial extension of a finite field.

To achieve this, we develop techniques for Toeplitz matrix operations,
for generalized convolutions, short product and power series division
and remainder whose output is also part of the input.
\end{abstract}

\clearpage
\tableofcontents

\section{Introduction}
Multiplication is one of the most fundamental arithmetic operations in
computer science and in particular in computer algebra and symbolic
computation.
In terms of arithmetic operations, for instance, from the early work
of~\cite{Karatsuba:1963:multiplication,SchoenhageStrassen1971,Strassen:1969:GENO}
to the most recent work of~\cite{Cantor:1991:mulpoly,Harvey:2022:JACM:nlogn,Alman:2025:asymmetry},
many
sub-quadratic (resp. sub-cubic) algorithms
were developed for polynomial (resp. matrix) multiplication.
But these fast algorithms usually come at the expense
of (potentially large) extra temporary space to perform the
computation.
On the contrary, classical, quadratic (resp. cubic)
algorithms, when computed sequentially, quite often require very few
(constant) extra registers.
It is an intriguing question whether there is a necessary trade-off
between space and time, that is whether fast algorithms always require
extra space to perform their computations. The use of extra space
could hinder the practical efficiency of the algorithms,
due for instance to cache misses.
For parallel computations, it is easier to get good
parallel speedups if the sub-computations run in place, to avoid
memory management.
Some initial lower bounds suggested that fast algorithms do indeed
require extra space~\cite{Abrahamson:1986:timespace}. But further
work then proposed simultaneously ``fast'' and ``in-place''
algorithms, for matrix or polynomial
operations~\cite{jgd:2009:WinoSchedule,Roche:2009:spacetime,Harvey:2010:issactft,Giorgi:2019:issac:reductions,Giorgi:2020:issac:inplace}.

We here extend the latter line of work to
\emph{accumulating} algorithms.
Actually, one of the main ingredients on non-accumulating algorithms is to
use the (free) space of the output as intermediate storage. This
makes it possible to avoid existing lower bounds.
But when the result has to be accumulated, \emph{i.e.}, if the output is also
part of the input, this free space does not exist.
To be able to design accumulating in-place algorithms we thus relax the in-place
model to allow algorithms to also modify their input, therefore to use
them as intermediate storage,
\emph{provided that they are restored to their initial state after
  completion of the procedure}.
This is in fact a natural possibility in many programming environments.
Furthermore, this restoration allows for recursive combinations of
such procedures, as the (non-concurrent) recursive calls will not
mess up the state of their callers.
We thus propose a generic technique transforming any bilinear
algorithm into an in-place algorithm under this model.
This directly applies to accumulating polynomial and matrix
multiplication algorithms, including fast ones such as Karatsuba's~\cite{Karatsuba:1963:multiplication} or Strassen's~\cite{Strassen:1969:GENO}.
Further, the technique actually generalizes to any linear
accumulation, \emph{i.e.}, not only bilinear formulae, provided that the input
of the accumulation can be itself reversibly computed in-place
(therefore also potentially in-place of some of its own input if
needed).

Then we use this technique to develop in-place modular methods for
dense univariate polynomials over a finite ring.
For instance, we compute in-place the remainder only of the Euclidean
division.
This means that, e.g., with respect
to~\cite[\algname~3]{Giorgi:2020:issac:inplace}, we obtain the
remainder without needing any space for the quotient.

As polynomials and Toeplitz matrices are indeed different representations
of the same objects, see,
e.g.,~\cite{Bini:1985:IPL:poldiv,Bini:1994:polymatcomp,Giorgi:2019:issac:reductions},
we develop as building blocks fast methods for Toeplitz matrix operations,
in-place \emph{with accumulation} or \emph{over-place},
as well as for generalized convolutions, short product and power
series division and remainder,
where the output is also part of the input.

As a direct application of these techniques  we finally obtain
in-place algorithms for the multiplication in a polynomial extension
of a finite field.

This paper extends the results
of~\cite{jgd:2024:bilin,jgd:2024:inplacerem} as follows:
\begin{itemize}
\item We describe an implementation of the first accumulating and
  in-place Strassen-like matrix multiplication algorithm and show that
  it compares favorably in practice with the standard not-in-place
  variants.
\item We describe an implementation of the first in-place variant of
  Karatsuba polynomial multiplication and show that it has very close
  performance to that of the state-of-the-art library NTL.
\item We propose fast in-place algorithms for most of the basic linear
  algebra subroutines (triangular multiplication and solving, Gaussian
  elimination, inverse, nullspace, rank updates) through reductions to
  matrix multiplication.
\item We propose fast in-place algorithms for more structured
  matrices, \emph{i.e.}, Toeplitz-like matrix-vector multiplication.
\item We give here a full interpretation of Toeplitz matrix-vector
  operations in terms of polynomial or power series operations, in
  order to propose polynomial only versions of all our in-place
  modular polynomial routines.
\item We improve on the algorithms for fast in-place polynomial short
  products. In particular, we improve their complexity bounds and
  remove the need of a call-stack for this operation.
\end{itemize}

Our paper is organized as follows. We give our model for in-place computations
and recall classical in-place algorithms
in~\cref{ssec:inplace}. We then introduce our main notations
in~\cref{ssec:notations}
and
recall the classical (quadratic or cubic) algorithms from the perspective
of in-place accumulating computations in~\cref{ssec:classical}.
We then detail in~\cref{sec:linacc} our novel technique for in-place
accumulation.
With this technique and further optimizations, we can also
derive new fast and in-place algorithms for the accumulating
multiplication of matrices, \cref{sec:strassen}, and of polynomials,
\cref{sec:inpaccpol}.
Then, in~\cref{sec:fconv,sec:toeplitz}, we derive novel in-place algorithms for
circulant and Toeplitz matrices.
Finally, in~\cref{sec:reminplace} we present fast in-place algorithms
computing just the polynomial remainder, and for an accumulated
modular multiplication.

\subsection*{Acknowledgements}
We are grateful to the anonymous reviewers for their comments and
suggestions, which have helped improve the quality of our article.

\subsection{Computational model}\label{ssec:inplace}
Our computational model is an \emph{algebraic RAM}. Inputs
and outputs are arrays of ring elements. Ring elements are
assumed to require bounded space.
(For simplicity, our algorithms are described over a
finite field $\F$, unless otherwise stated.)
The machine is made of \emph{algebraic registers} that each contain
one ring element, and \emph{pointer registers} that each contain
one pointer, that is one integer. Atomic operations are ring
operations on the algebraic registers and basic pointer arithmetic.
We assume that the pointer registers are large enough to store the
length of the input/output arrays.

Both inputs and outputs have read/write permissions.
But algorithms are only allowed to modify their inputs
\textbf{if their inputs are restored to their initial
state} afterwards.
In this model, we call \emph{in-place} an algorithm using only
\textbf{the space of its inputs, its outputs, and at most $\bigO{1}$ extra
space}. For recursive algorithms, some space may be required to store the recursive
call stack. (This stack is only made of pointers and its size is bounded
by the recursion depth of the algorithms. In practice, it is managed by the compiler.)
Nonetheless, we call \emph{in-place} a recursive algorithm whose only
extra space is the call stack.
In our complexity summaries (\Cref{tab:kara,tab:fft}), we include the
size of the stack.

The main limitations of this model are for black-box inputs, or for
inputs whose representations share some data. It is also not suitable
for rings whose elements have unbounded length such as the ring of integers.
A model with read-only inputs would be more powerful, but mutable
inputs turn out to be necessary in our case.
In particular, the algorithms we describe are \emph{in-place
with accumulation}. The archetypical example is a multiply-accumulate
operation $a \pe b\times c$. For such an algorithm, the condition is
that $b$ and $c$ are restored to their initial states at the end of
the computation, while $a$ (which is also part of the input) is replaced
by $a+bc$.
As a variant, we describe \emph{over-place} algorithms, that
replace (parts of) the input by the output (e.g., $\vec{a}\gets{b}\cdot\vec{a}$).
Similarly, the whole input can be modified, provided that the
parts of the input that are not the output are restored afterwards.
In the following we signal by a ``\algorithmicreadonly'' tag the parts
of the input that the algorithm is not allowed to modify (the other
parts are modifiable as long as they are restored).
Note that in-place algorithms with accumulation are a special case of
over-place algorithms.
Our model is somewhat similar to catalytic machines and transparent
space~\cite{Buhrman:2014:STOC:catalytic}, but
using only the input and output as catalytic space. Also, we do preserve the
(not in-place) time complexity, up to a (quasi)-linear overhead.
We refer to~\cite{Buhrman:2014:STOC:catalytic,Roche:2009:spacetime,Giorgi:2019:issac:reductions}
for more details.
Another closely related model is Goldreich's global storage
model~\cite{Goldreich:2008:CCCP,Goldreich:2024:treeeval} that uses one global space for
inputs and outputs, and some local space as extra space for the computations.

\begin{remark}
    The vocabulary for algorithms that use no extra space is diverse and
    sometimes inconsistent between papers. We summarize our three main
    notions:
    \begin{itemize}
        \item An algorithm is \emph{in-place} if it uses $O(1)$ extra space,
            in addition to the input and output space. In this paper, we still
            call in-place recursive algorithms that require a call stack. It
            will be easily checked that these call stacks are made of $O(\log n)$
            pointers only where $n$ is the input size. Another term for in-place
            is \emph{constant-space}.
        \item An algorithm is \emph{over-place} if it replaces (part of) its input
            by the output. This does not assume anything on the extra space required
            for the algorithm. Some authors use \emph{in-place} for this notion.
        \item An \emph{accumulating} algorithm is a special case of an over-place
            algorithm, where the result of some computation is accumulated into part
            of the input. A typical example is a computation $a \pe b\times c$.
    \end{itemize}
    We draw the attention of the reader to the fact that ``in-place'' can therefore
    refer to either the first or the second concept in the literature.
\end{remark}

\subsection{Notations}\label{ssec:notations}

Polynomials over a finite field $\F$ are denoted by capital letters
$A$, $B$, \dots{} A degree-$n$ polynomial $A\in\F[X]$ has $(n+1)$
coefficients denoted $a_0,\ldots,a_n$. Given $A$, $B\in\F[X]$, $A\bdiv
B$ and $A\bmod B$ denote the quotient and remainder in the Euclidean
division of $A$ by $B$, respectively. In particular,
$A = (A\bdiv{B})\times B + (A\bmod B)$, with $\deg(A\bmod B) < \deg(B)$. Matrices
over $\F$ are also denoted by capital letters $A$, $B$, \dots{}
Entries of a matrix $A\in \F^{m\times n}$ are denoted $A_{ij}$ (or
$A_{i,j}$ for better readability), $1\le i\le m$, $1\le j\le n$.

In algorithms, a degree-$n$ polynomial $A$ is stored in a size-$(n+1)$ array of coefficients. By a slight abuse of notation, a polynomial is identified with its array representation and similarly the notation $a_i$ denotes both the coefficient of degree $i$ of $A$ and the $i$th cell of its array representation. This is similar for matrices. There is no real distinction between inputs and outputs: Every array manipulated by the algorithm is part of the input. Instead of an output, to each algorithm is associated a ``\algorithmicensure'' which indicates the array that has been modified. (Other arrays are implicitly assumed to be finally restored.)

We denote by $\M(n)$ a \emph{multiplication time}~\cite{MCA2013} for $\F[X]$ such that two degree-$n$ polynomials over $\F$ can be multiplied using at most $\M(n)$ operations in $\F$ by a \emph{bilinear algorithm}. The restriction to bilinear algorithms is benign since all known \emph{algebraic} algorithms are bilinear. As customary, we assume that $\M(n)/n$ is non-decreasing and $\M(mn) \le m^2\M(n)$. One can take $\M(n) = O(n\log n\log\log n)$~\cite{Cantor:1991:mulpoly}, \emph{cf.} also~\cite{Harvey:2022:JACM:nlogn} for more precise bounds.

Similarly, we denote by $\MM(m;k;n)$ the cost of a bilinear algorithm to multiply
an $m{\times}k$ matrix by a $k{\times}n$ one. We write $\MM(m)$ if $m=k=n$.
We also denote by $\omega>2$, the exponent of the dominant term
of this cost when the matrices are square.\footnote{If we let
  $\{n_1,n_2,n_3\}=\{m,k,n\}$ with $n_1\leq{n_2}\leq{n_3}$, then by
  cutting the matrices to the smallest dimension, it is
  straightforward to see that we therefore always have
  $\MM(m;k;n)\leq\lceil\frac{m}{n_1}\rceil\lceil\frac{k}{n_1}\rceil\lceil\frac{n}{n_1}\rceil\MM(n_1)=\bigO{n_1^{\omega-2}n_2n_3}$.} Strassen's algorithm allows taking $\omega = \log(7) \simeq 2.807$ \cite{Strassen:1969:GENO} while the most recent results show that one can take $\omega<2.371339$~\cite{Alman:2025:asymmetry}.

Finally, for a vector $\vec{c}\in\F^n$, we denote by $\cev c\in\F^n$ the \emph{reversed} vector defined as $(\cev c)_i = (\vec c)_{n-i+1}$. In algorithms, instructions such as $\cev c\pe \vec a$ means that the vector $\vec a$ is accumulated into $\vec c$ but in reversed order.

\subsection{Classical algorithms}\label{ssec:classical}

Classical algorithms for matrix and polynomial operations can be
performed in-place, without any call stack, as recalled
in~\cref{alg:classicmulpoly,alg:classicmulmat}.
\begin{algorithm}[htbp]\caption{Quadratic i-p. accumulating polynomial multiplication.}\label{alg:classicmulpoly}
\begin{algorithmic}[1]
\REQUIRE $A(X)$, $B(X)$, $C(X)\in\F[X]$ of respective degrees  $m$, $n$, $m+n$.
\READONLY $A$, $B$.
\ENSURE $C(X)\pe{A(X)B(X)}$
\FOR{$0\le i\le m$, $0\le j\le n$}
\STATE $c_{i{+}j}\pe{a_ib_j}$;
\ENDFOR
\end{algorithmic}
\end{algorithm}

\begin{algorithm}[htbp]\caption{Cubic i-p. accumulating matrix multiplication.}\label{alg:classicmulmat}
\begin{algorithmic}[1]
    \REQUIRE $A\in\F^{m\times\ell}$, $B\in\F^{\ell\times n}$, $C\in\F^{m\times n}$.
\READONLY $A$, $B$.
\ENSURE $C\pe{AB}$
\FOR{$1\le i\le m$, $1\le j\le n$, $1\le k\le\ell$}
\STATE $C_{ij}\pe{A_{ik}B_{kj}}$;
\ENDFOR
\end{algorithmic}
\end{algorithm}

Also classical, quadratic, algorithms for polynomial
remaindering and triangular matrix operations can be performed
in-place, as recalled in \cref{alg:classical,alg:quadrem}.

For any field $\F$ we have for instance the following
over-place algorithms for triangular matrix operations, given
in~\cref{alg:classical}.

\begin{algorithm}[htbp]\caption{Over-place quadratic triangular matrix
    operations\\ {\footnotesize (left: matrix-vector multiplication;
      right: triangular system solve)}}\label{alg:classical}
\begin{algorithmic}[1]
\REQUIRE $U\in\F^{m{\times}m}$ upper triangular and $\vec{v}\in\F^m$
\READONLY $U$
\end{algorithmic}
\begin{minipage}{.4\columnwidth}
\begin{algorithmic}[1]
\ENSURE $\vec{v}\leftarrow{U\cdot{}\vec{v}}$
\FOR{$i=1$ \To $m$}
\FOR{$j=1$ \To $i-1$}
\STATE $v_j \pe U_{ji} v_i$
\ENDFOR
\STATE $v_i\leftarrow{U_{ii}v_i}$;
\ENDFOR
\end{algorithmic}
\end{minipage}\hfill
\begin{minipage}{.53\columnwidth}
\begin{algorithmic}[1]
\ENSURE $\vec{v}\leftarrow{U^{-1}\cdot{}\vec{v}}$
\FOR{$i=m$ \DownTo $1$}
\FOR{$j=m$ \DownTo $i+1$}
\STATE $v_i \me U_{ij} v_j$
\ENDFOR
\STATE $v_i\leftarrow{U_{ii}^{-1}v_i}$;\hfill\COMMENT{if $U_{ii}\in\F^*$}
\ENDFOR
\end{algorithmic}
\end{minipage}
\smallskip
\end{algorithm}

The classical long-division algorithm provides a quadratic in-place algorithm
for computing the remainder of two polynomials without computing the quotient, see~\cref{alg:quadrem}.

\begin{algorithm}[htbp]\caption{In-place quadratic polynomial remainder.}\label{alg:quadrem}
\begin{algorithmic}[1]
\REQUIRE $A(X)$, $B(X)$, $R(X)$ in $\F[X]$, of respective degrees $N$, $M$ and $M-1$.
\READONLY $A(X)$, $B(X)$.
\ENSURE $R(X)=A(X)\mod{B(X)}$.
\STATE Let $\bar{B}=B\bmod X^{M}$ and
$n=\max(N-M,-1)$
\STATE $R\gets A\bdiv X^{n+1}$;
\FOR{$i=n$ \DownTo $0$}
\STATE $q\leftarrow{r_{M-1}\cdot{}b_M^{-1}}$;\hfill\COMMENT{leading coefficients of $R$ and $B$}
\STATE $R\gets a_i + X\cdot R\bmod X^M$;
\STATE $R \me{q\cdot{}\bar{B}}$;
\ENDFOR
\end{algorithmic}
\end{algorithm}

\Cref{alg:quadrem} can be made over-place of its input $A$ by
considering that $R$ is just a ``pointer'' to some position in
$A$ (with $q$ in $r_{M}$): This is
then close to the in-place quadratic version given
in~\cite{Monagan:1993:disco}.

\section{In-place linear accumulation}\label{sec:linacc}
Karatsuba polynomial
multiplication~\cite{Karatsuba:1963:multiplication}
and Strassen matrix multiplication~\cite{Strassen:1969:GENO}
are famous optimizations of bilinear formulae on their inputs: Results
are linear combinations of products of bilinear combinations of the inputs.
To compute recursively such a formula in-place, we perform
each product one at a time. For each product, both factors are then
linearly combined in-place into one of the inputs beforehand and
restored afterwards. The product of both entries is at that point
accumulated in one part of the output and then distributed to the
other parts.
The difficulty is to perform this distribution in-place, {\em without
recomputing the product}. Our idea is to pre-subtract one output from the
other, then accumulate the product to one output, and finally re-add the
newly accumulated output to the other one: Overall both outputs just
have accumulated the product, in-place. Potential constant factors can
also be dealt with pre-divisions and post-multiplications.
Basically we need two kinds of in-place operations, and their
combinations.
First, as shown in~\cref{eq:basemul}, an in-place accumulation of a
quantity multiplied by a (known in advance) invertible constant:
\begin{equation}\label{eq:basemul}
\left\lbrace{}c\de\mu;~c\pe m;~c\fe\mu;\right\rbrace~\text{computes
  in-place}~c\gets{c+\mu\cdot{m}}.
\end{equation}
Second, as shown in~\cref{eq:basedist}, an in-place distribution of
the same quantity, without recomputation, to several outputs:
\begin{equation}\label{eq:basedist}
\left\lbrace{}d\me{c};~c\pe{m};~d\pe{c};\right\rbrace~\text{computes
  in-place}~\left\{\begin{array}{@{}l@{\,}l@{}}
c&\gets{c+m};\\
d&\gets{d+m}.\\
\end{array}
\right.
\end{equation}

\Cref{ex:bilin} shows how to combine several of these operations,
while also linearly combining parts of the input.
\begin{example}\label{ex:bilin}
  Suppose that for some inputs/outputs $a$, $b$, $c$, $d$, $r$, $s$, one wants to compute an
  intermediate product $p=(a+3b)*(c+d)$ only once and then distribute
  and accumulate that product to two of its outputs (or results),
  such that we have both $r\gets{r+5p}$ and $s\gets{s+2p}$.
  To perform this in-place, first accumulate $a\pe{3b}$ and $c\pe{d}$,
  then pre-divide $r$ by $5$, as in~\cref{eq:basemul}.
  Now we directly have $p=ac$ that can be computed once,
  and then accumulated to $r$, and to $s$, if the latter is prepared:
  divide it by $2$, and pre-subtract $r$ or, equivalently,
  pre-subtract $2r$. This is
  $s\me{2r}$ followed by $r\pe{ac}$. After this, we can
  reciprocate (or unroll) the precomputations: This distributes
  the product to the other result and restores the read-only inputs to
  their initial state.
  This is summarized as follows:
\begin{center}
  \fbox{\renewcommand{\baselinestretch}{1}
      \ensuremath{\ \left\lbrace\begin{aligned}
	    a\pe{3b}; &\quad c\pe{d}; & r\de{5}\phantom{r};\\[-5pt]
	    s\me{2r}; &\quad r\pe{ac}; & s\pe{2r}; \\[-5pt]
	     a\me{3b};&\quad c\me{d}; &r\fe{5}\phantom{r};
	  \end{aligned}\right\rbrace
	\begin{array}{l}
	  \text{computes in-place:}\\[-0pt]
	  \left\lbrace\begin{aligned}
	    r& \gets{r+5(a+3b)(c+d)};\\[-5pt]
	    s& \gets{s+2(a+3b)(c+d)}.\\[-1pt]
	  \end{aligned}\right.
	\end{array}
      }}
\end{center}
\end{example}

\Cref{alg:bilin} shows how to implement this in general, taking into
account the constant (or read-only) multiplicative coefficients of all
the linear combinations. We suppose that inputs are in three distinct
sets: left-hand sides, $\vec{a}$, right-hand sides, $\vec{b}$, and those
accumulated to the results, $\vec{c}$.
We denote by $\odot$ the point-wise multiplications of
left-hand sides by right-hand sides (their Hadamard product).
Then~\cref{alg:bilin} computes $\vec{c}\pe\mat{\mu}\vec{m}$, for
$\vec{m}=(\mat{\alpha}\vec{a})\odot(\mat{\beta}\vec{b})$, with
$\mat{\alpha}$, $\mat{\beta}$ and $\mat{\mu}$ matrices of
constants.
These matrices define the \textsc{hm} \emph{representation} of a
bilinear algorithm, as in~\cite{jgd:2024:bilin,jgd:2024:accurate} and
references therein. They can also be denoted by
$\mat{L}=\mat{\alpha}$, $\mat{R}=\mat{\beta}$, $\mat{P}=\mat{\mu}$ as
  they act respectively on the left-hand side, right-hand side and
  post (or product) side.
\begin{equation}\label{eq:LRP}
\vec{c}\pe \mat{P}\cdot(\mat{L}\cdot\vec{a})\odot(\mat{R}\cdot\vec{b})
\end{equation}

\begin{algorithm}[!ht]
  \caption{In-place bilinear formula.}\label{alg:bilin}
  \begin{algorithmic}[1]
    \REQUIRE $\vec{a}\in\F^m$, $\vec{b}\in\F^n$, $\vec{c}\in\F^s$;
    $\mat{\alpha}\in\F^{t{\times}m}$, $\mat{\beta}\in\F^{t{\times}n}$,
    $\mat{\mu}\in\F^{s{\times}t}$.
    \READONLY$\mat{\alpha}$, $\mat{\beta}$, $\mat{\mu}$ (all $3$ without zero-rows).
    \ENSURE $\vec{c}\pe\mat{\mu}\vec{m}$, for
    $\vec{m}=(\mat{\alpha}\vec{a})\odot(\mat{\beta}\vec{b})$.
    \FOR{$\ell=1$ \To $t$}
    \STATE\label{lin:alpha}Find one~$i$ s.t. $\alpha_{\ell,i}\neq{0}$;
    $a_i\fe\alpha_{\ell,i}$;
    \ForDoEnd[lin:foralpha]{$\lambda=1$ \To $m$, $\lambda\neq{i}$,
      $\alpha_{\ell,\lambda}\neq{0}$ }{ $a_i\pe\alpha_{\ell,\lambda}a_\lambda$}
    \STATE\label{lin:beta}Find one~$j$ s.t. $\beta_{\ell,j}\neq{0}$;
    $b_j\fe\beta_{\ell,j}$;
    \ForDoEnd[lin:forbeta]{$\lambda=1$ \To $n$, $\lambda\neq{j}$,
      $\beta_{\ell,\lambda}\neq{0}$}{ $b_j\pe\beta_{\ell,\lambda}b_\lambda$}
    \STATE\label{lin:mu}Find one~$k$ s.t. $\mu_{k,\ell}\neq{0}$;
    $c_k\de\mu_{k,\ell}$;
    \ForDoEnd[lin:formu]{$\lambda=1$ \To $s$, $\lambda\neq{k}$,
      $\mu_{\lambda,\ell}\neq{0}$}{$c_\lambda\me\mu_{\lambda,\ell}{c_k}$}
    \STATE\label{lin:product}$c_k\pe{a_i\cdot{b_j}}$\hfill\COMMENT{this is the product $m_\ell$, computed only once}
    \ForDoEnd[lin:distribmu]{$\lambda=1$ \To $s$, $\lambda\neq{k}$,
      $\mu_{\lambda,\ell}\neq{0}$}{$c_\lambda\pe\mu_{\lambda,\ell}{c_k}$}
    \hfill\COMMENT{undo~\ref{lin:formu}}
    \STATE\label{lin:fmuk}$c_k\fe\mu_{k,\ell}$;\hfill\COMMENT{undo~\ref{lin:mu}}
    \ForDoEnd[lin:distribbeta]{$\lambda=1$ \To $n$, $\lambda\neq{j}$,
      $\beta_{\ell,\lambda}\neq{0}$}{$b_j\me\beta_{\ell,\lambda}b_\lambda$}
    \hfill\COMMENT{undo~\ref{lin:forbeta}}
    \STATE\label{lin:dbetaj}$b_j\de\beta_{\ell,j}$;\hfill\COMMENT{undo~\ref{lin:beta}}
    \ForDoEnd[lin:distribalpha]{$\lambda=1$ \To $m$, $\lambda\neq{i}$,
      $\alpha_{\ell,\lambda}\neq{0}$}{$a_i\me\alpha_{\ell,\lambda}a_\lambda$}
    \hfill\COMMENT{undo~\ref{lin:foralpha}}
    \STATE\label{lin:dalphai}$a_i\de\alpha_{\ell,i}$;\hfill\COMMENT{undo~\ref{lin:alpha}}
    \ENDFOR
  \end{algorithmic}
\end{algorithm}

\begin{remark}\label{rk:parallelism}
  \Cref{lin:alpha,lin:foralpha,lin:beta,lin:forbeta,lin:mu,lin:formu,lin:distribmu,lin:fmuk,lin:distribbeta,lin:dbetaj,lin:distribalpha,lin:dalphai}   of~\cref{alg:bilin} are acting on independent parts of the
  input, $\vec{a}$ and $\vec{b}$, and of the output $\vec{c}$.
  If needed they could therefore be computed
  in parallel
  or in different orders,
  and even potentially grouped or factorized across the main loop (on
  $\ell$).
A C++ implementation of~\cref{alg:bilin}
(\href{https://github.com/jgdumas/plinopt/blob/main/trilplacer.cpp}{\texttt{trilplacer}})
is available in the
\plinopt~library~\cite{Dumas:2024:plinopt}.
\end{remark}

To simplify the counting of operations,
we denote by \ADD both the addition or subtraction of elements, $\pe$
or $\me$; by \MUL the (tensor) product of elements, $\odot$;
and by \SCA the scaling by constants, $\fe$ or $\de$.
We also denote by $\#x$ (resp. $\sharp{x}$) the number of
non-zero (resp. $\not\in\{0,1,-1\}$) elements in a matrix $x$.

\begin{theorem}\label{thm:bilin}
\Cref{alg:bilin} is correct, in-place, and requires
$t$ \MUL,
$2(\#\alpha+\#\beta+\#\mu)-5t$ \ADD and
$2(\sharp\alpha+\sharp\beta+\sharp\mu)$ \SCA operations.
\end{theorem}
\begin{proof}
First, as the only used operations ($\pe$, $\me$, $\fe$, $\de$) are
in-place ones, the algorithm is in-place.
Second, the algorithm is correct both for the input and the
output:
the input is well restored, as
$(\alpha_{\ell,i}a_i+\sum\alpha_{\ell,\lambda}a_\lambda-\sum\alpha_{\ell,\lambda}a_{\lambda})/\alpha_{\ell,i}=a_i$
and
$(\beta_{\ell,j}b_j+\sum\beta_{\ell,\lambda}b_\lambda-\sum\beta_{\ell,\lambda}b_\lambda)/\beta_{\ell,j}=b_j$;
the output is correct as
$c_\lambda-\mu_{\lambda,\ell}c_k/\mu_{k,\ell}+\mu_{\lambda,\ell}(c_k/\mu_{k,\ell}+a_ib_j)=c_\lambda+\mu_{\lambda,\ell}a_ib_j$
and
$(c_k/\mu_{k,\ell}+a_ib_j)\mu_{k,\ell}=c_k+\mu_{k,\ell}a_ib_j$.
Third, for the number of operations,
\cref{lin:alpha,lin:foralpha} require one multiplication by a constant for each
non-zero element $a_{\lambda}$ in the row and one less addition.
But multiplications and divisions by $1$ are no-op, and by $-1$ can be
dealt with subtraction. This is $\#\alpha-t$ additions
and $\sharp\alpha$ constant multiplications.
\cref{lin:beta,lin:forbeta} (resp. \cref{lin:mu,lin:formu}) are
similar for each non-zero element in $b_{\lambda}$ (resp. in
$\mu$). Finally, \cref{lin:product} performs $t$ multiplications of
elements and $t$ additions. The remaining lines double the number of
\ADD and \SCA.
This is $t+2(\#\alpha+\#\beta+\#\mu-3t)=2(\#\alpha+\#\beta+\#\mu)-5t$ \ADD.
\end{proof}

\begin{remark} Similarly, slightly more generic accumulation
  operations of the form
  $\vec{c}\gets\vec{\gamma}\odot\vec{c}+\mat{\mu}\vec{m}$, for a
  vector $\gamma\in\F^{s}$, can also be computed in-place: Precompute
  first $\vec{c}\gets\vec{\gamma}\odot\vec{c}$, then
  call~\cref{alg:bilin}.
\end{remark}

For instance, to use~\cref{alg:bilin} with matrices or polynomials,
each product $m_\ell$ is in fact computed recursively.
Further, in an actual implementation of a fixed formula, one can
combine more efficiently the pre- and post-computations over
the main loop on $\ell$, as in~\cref{rk:parallelism}.
See~\cref{sec:strassen,sec:inpaccpol} for examples of recursive
calls, together with sequential optimizations and combinations.

In fact the method for accumulation, computing each bilinear
multiplication once is generalizable.
With the notations of~\cref{alg:bilin}, any algorithm of the form
$\vec{c}\pe\mat{\mu}\vec{m}$ can benefit from this technique,
provided that each $m_j$ can be obtained from a function
that can be computed in-place.
Let $F_j:\Omega\to\F$ be such a function on some inputs from a
space $\Omega$, for which an in-place algorithm exists.
Then we can accumulate it in-place, \emph{if it satisfies the following
constraint}: That it is not using its output space as an available
intermediary memory location.
Further, this function can be in-place in different models:
it can follow our model of~\cref{ssec:inplace}, if there is a way to
put its input back into their initial states, or some other model,
again provided that it follows the above constraint.
Then, the idea is just to keep from~\cref{alg:bilin}
the~\cref{lin:mu,lin:formu,lin:product,lin:distribmu,lin:fmuk},
replacing~\cref{lin:product} by the in-place call to $F_j$,
potentially surrounding that call by manipulations on the inputs of
$F_j$ (just like the one performed on $\vec{a}$ and $\vec{b}$
in~\cref{alg:bilin}).
We give examples of the application of the generalized method
of~\cref{thm:general} to non-bilinear formulae in~\cref{ssec:aat},
and we can thus show that:
\begin{theorem}\label{thm:general}
Let $\vec{c}\in\F^s$ and $\mat{\mu}\in\F^{s{\times}t}$, without zero-rows.
Let $\vec{F}=(F_j:\Omega\to\F)_{j=1..t}$ be a collection of functions
and $\omega\in\Omega$.
If all these functions are computable in-place, without using their
output space as an
intermediary memory location,
then there exists an in-place algorithm computing
$\vec{c}\pe\mat{\mu}\vec{F}(\omega)$ in-place, requiring a single call to
each $F_j$,
together with
$(2\#\mu-t)$ \ADD and
$2\sharp\mu$ \SCA ops.
\end{theorem}

\section{In-place Strassen matrix multiplication with accumulation}\label{sec:strassen}

Considered as~${\matrixsize{2}{2}}$ matrices, the matrix
product with accumulation ~${\mat{C}\pe\MatrixProduct{A}{B}}$ could be computed using
Strassen-Winograd (S.-W.) algorithm by performing the following computations:
\begin{gather}
\begin{array}{ll}
\mathcolor{\triadone}{\rho_{1}}\gets{\mathcolor{\triadone}{a_{11}}\mathcolor{\triadone}{b_{11}}},
\quad
\mathcolor{\triadthree}{\rho_{3}}\gets{(\mathcolor{\triadthree}{-a_{11}-a_{12}+a_{21}+a_{22}})\mathcolor{\triadthree}{b_{22}}},
\\
\mathcolor{\triadtwo}{\rho_{2}}\gets{\mathcolor{\triadtwo}{a_{12}}\mathcolor{\triadtwo}{b_{21}}},
\quad
\mathcolor{\triadfour}{\rho_{4}}\gets{\mathcolor{\triadfour}{a_{22}}(\mathcolor{\triadfour}{-b_{11}+b_{12}+b_{21}-b_{22}})},
\\
\mathcolor{\triadfive}{\rho_{5}}\gets{(\mathcolor{\triadfive}{a_{21}+a_{22}})(\mathcolor{\triadfive}{-b_{11}+b_{12}})},
\quad
\mathcolor{\triadsix}{\rho_{6}}\gets{(\mathcolor{\triadsix}{-a_{11}+a_{21}})(\mathcolor{\triadsix}{b_{12}-b_{22}})},
\\
\mathcolor{\triadseven}{\rho_{7}}\gets{(\mathcolor{\triadseven}{-a_{11}+a_{21}+a_{22}})(\mathcolor{\triadseven}{-b_{11}+b_{12}-b_{22}})},
\end{array}\nonumber\\
\label{eq:StrassenWinogradMultiplicationAlgorithm}
\begin{bmatrix} c_{11} &c_{12} \\ c_{21} &c_{22} \end{bmatrix}
\pe
\begin{bmatrix}
\mathcolor{\triadone}{\rho_{1}} + \mathcolor{\triadtwo}{\rho_{2}} &
\mathcolor{\triadone}{\rho_{1}} - \mathcolor{\triadthree}{\rho_{3}} + \mathcolor{\triadfive}{\rho_{5}} - \mathcolor{\triadseven}{\rho_{7}}\\
\mathcolor{\triadone}{\rho_{1}} + \mathcolor{\triadfour}{\rho_{4}} +
\mathcolor{\triadsix}{\rho_{6}} - \mathcolor{\triadseven}{\rho_{7}} &
\mathcolor{\triadone}{\rho_{1}}+\mathcolor{\triadfive}{\rho_{5}} +
\mathcolor{\triadsix}{\rho_{6}} - \mathcolor{\triadseven}{\rho_{7}}
\end{bmatrix}.
\end{gather}
This algorithm uses $7$ multiplications of half-size matrices and $24+4$
additions (that can be factored into only $15+4$~\cite{Winograd:1977:complexite}:
$4$ involving $A$, $4$ involving $B$ and $7$ involving the products,
plus $4$ for the accumulation).
This can be used recursively on matrix blocks, halved at each
iteration, to obtain a sub-cubic algorithm. To save on operations, it
is of course interesting to compute the products only once, that is
store them in extra memory chunks.
To date, up to our knowledge, the best versions that reduced this
extra memory space (also overwriting the
input matrices but not putting them back in place) were proposed
in~\cite{jgd:2009:WinoSchedule}:
their best sub-cubic accumulating product used $2$ temporary blocks
per recursive level, thus a total of extra memory required to
be~$\frac{2}{3}n^2$.

\subsection{In-place accumulating matrix multiplication with 7
  recursive calls and 18 additions}\label{app:inplsw}

With~\cref{alg:bilin} we instead obtain an in-place sub-cubic algorithm for
accumulating matrix multiplication, with $\bigO{1}$ extra temporary field element.
From~\cref{eq:StrassenWinogradMultiplicationAlgorithm} indeed (see
also the representation
in~\cite{Hopcroft:1973:duality,Bshouty:1995:minwinoadd}),
we can extract
the \textsc{hm} matrices
\begin{equation}\label{eq:alphabetamu}\setlength\arraycolsep{3pt}
\mu=\begin{bmatrix}
\mathcolor{\triadone}{1}&\mathcolor{\triadtwo}{1}&\mathcolor{\triadthree}{0}&\mathcolor{\triadfour}{0}&\mathcolor{\triadfive}{0}&\mathcolor{\triadsix}{0}&\mathcolor{\triadseven}{0}\\
\mathcolor{\triadone}{1}&\mathcolor{\triadtwo}{0}&\mathcolor{\triadthree}{-1}&\mathcolor{\triadfour}{0}&\mathcolor{\triadfive}{1}&\mathcolor{\triadsix}{0}&\mathcolor{\triadseven}{-1}\\
\mathcolor{\triadone}{1}&\mathcolor{\triadtwo}{0}&\mathcolor{\triadthree}{0}&\mathcolor{\triadfour}{1}&\mathcolor{\triadfive}{0}&\mathcolor{\triadsix}{1}&\mathcolor{\triadseven}{-1}\\
\mathcolor{\triadone}{1}&\mathcolor{\triadtwo}{0}&\mathcolor{\triadthree}{0}&\mathcolor{\triadfour}{0}&\mathcolor{\triadfive}{1}&\mathcolor{\triadsix}{1}&\mathcolor{\triadseven}{-1}
\end{bmatrix}\quad
\alpha=\begin{bmatrix}
\mathcolor{\triadone}{1}&\mathcolor{\triadone}{0}&\mathcolor{\triadone}{0}&\mathcolor{\triadone}{0}\\
\mathcolor{\triadtwo}{0}&\mathcolor{\triadtwo}{1}&\mathcolor{\triadtwo}{0}&\mathcolor{\triadtwo}{0}\\
\mathcolor{\triadthree}{-1}&\mathcolor{\triadthree}{-1}&\mathcolor{\triadthree}{1}&\mathcolor{\triadthree}{1}\\
\mathcolor{\triadfour}{0}&\mathcolor{\triadfour}{0}&\mathcolor{\triadfour}{0}&\mathcolor{\triadfour}{1}\\
\mathcolor{\triadfive}{0}&\mathcolor{\triadfive}{0}&\mathcolor{\triadfive}{1}&\mathcolor{\triadfive}{1}\\
\mathcolor{\triadsix}{-1}&\mathcolor{\triadsix}{0}&\mathcolor{\triadsix}{1}&\mathcolor{\triadsix}{0}\\
\mathcolor{\triadseven}{-1}&\mathcolor{\triadseven}{0}&\mathcolor{\triadseven}{1}&\mathcolor{\triadseven}{1}
\end{bmatrix}\quad
\beta=\begin{bmatrix}
\mathcolor{\triadone}{1}&\mathcolor{\triadone}{0}&\mathcolor{\triadone}{0}&\mathcolor{\triadone}{0}\\
\mathcolor{\triadtwo}{0}&\mathcolor{\triadtwo}{0}&\mathcolor{\triadtwo}{1}&\mathcolor{\triadtwo}{0}\\
\mathcolor{\triadthree}{0}&\mathcolor{\triadthree}{0}&\mathcolor{\triadthree}{0}&\mathcolor{\triadthree}{1}\\
\mathcolor{\triadfour}{-1}&\mathcolor{\triadfour}{1}&\mathcolor{\triadfour}{1}&\mathcolor{\triadfour}{-1}\\
\mathcolor{\triadfive}{-1}&\mathcolor{\triadfive}{1}&\mathcolor{\triadfive}{0}&\mathcolor{\triadfive}{0}\\
\mathcolor{\triadsix}{0}&\mathcolor{\triadsix}{1}&\mathcolor{\triadsix}{0}&\mathcolor{\triadsix}{-1}\\
\mathcolor{\triadseven}{-1}&\mathcolor{\triadseven}{1}&\mathcolor{\triadseven}{0}&\mathcolor{\triadseven}{-1}
\end{bmatrix}
\end{equation}

All coefficients being $1$ or $-1$ the resulting in-place algorithm
can compute the accumulation $C\pe{AB}$ without constant
multiplications.
It thus requires $7$ recursive calls and, from~\cref{thm:bilin},
at most $2(\#\alpha+\#\beta+\#\mu-3t)=2(14+14+14-3*7)=42$ block
additions.
Just like the $24$ additions
of~\cref{eq:StrassenWinogradMultiplicationAlgorithm} can be factored
into $15$, one can also optimize the in-place algorithm.
For instance, looking at $\alpha$ we see that performing the products
in the order $\rho_{6}$, $\rho_{7}$, $\rho_{3}$, $\rho_{5}$ and
accumulating in $a_{21}$ enables to perform all additions/sub\-trac\-tions
in $A$ with only $6$ operations (this is in fact optimal, see~\cref{prop:six}).
This is similar for $\beta$ if the order
$\rho_{6}$, $\rho_{7}$, $\rho_{4}$, $\rho_{5}$ is used and
accumulation is in $b_{12}$.
Thus ordering for instance $\rho_{6}$, $\rho_{7}$, $\rho_{4}$,
$\rho_{3}$, $\rho_{5}$ will reduce the number of block additions to $26$.
Now looking at $\mu$ (more precisely at its transpose,
see~\cite{Kaminski:1988:transpose}), a similar reduction can be
obtained, e.g., if one of the orders
($\rho_{6}$, $\rho_{7}$, $\rho_{1}$, $\rho_{5}$)
or
($\rho_{5}$, $\rho_{7}$, $\rho_{1}$, $\rho_{6}$) is used
and accumulation is in $c_{22}$.

Therefore, using the ordering
$\rho_{6},\rho_{7},\rho_{1},\rho_{4},\rho_{3},\rho_{5},\rho_{2}$
requires only $18$ additions (plus $7$ accumulations in $C$).
This is shown in~\cref{alg:ipsw}, which can be obtained (after
potentially trying several random optimizations to get the optimal
cost) via the \plinopt~library with:
\begin{verbatim}
./bin/trilplacer data/2x2x2_7_Winograd_{L,R,P}.sms
\end{verbatim}

This strategy enables us to reduce the
number of additions obtained when calling~\cref{alg:bilin}, from $42+7$
to $18+7$: Mostly remove successive additions or subtractions that are
reciprocal on either sub-matrices. This optimized version is given
in~\cref{alg:ipsw} and reaches the minimal possible number of
extra additions/sub\-trac\-tions, as shown in~\cref{thm:eighteen}.

\begin{algorithm}[!ht]
\caption{In-place accumulating S.-W. matrix multiplication.}\label{alg:ipsw}
    \begin{algorithmic}[1]
\REQUIRE
$A=\begin{bmatrix} A_{11} & A_{12} \\  A_{21} &A_{22}\end{bmatrix}$,
$B=\begin{bmatrix} B_{11} & B_{12} \\  B_{21} &B_{22}\end{bmatrix}$,
$C=\begin{bmatrix} C_{11} & C_{12} \\  C_{21} &C_{22}\end{bmatrix}$.
\ENSURE $C\pe{AB}$.
\STATE $A_{21} \me A_{11}$; $B_{12} \me B_{22}$; $C_{21} \me C_{22}$;
\STATE \colorbox{cyan!10}{$C_{22} \pe A_{21} * B_{12}$;}
\STATE $A_{21} \pe A_{22}$; $B_{12} \me B_{11}$; $C_{12} \me C_{22}$;
\STATE \colorbox{cyan!10}{$C_{22} \me A_{21} * B_{12}$;}
\STATE $C_{11} \me C_{22}$;
\STATE \colorbox{cyan!10}{$C_{22} \pe A_{11} * B_{11}$;}
\STATE $C_{11} \pe C_{22}$; $B_{12} \pe B_{21}$; $C_{21} \pe C_{22}$;
\STATE \colorbox{cyan!10}{$C_{21} \pe A_{22} * B_{12}$;}
\STATE $B_{12} \pe B_{22}$; $B_{12} \me B_{21}$; $A_{21} \me A_{12}$;
\STATE \colorbox{cyan!10}{$C_{12} \me A_{21} * B_{22}$;}
\STATE $A_{21} \pe A_{12}$; $A_{21} \pe A_{11}$;
\STATE \colorbox{cyan!10}{$C_{22} \pe A_{21} * B_{12}$;}
\STATE $C_{12} \pe C_{22}$; $B_{12} \pe B_{11}$; $A_{21} \me A_{22}$;
\STATE \colorbox{cyan!10}{$C_{11} \pe A_{12} * B_{21}$;}
\end{algorithmic}
\end{algorithm}

Any bilinear algorithm for matrix multiplication (see, e.g.,
\url{https://fmm.univ-lille.fr/}) can in fact be dealt with similarly.

The obtained number of temporary blocks of dimensions
$\frac{n}{2}\times\frac{n}{2}$ required for the computation
in~\cref{alg:ipsw} is compared to that of
previously known accumulating algorithms in~\cref{tab:inpsw}.
\begin{table}[!ht]\centering
\caption{Reduced-memory accumulating S.-W. multiplication.}\label{tab:inpsw}
\begin{tabular}{lccc}
\toprule
Alg. & Temp. blocks & inputs & accumulation \\
\midrule
\cite{HussLederman:1996:ISA} & $3$ & {\color{teal} read-only} & {\xyes}\\
\cite{jgd:2009:WinoSchedule} & $2$ & {\color{teal} read-only} & {\xyes}\\
\cref{alg:ipsw} & $0$ & mutable & {\xyes}\\
\bottomrule
\end{tabular}
\end{table}

Now for the number of operations and practical efficiency.
Using $18$ extra additions, without thresholds and for powers of two,
the dominant term of the overall arithmetic cost is $8n^{\log_2(7)}$,
for the in-place version.
This is roughly a third more operations than the $6n^{\log_2(7)}$
dominant term of the cost for the version using extra temporaries.

\begin{figure}[!ht]
\caption{Optimal In-place accumulating Strassen-Winograd MM.}\label{fig:fflas}
\includegraphics[width=.9\textwidth]{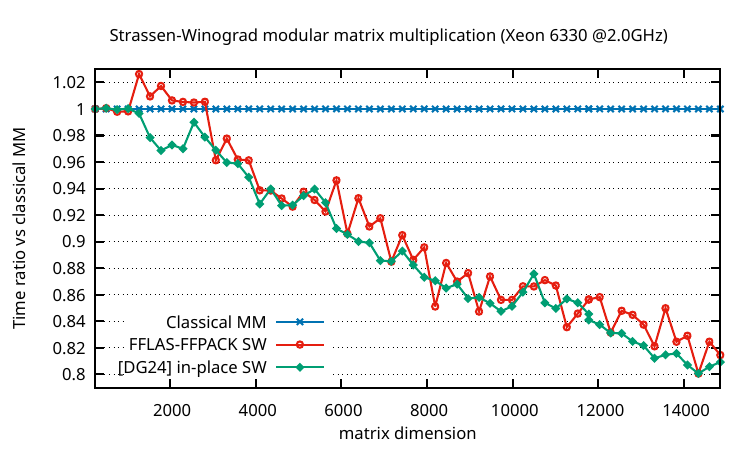}
\end{figure}

But it turns out that in practice, there is no penalty of using these
extra operations to reduce memory, in terms of performance:
\Cref{fig:fflas} compares the time of the reference
\href{https://github.com/linbox-team/fflas-ffpack}{FFLAS}\footnote{\url{http://linbox-team.github.io/fflas-ffpack}.}
Strassen-Winograd matrix multiplication to that of~\cref{alg:ipsw}.
The multiplications are performed modulo $131071$ on a single core of
a Xeon 6330 CPU \symbol{64}2.00GHz.
The reference implementation first computes the product $T=AB$ in a
temporary block $T$ using $2$ extra half-dimension blocks per
recursive call via~\cite{Douglas:1994:Gemmw}, and then accumulates $C\pe{T}$.
This is a total of $n^2+\sum_{i=1}^{\log_2(n)}
2\left(\frac{n}{2^i}\right)^2=\frac{5}{3}n^2$ extra memory usage.
\Cref{fig:fflas} shows that using extra memory does not provide a
faster routine: \Cref{alg:ipsw} indeed shows the same kind of
acceleration with respect to the conventional algorithm and seems even
less sensitive to particularities of the dimensions of the matrices.

\subsection{In-place additive complexity}\label{ssec:minimal}
We now prove that $18$ additions is the minimal number of additions
required by an in-place algorithm resulting
from any bilinear algorithm for matrix multiplication using only $7$
multiplications.
For this we consider elementary operations on
variables (similar to elementary linear algebra operators):
\emph{variable-switching} (swapping variable $i$ and variable $j$);
\emph{variable-multiplying} (multiplying a variable by a constant);
\emph{variable-addition} (adding one variable, potentially multiplied
by a constant, to another variable).
An \emph{elementary program} is a program using only these
three kind of operations.
Now, the in-place implementation of a linear function on
its input, for $\mat{\alpha}\in\F^{t{\times}m}$ and $\vec{a}\in\F^m$,
is the computation of each of the $t$ coefficients of
$\mat{\alpha}\vec{a}$, using only elementary operations and
only the variables of $\vec{a}$ as temporary variables.
We start by proving in~\cref{lem:nocannorzero}
that in any bilinear algorithm for matrix
multiplication using only $7$ multiplications, the columns of the
associated matrices $\mat{\alpha},\mat{\beta},\mat{\mu}$
(as in~\cref{eq:alphabetamu})
cannot contain too many zeroes.
\begin{lemma}\label{lem:nocannorzero}
If
$(\mat{\alpha},\mat{\beta},\mat{\mu})\in\F^{7{\times}4}\times\F^{7{\times}4}\times\F^{4{\times}7}$
is the \emph{\textsc{hm}} representation of a bilinear algorithm for matrix
multiplication, then none of $\mat{\alpha},\mat{\beta},\Transpose{\mat{\mu}}$
contains a zero column vector, nor a multiple of a standard basis vector.
\end{lemma}
\begin{proof}
The dimensions of the matrices indicate that the multiplicative
complexity of the algorithm is $7$.
From~\cite{Groote:1978:optimal} we know that all such bilinear
algorithms can be obtained from one another.
Following~\cite[Lemma~6]{Bshouty:1995:minwinoadd}, then any associated
$\mat{\alpha},\mat{\beta},\Transpose{\mat{\mu}}$ matrix is some row or column
permutation, or the multiplication by some $G\otimes{H}$ (the Kronecker
product of two invertible $2{\times}2$ matrices), of the matrices
of~\cref{eq:alphabetamu}.
By duality~\cite{Hopcroft:1973:duality}, see
also~\cite[Eq. (3)]{Bshouty:1995:minwinoadd},
it is also sufficient to consider any one of the $3$ matrices.
We thus let
$K=G\otimes{H}$.
Then any column of $K$ is of the form
$\Transpose{\begin{bmatrix} ux , uy , vx , vy\end{bmatrix}}$,
where $\begin{smatrix} u\\v\end{smatrix}$ is a column of $G$ and
$\begin{smatrix}x\\y\end{smatrix}$ is a column of $H$.
Further as $G$ is invertible,
$u$ and $v$
cannot be both zero simultaneously
and, similarly,
$x$ and $y$
cannot be both zero simultaneously.
Now consider for instance
$\mat{\alpha}\cdot{K}$, with
$\mat{\alpha}$ of~\cref{eq:alphabetamu}.
Then any column $\vec{\theta}$ of $\mat{\alpha}\cdot{K}$
is of the form:
\[\gTranspose{1.5pt}{\begin{bmatrix}ux, uy, -ux-uy+vx+vy, vy, vx+vy, -ux+vx,-ux+vx+vy\end{bmatrix}}.\]
For such a column to be a multiple of a standard basis vector
or the zero vector, at least $6$ of its $7$ coefficients must be zero.
For instance, this means that at least two out of rows $1$, $2$ and $4$
must be zero: or that at least two of $ux$, $uy$ or $vy$ must be
zero. This limits us to three cases: (1) $u=0$, (2) $y=0$ or (3) $x=v=0$.
If $u=0$, then
$\vec{\theta}=v\gTranspose{1pt}{\begin{bmatrix}0,0,x+y,y,x+y,x,x+y\end{bmatrix}}$;
at least one of rows $4$ or $6$ has to be zero, thus,
w.l.o.g. suppose $x=0$, we obtain that
$\vec{\theta}=vy\gTranspose{1pt}{\begin{bmatrix}0,0,1,1,1,0,1\end{bmatrix}}$
with none of $v$ nor $y$ being zero (otherwise $G$ or $H$ is not
invertible);
such a column cannot be a multiple of a standard basis vector
nor the zero vector.
Similarly, if $y=0$, then
$\vec{\theta}=x\gTranspose{1pt}{\begin{bmatrix}u,0,-u+v,0,v,-u+v,-u+v\end{bmatrix}}$;
at least one of rows $1$ or $5$ has to be zero, thus,
w.l.o.g. we can suppose that $v=0$. We then obtain that
$\vec{\theta}=ux\gTranspose{1pt}{\begin{bmatrix}1,0,-1,0,0,-1,-1\end{bmatrix}}$;
such a column cannot be a multiple of a standard basis vector
nor the zero vector.
Finally, if $x=v=0$, then
$\vec{\theta}=uy\gTranspose{1pt}{\begin{bmatrix}0,1,-1,0,0,0,0\end{bmatrix}}$;
again that column cannot be a multiple of a standard basis vector
nor the zero vector.
\end{proof}
Now we show that any in-place elementary algorithm requires at least $1$
extra operation to put back the input in its initial state.
\begin{lemma}\label{lem:plusone}
Let $\vec{a}\in\F^m$ and $\mat{\alpha}\in\F^{t{\times}m}$ with at
least one row which is neither the zero row, nor a standard basis vector.
Now suppose that, without any constraints in terms of temporary
registers, $k$ is the minimal number of elementary operations required
to compute $\mat{\alpha}\vec{a}$.
Then any algorithm computing all $t$ values of
$\mat{\alpha}\vec{a}$, in-place of $\vec{a}$, requires at least $k+1$
elementary operations.
\end{lemma}
\begin{proof}
Consider an in-place algorithm realizing $\mat{\alpha}\vec{a}$ in $f$
operations.
Any zero or standard basis vector row can
be realized without any operations on $\vec{a}$.
Now take this algorithm at the moment where the last of the other rows of
$\mat{\alpha}$ are realized (at that point all the $t$ values are
realized). Then this last realization (a non-trivial linear combination of the
initial values of $\vec{a}$) has to have been stored in one entry 
of $\vec{a}$, say $a_i$.
Therefore, at this point, the in-place algorithm
has to perform at least one more operation to put back $a_i$ to its
initial state.
Therefore, by replacing all the in-place computations by operations on
extra registers and omitting the operation(s) that restore this $a_i$,
we obtain an algorithm with less than $f-1$ elementary operations that
realizes $\mat{\alpha}\vec{a}$ and thus: $(f-1)\geq{k}$.
\end{proof}

\begin{proposition}\label{prop:six}
For the in-place realization of each of the two linear operators
$\mat{\alpha}$ and $\mat{\beta}$, of any bilinear matrix
multiplication algorithm using only $7$ multiplications,
and the restoration of the initial states of their input,
at least $6$ operations are needed.
\end{proposition}
\begin{proof}
A bilinear matrix multiplication algorithm has to compute
$\mat{\alpha}\vec{a}$, with $\vec{a}$ the entries of the left input of
the matrix multiplication, while $\mat{\beta}$ deals with the right
input.
These $\mat{\alpha}$ and $\mat{\beta}$ matrices cannot contain a
($4$-dimensional) zero row:
otherwise there would exist an algorithm using less than $6$
multiplications, but $7$ is minimal~\cite{Winograd:1971:minseven}.
If $\mat{\alpha}$ or $\mat{\beta}$ contain at least $5$ rows that are not
standard basis vectors, then they require at
least $5$ non-trivial operations to be computed, and therefore at
least $6$ elementary operations with an in-place algorithm,
by~\cref{lem:plusone}.
The matrices also cannot contain more than $3$ multiples of standard
basis vectors, by~\cite[Lemma~8]{Bshouty:1995:minwinoadd}.
There thus remains now only to consider matrices with exactly $3$ rows
that are multiples of standard basis vectors.
Let $\mat{M}$ be the $4{\times}4$ sub-matrix obtained from
$\mat{\alpha}$ (or $\mat{\beta}$) by removing those $3$ standard basis vectors.
By~\cref{lem:nocannorzero}, no column of $\mat{M}$ can be the zero column:
otherwise a $7$-dimensional column of $\mat{\alpha}$ (or
$\mat{\beta}$) would be either a multiple of a standard basis vector,
or the zero vector.
This means that every variable of $\vec{a}$ has to be used at least
once to realize the $4$ operations of $\mat{M}\vec{a}$.
Now suppose that there exists an in-place algorithm realizing
$\mat{M}\vec{a}$ in $5$ elementary operations.
Any operations among these $5$ that, as its results, puts back a
variable into its initial state, does not realize any row of
$\mat{M}\vec{a}$ (because putting back a variable to its initial state
is the trivial identity on this initial variable, and this would be
represented by a $4$-dimensional standard basis vector, which
$M$ do not contain, by construction).
Therefore, at most one among these $5$ operations puts back a variable
of $\vec{a}$ into its initial state (otherwise $\mat{M}\vec{a}$, and
therefore  $\mat{\alpha}\vec{a}$ or $\mat{\beta}\vec{a}$, would be
realizable in strictly less than $4$ operations).
Thus, at most one variable of $\vec{a}$ can be modified during
the algorithm (otherwise the algorithm would not be able to put back
all its input variables into their initial state).

W.l.o.g suppose this only modified variable is $a_1$.
Finally, as all the other $3$ variables must be used in at least one
of the $5$ elementary operations, at least $3$ operations are
of the form $a_1\pe\lambda_i{a_i}$ for $i=2,3,4$ and some constants $\lambda_i$.
After those, to put back $a_1$ into its initial state, each one
of these $3$ independent variables, $a_2$, $a_3$ and $a_4$, must be ``removed''
from $a_1$ at some point of the elementary program.
But, with a total of $5$ operations, there remains only $2$ other possible
elementary operations, each one of those modifying only~$a_1$.
Therefore not all $3$ variables can be removed and thus no in-place
algorithm can use only $5$ operations.
\end{proof}
Finally, it remains to consider the linear combinations of the $7$
multiplications to conclude that~\cref{alg:ipsw} realizes the minimal
number of operations for any in-place algorithm with $7$
multiplications.
\begin{theorem}\label{thm:eighteen}
  At least $25$ additions are required to compute in-place
  any bilinear matrix multiplication algorithm using
  only $7$ multiplications and to restore
  its input matrices to their initial states afterwards.
\end{theorem}
\begin{proof} \Cref{prop:six} shows that at least $6$ operations are
  required to realize $\alpha$ (or $\beta$).
  For $\mu$, we in fact compute $\vec{c}\pe\mu\vec{\rho}$,
  so we need to consider the matrix
  $P=\begin{smatrix}I_4&\mu\end{smatrix}\in\F^{4{\times}11}$
  and
  the vector
  $\vec{\xi}=\begin{smatrix}\vec{c}\\\vec{\rho}\end{smatrix}$.
  Consider now an elementary program that realizes $P\vec{\xi}$,
  in-place of $\vec{c}$ only. This implies for instance that if
  $\vec{\rho}$ is zero, $\vec{c}$ should ultimately be put back to its initial
  state.
  Finally, consider the transposed program
  $\Transpose{P}\vec{\underline{c}}$: it must be in-place of
  $\vec{\underline{c}}$, while putting back $\vec{\underline{c}}$ to
  its initial state afterwards.
  By \cref{prop:six}, $\Transpose{\mu}$, thus
  $\Transpose{P}\in\F^{11{\times}4}$, requires at least $6$ elementary
  operations to be performed.
  By Tellegen's transposition principle, see
  also~\cite[Theorem~7]{Kaminski:1988:transpose}, computing the
  transposed program requires at least $6+(11-4)=13$ operations.
  This gives a total of at least $6+6+13=25$ additions.
\end{proof}

\Cref{thm:eighteen} thus shows that our~\cref{alg:ipsw} with $18$
elementary additions and $7$ from the $7$ recursive calls (for a
total of $18+7=25$ additions) is an optimal in-place
bilinear matrix multiplication algorithm using
only $7$ multiplications.

To go beyond our minimality result for operations, one could
try an alternate basis of~\cite{Karstadt:JACM:2020:MMfaster}.
But an argument similar to that of~\cref{prop:six} shows
that alternate basis does not help for the in-place case.

\begin{proposition}\label{prop:sixschwartz}
For the in-place realization of each of the linear operators
arising from the sparsification of any bilinear matrix
multiplication algorithm using only $7$ multiplications,
and for the restoration of the initial states of their input,
at least $6$ operations are needed.
\end{proposition}
\begin{proof}
The alternate basis method of~\cite{Karstadt:JACM:2020:MMfaster}
consists in sparsifying the matrices of~\cref{eq:alphabetamu}, via
right multiplication by $4{\times}4$ invertible matrices.
The sparsest obtained matrices are given in:
\begin{equation}\label{eq:schwartz}
\begin{bmatrix}
1&1&0&0&0&0&0\\
0&0&0&-1&1&0&0\\
0&0&1&0&0&1&0\\
1&0&0&0&1&1&-1\\
\end{bmatrix}\!\!,\quad
\begin{bmatrix}
0&0&1&0\\
1&0&0&0\\
-1&0&0&1\\
0&1&0&1\\
0&0&1&1\\
0&-1&0&0\\
0&0&0&1\\
\end{bmatrix}\!\!,\quad
\begin{bmatrix}
1&0&0&0\\
0&1&0&0\\
0&0&1&-1\\
0&1&0&1\\
0&0&1&0\\
1&0&0&1\\
0&0&0&1\\
\end{bmatrix}\!\!.
\end{equation}
These sparse matrices are obtained from the ones
in~\cref{eq:alphabetamu} via the following respective change of bases:
\begin{equation}
\begin{bmatrix}
1&0&0&0\\
0&0&-1&1\\
0&-1&0&1\\
0&0&0&1\\
\end{bmatrix}\!\!,\quad
\begin{bmatrix}
0&1&0&0\\
1&0&-1&0\\
1&0&0&0\\
-1&0&1&1\\
\end{bmatrix}\!\!,\quad
\begin{bmatrix}
1&0&0&0\\
0&0&1&0\\
-1&1&0&0\\
-1&1&0&-1\\
\end{bmatrix}\!\!.
\end{equation}

We then follow the same line of reasoning as in~\cref{prop:six}, where
we mostly need to adapt~\cref{lem:nocannorzero}.
W.l.o.g, we consider a $4{\times}4$ transformation $M$ of the middle
matrix in~\cref{eq:schwartz}.

If the resulting product matrix has only $3$ rows that are multiple of
a standard basis vector, then $6$ multiplications are minimal
by~\cref{prop:six}.
The only other possibility is thus, as in~\cref{eq:schwartz}, that it
contains exactly $4$ rows that are multiples of a standard basis vector.

Now, each one of the four columns of the resulting product has the form:
\[\Transpose{\begin{bmatrix}c&a&d-a&b+d&c+d&-b&d\end{bmatrix}}.\]
But such a column can never be a zero-column, nor a multiple of a
standard basis vector.

Therefore, as in~\cref{prop:six} all $4$ variables must appear in the
$3$ rows that are not standard basis vectors and at most one variable can be
modified.
So again at least $3$ operations are of the form
$a_1\pe\lambda_i{a_i}$ for $i=2,3,4$ and some constants $\lambda_i$.
And then again at least $3$ operations are required to put back $a_1$
in its initial state and no in-place
algorithm can use strictly less than $6$ operations.
\end{proof}

\subsection{Fast over-place linear algebra}\label{ssec:ffpack}
In~\cite{jgd:2008:toms,jgd:2013:pluq}, many fast dense linear
algebra block recursive routines (system solving, factorizations,
etc.) were designed to use as extra memory almost only that of the
accumulating matrix multiplication that they used.
With our new in-place technique we now have fast in-place accumulating
matrix multiplications, from any fast bilinear multiplication.
We can therefore now combine these results to get fully in-place fast
linear algebra.

In the following, we express the complexities in terms of the
(bilinear) matrix multiplication time $\MM(m;k;n)$ as defined in
\cref{ssec:notations}. Indeed, \Cref{thm:bilin} shows that any bilinear
matrix multiplication algorithm gives rise to an in-place accumulating
variant with same asymptotic complexity. And \cref{app:inplsw,ssec:minimal}
provide a finer analysis of the algorithm obtained from the original
Strassen-Winograd multiplication algorithm.

\subsubsection{Over-place TRMM and TRSM}
This is the case for instance for triangular matrix multiplication
(TRMM) and triangular system solving (TRSM), as shown
in~\cref{alg:TRMM,alg:TRSM}, both adapted from~\cite[\S~4]{jgd:2008:toms}.

\begin{algorithm}[!ht]
  \caption{Over-place TRMM.}
  \label{alg:TRMM}
  \begin{algorithmic}[1]
    \REQUIRE $T\in\F^{m{\times}m}$ upper-triangular, $B\in\F^{m{\times}n}$;
    \ENSURE $B\leftarrow{T\cdot{B}}$.
    \IF{$m\leq\threshold$}
    \STATE Apply the quadratic in-place TRMM.
    \ELSE
    \Statex\COMMENT{split the matrices as: $T=\begin{bmatrix}T_1&T_2\\&T_3\end{bmatrix}$,
      $B=\begin{bmatrix}B_1\\B_2\end{bmatrix}$}
    \STATE $B_1\leftarrow{TRMM(T_1,B_1)}$ \hfill\COMMENT{recursive call}
    \STATE $B_1\pe{T_2\cdot{B_2}}$ \hfill\COMMENT{\cref{alg:ipsw}}
    \STATE $B_2\leftarrow{TRMM(T_3,B_2)}$ \hfill\COMMENT{recursive call}
    \ENDIF
  \end{algorithmic}
\end{algorithm}

\begin{algorithm}[!ht]
  \caption{Over-place TRSM.}
  \label{alg:TRSM}
  \begin{algorithmic}[1]
    \REQUIRE $T\in\F^{m{\times}m}$ upper-triangular, $B\in\F^{m{\times}n}$;
    \ENSURE $B\leftarrow{T^{-1}B}$.
    \IF{$m\leq\threshold$}
    \STATE Apply the quadratic in-place TRSM.
    \ELSE
    \Statex\COMMENT{split the matrices as: $T=\begin{bmatrix}T_1&T_2\\&T_3\end{bmatrix}$,
      $B=\begin{bmatrix}B_1\\B_2\end{bmatrix}$}
    \STATE $B_2\leftarrow{TRSM(T_3,B_2)}$ \hfill\COMMENT{recursive call}
    \STATE $B_1\me{T_2\cdot{B_2}}$ \hfill\COMMENT{\cref{alg:ipsw}}
    \STATE $B_1\leftarrow{TRSM(T_1,B_1)}$ \hfill\COMMENT{recursive call}
    \ENDIF
  \end{algorithmic}
\end{algorithm}

\subsubsection{Over-place PLUQ}
Now, combining matrix multiplication with triangular solving, we can
extend our method to the PLUQ Gaussian factorization
of~\cite[Algorithm~1]{jgd:2013:pluq}: The latter indeed uses only
recursive calls, matrix multiplication, TRSM and permutations.
If the matrix has generic rank profile, then $P=Q=I_m$ and by using our
fast in-place matrix multiplication and fast over-place TRSM, this
gives a fast over place PLUQ factorization. Otherwise, it still does not
need any arithmetic registers, but has still to store the permutation matrices,
as $O(n)$ extra pointer registers.

\begin{remark}\label{rk:ippluq}
  If the field has more than $mn$ elements, then each
  arithmetic register could in principle store two indices (the two
  coordinates addressing any matrix entry).
  In this case, it is possible to store both $P$ and $Q$ permutation
  matrices within the elements of the initial matrix.
  Consider for instance a (block recursive) factorization revealing
  the row rank profile. At any point of the factorization, if some
  permutation is required, it means that there is a zero at the
  location of a pivot and that a non-zero pivot in the same row has to
  be chosen instead, if it exists, or that one has to use another
  row.
  By the rules of Gaussian elimination, this row with at least one
  zero, will not be modified anymore during the factorization.
  In other words the zero at the pivot location will remain a zero
  coefficient within the final LU factorization.
  The idea is thus as follows: Use only two extra pointer registers to
  store the coordinates of the first (row-major) pivot not in the
  default position. This induces the location of a zero element in the
  factorization. Then store the coordinates of the second pivot not in the
  default position instead of this arithmetic zero, and so on.
  By this technique, only two extra pointer registers are needed to
  recover all the information of both the $P$ and $Q$ matrices.
  Note that in an actual implementation of this trick, the coordinates
  could change along the course of the elimination, at least
  virtually, following the subsequent permutations.
  All the coordinates would then have to be updated
  accordingly, thus potentially all along this path of coordinates.
\end{remark}

\subsubsection{Over-place KERN, INVT and INV}\label{ssec:inv}
Further, we now also consider the null-space (KERN), triangular inverse
(INVT) and inverse (INV) algorithms of~\cite[\S~6]{jgd:2008:toms}.
Those make use of only recursive calls, TRSM, PLUQ and accumulating
matrix multiplication. Therefore, again combining with our technique,
one obtains fast over-place algorithms for these three other routines.

\subsection{Fast in-place square and rank-k update}\label{ssec:aat}
Now, some non-bilinear algorithm can also be
transformed into accumulating and in-place algorithms: This is true
for instance for both the square of a matrix or the multiplication of
a matrix by its transpose.

\subsubsection{In-place SQUARE}
Both are shown next, and, first on the symmetric algorithm for the
square of matrices given in~\cite{Bodrato:ISSAC2010}. The resulting
in-place algorithm is given in~\cref{alg:square}.
\begin{algorithm}[!ht]
\caption{In-place accumulating S.-W. matrix-square.}\label{alg:square}
\begin{algorithmic}[1]
\REQUIRE
$A=\begin{bmatrix} A_{11} & A_{12} \\  A_{21} &A_{22}\end{bmatrix}$,
$C=\begin{bmatrix} C_{11} & C_{12} \\  C_{21} &C_{22}\end{bmatrix}$.
\ENSURE $C\pe{A^2}$.
\STATE $A_{22} \me A_{21}$; $C_{12} \pe C_{22}$;
\STATE \colorbox{cyan!10}{$C_{22} \pe A_{22} * A_{22}$;}
\STATE $A_{22} \pe A_{12}$; $A_{22} \me A_{11}$;
\STATE \colorbox{cyan!10}{$C_{12} \me A_{22} * A_{12}$;}
\STATE \colorbox{cyan!10}{$C_{21} \me A_{21} * A_{22}$;}
\STATE $C_{21} \me C_{22}$; $A_{22} \pe A_{11}$;
\STATE \colorbox{cyan!10}{$C_{22} \me A_{22} * A_{22}$;}
\STATE $C_{11} \pe C_{22}$;
\STATE \colorbox{cyan!10}{$C_{22} \me A_{12} * A_{21}$;}
\STATE $A_{22} \pe A_{21}$; $C_{12} \me C_{22}$; $C_{11} \me C_{22}$;
\STATE \colorbox{cyan!10}{$C_{22} \pe A_{22} * A_{22}$;}
\STATE $A_{22} \me A_{12}$; $C_{21} \pe C_{22}$;
\STATE \colorbox{cyan!10}{$C_{11} \pe A_{11} * A_{11}$;}
\end{algorithmic}
\end{algorithm}

\subsubsection{In-place SYRK}
Second, thanks to~\cref{alg:ipsw} and with some care on transposes,
the same technique can be adapted to, e.g.,
\cite[Alg.~12]{Dumas:2023:adjoint},
which performs the multiplication of a matrix by its transpose.
With an accumulation, this is a classical \emph{Symmetric Rank-k
  Update} (or SYRK): $C\leftarrow{\alpha{A}\Transpose{A}+\beta{C}}$.

Following the notations of the latter algorithm (which is not a
bilinear algorithm on its single input matrix), the in-place
accumulating version is shown in~\cref{alg:ipaat},
using any (fast to apply) skew-unitary $Y\in\F^{n{\times}n}$.
It has been obtained automatically by the method
of~\cref{thm:general}, and it thus preserves the need of only $5$
multiplications $P_1$ to ${P_5}$. It has then been scheduled to
reduce the number of extra operations.

\Cref{alg:ipaat} requires
$3$ recursive calls,
$2$ multiplications of two independent half matrices,
$4$ multiplications by a skew-unitary half matrix,
$8$ additions (of half inputs),
$12$ semi-additions (of half triangular outputs).
Provided that the multiplication by the skew-unitary matrix can be
performed in-place in negligible time,
this gives a dominant term of the complexity bound
for~\cref{alg:ipaat} of a fraction $\frac{2}{2^\omega-3}$ of the
cost of the full in-place algorithm.
This is a factor $\frac{1}{2}$, when~\cref{alg:ipsw} is used for the
two block multiplications of independent matrices ($P_4$ and $P_3$).

\begin{algorithm}[!ht]
\caption{In-place accumulating mult. by its transpose (SYRK).}\label{alg:ipaat}
\begin{algorithmic}[1]
\REQUIRE $A=\begin{bmatrix} A_{11} & A_{12} \\  A_{21}
  &A_{22}\end{bmatrix}\in\F^{m{\times}2n}$;
symmetric $C=\begin{bmatrix} C_{11} & \Transpose{C_{21}} \\  C_{21}
  &C_{22}\end{bmatrix}\in\F^{m{\times}m}$.
\ENSURE $\Low{C}\pe\Low{A\cdot\Transpose{A}}$. \hfill\COMMENT{update
  bottom left triangle}
\STATE $\Low{C_{22}} \me \Low{C_{11}}$;   $\Low{C_{21}} \me \Low{C_{11}}$;
\STATE  $\Up{C_{21}} \me \Low{C_{11}}^{\intercal}$;
\STATE \colorbox{cyan!10}{$\Low{C_{11}} \pe \Low{A_{11}*A_{11}^{\intercal}}$;} \hfill\COMMENT{$P_1$ Rec.}
\STATE  $\Up{C_{21}} \pe \Low{C_{11}}^{\intercal}$;
\STATE $\Low{C_{21}} \pe \Low{C_{11}}$;   $\Low{C_{22}} \pe \Low{C_{11}}$;
\STATE \colorbox{cyan!10}{$\Low{C_{11}} \pe \Low{A_{12}*A_{12}^{\intercal}}$;} \hfill\COMMENT{$P_2$ Rec.}
\STATE $A_{11} \fe Y$;   $A_{21} \fe Y$; $A_{11} \me A_{21}$; $A_{21} \me A_{22}$;
\STATE $\Low{C_{22}} \me \Low{C_{21}}$;   $\Low{C_{22}} \me \Low{C_{21}^{\intercal}}$;
\STATE \colorbox{cyan!10}{$C_{21} \pe A_{11}*A_{21}^{\intercal}$;} \hfill\COMMENT{$P_4$ ({e.g.}, \cref{alg:ipsw})}
\STATE $\Low{C_{22}} \pe \Low{C_{21}^{\intercal}}$;
\STATE $A_{21} \me A_{11}$;
\STATE  $\Up{C_{21}} \me \Low{C_{21}}^{\intercal}$;
\STATE \colorbox{cyan!10}{$\Low{C_{21}} \pe \Low{A_{21}*A_{21}^{\intercal}}$;} \hfill\COMMENT{$P_5$ Rec.}
\STATE  $\Up{C_{21}} \pe \Low{C_{21}}^{\intercal}$;   $\Low{C_{22}} \pe \Low{C_{21}}$;
\STATE $A_{21} \pe A_{12}$;
\STATE \colorbox{cyan!10}{$C_{21} \pe  A_{22}*A_{21}^{\intercal}$;} \hfill\COMMENT{$P_3$ ({e.g.}, \cref{alg:ipsw})}
\STATE $A_{21} \me A_{12}$; $A_{21} \pe A_{11}$; $A_{21} \pe A_{22}$; $A_{11} \pe A_{21}$;
\STATE $A_{21} \fe Y^{-1}$; $A_{11} \fe Y^{-1}$;
\end{algorithmic}
\end{algorithm}

Now, the skew-unitary matrices used in~\cite{Dumas:2023:adjoint}, are
either a multiple of the identify matrix, or the Kronecker product of
$\begin{smatrix}a&b\\-b&a\end{smatrix}$ by the identity matrix,
for $a^2+b^2=-1$ and $a\neq{0}$.
The former is easily performed in-place in time \bigO{n^2}.
For the latter,
the multiplication $\begin{smatrix}a&b\\-b&a\end{smatrix}\vec{u}$
can be realized in place by the algorithm:
$u_1\fe{a}$; $u_1\pe{b{\cdot}u_2}$; $u_2\fe(a+b^2a^{-1})$;
$u_2\pe{\left(-ba^{-1}\right){\cdot}u_1}$. (This corresponds
to~\cref{eq:twobytwomul} in \cref{ssec:consecutive}.)

\subsubsection{In-place SYR2K and symmetric factorization}
The next step is to be able to perform an over place symmetric
factorization. For this, the algorithm of~\cite{jgd:2018:ldlt} uses
recursive calls, accumulating matrix multiplication, TRMM, TRSM, SYRK,
diagonal scaling, permutations and also the symmetric rank 2k update.
This latter SYR2K computes the upper or lower triangular part of
$\Low{C}\pe\Low{A\cdot\Transpose{B}+B\cdot\Transpose{A}}$
for any $A\in\F^{m{\times}k}$ and $B\in\F^{m{\times}k}$.
If the symmetric matrix $C$ can actually use a square space,
a fast way to compute the latter is to start by zeroing the  upper
triangular part of $C$.
Then compute one accumulated multiplication and end by accumulating
the upper part of the result on the lower part ($\Up{C}=0$;
$C\pe{A\cdot\Transpose{B}}$; $\Low{C}\pe\Up{C}$).
That way the cost of the computation is dominated by exactly one
matrix multiplication $\MM(m;k;m)$.
But this requires some space not needed in a symmetric matrix, either
its upper or its lower part.
Rather, \cref{alg:SYR2K} is fully in-place and deals with
only one of the two triangles of $C$.
The drawback is that the computational cost is modified to
be dominated by $\frac{2}{2^{\omega-1}-2}\MM(m;k;m)$. The latter is
just again $\MM(m;k;m)$ if $\omega=3$, but the ratio increases to more
than $1$ for smaller~$\omega>2$.

\begin{algorithm}[!ht]
  \caption{Over-place SYR2K.}
  \label{alg:SYR2K}
  \begin{algorithmic}[1]
    \REQUIRE $A\in\F^{m{\times}k}$; $B\in\F^{m{\times}k}$; symmetric
    $C\in\F^{m{\times}m}$.
    \ENSURE $\Low{C}\pe\Low{A\cdot\Transpose{B}+B\cdot\Transpose{A}}$.
    \IF{$m\leq\threshold$}
    \STATE Apply the quadratic in-place SYR2K.
    \ELSE
    \Statex\COMMENT{split the matrices as:
      $A=\begin{bmatrix}A_1\\A_2\end{bmatrix}$,
      $B=\begin{bmatrix}B_1\\B_2\end{bmatrix}$, $C=\begin{bmatrix}C_1& \\C_2&C_3\end{bmatrix}$}
    \STATE $\Low{C_1}\pe\Low{A_1\cdot\Transpose{B_1}+B_1\cdot\Transpose{A_1}}$ \hfill\COMMENT{recursive call}
    \STATE $\Low{C_3}\pe\Low{A_2\cdot\Transpose{B_2}+B_2\cdot\Transpose{A_2}}$ \hfill\COMMENT{recursive call}
    \STATE $C_2\pe{A_2\cdot\Transpose{B_1}}$ \hfill\COMMENT{\cref{alg:ipsw}}
    \STATE $C_2\pe{B_2\cdot\Transpose{A_1}}$ \hfill\COMMENT{\cref{alg:ipsw}}
    \ENDIF
  \end{algorithmic}
\end{algorithm}

Now, with~\cref{alg:ipsw,alg:TRMM,alg:TRSM,alg:SYR2K} one can directly
implement the in-place version of TRSSYR2K
from~\cite[Algorithm~1]{jgd:2018:ldlt}.
This with the in-place versions of GEMM, TRMM, TRSM, SYRK and SYR2K
of~\cref{alg:ipsw,alg:TRMM,alg:TRSM,alg:ipaat,alg:SYR2K},
now provide all the sub-routines for symmetric indefinite triangular
factorization~\cite[Algorithm~2~and~3]{jgd:2018:ldlt}.

To conclude this section, we recall in~\cref{tab:ffpack} the
different over-place linear algebra routines obtained, together with
the dominant term of their associated complexity bound.

For the first algorithms, the dominant term of the bound is
that of~\cite{jgd:2008:toms}.
For SYRK, we use the improvement of~\cite{Dumas:2023:adjoint}.
Then~\cref{alg:SYR2K} for SYR2K requires two recursive calls and two
rectangular multiplications. This a cost bounded as
$T_n(m)\leq{}2T_n(\frac{m}{2})+2\MM(\frac{m}{2};n;\frac{m}{2})$ and thus dominated by
$\displaystyle\frac{2}{2^{\omega}-4}\MM(m;n;m)$.

Finally, for the symmetric factorization, for the sake of simplicity,
we consider here only the generic rank profile case. Then the
factorization reduces to two recursive calls, one TRSM call and one SYRK
call on matrices of half-dimension~\cite[\S~6.4]{jgd:2008:toms}.
This is a cost bounded by
$T(m)\leq{}2T(\frac{m}{2})+TRSM(\frac{m}{2})+SYRK(\frac{m}{2})=2T(\frac{m}{2})+\left(\frac{2}{2^\omega-4}+\frac{2}{2^\omega-3}\right)\MM(\frac{m}{2})+\littleo{n^\omega}$.
The latter is $T(m)\leq\frac{2(2^{\omega+1}-7)}{(2^w-3)(2^w-4)(2^w-2)}=\frac{2^{\omega+2}-14}{8^{\omega}-9{\cdot}4^\omega+26{\cdot}2^\omega-24}$.

\begin{table}[!ht]\centering
\begin{threeparttable}
\caption{Over-place fast linear algebra algorithms.}\label{tab:ffpack}
\begin{tabular}{lccc}
\toprule
\multirow{2}{*}{Algorithm} & \multicolumn{2}{c}{Memory registers} & Complexity bound \\
& Algebraic& Pointer & dominant term\\
\midrule
TRMM, TRSM & $\bigO{1}$ & $\bigO{\log{n}}$ & $\displaystyle\frac{2}{2^{\omega}-4}\left\lceil\frac{n}{m}\right\rceil\MM(m)$\\[10pt]
INVT & $\bigO{1}$ & $\bigO{\log{n}}$ & $\displaystyle\frac{4}{(2^{\omega}-2)(2^\omega-4)}\MM(n)$\\[10pt]
PLUQ, KERN & $\bigO{1}$ & $\bigO{n}$\tnote{$\star$} & $\bigO{mnr^{\omega-2}}$\\[10pt]
INV & $\bigO{1}$ & $\bigO{n}$\tnote{$\star$} &
$\displaystyle\frac{3{\times}2^\omega}{(2^{\omega}-2)(2^\omega-4)}\MM(n)$\\[10pt]
SYRK & $\bigO{1}$ & $\bigO{\log{n}}$ &
$\displaystyle\frac{2}{2^{\omega}-3}\MM(m;n;m)$\\[10pt]
SYR2K & $\bigO{1}$ & $\bigO{\log{n}}$ & $\displaystyle\frac{2}{2^{\omega}-4}\MM(m;n;m)$\\[10pt]
$PLD\Transpose{L}\Transpose{P}$ & $\bigO{1}$ & $\bigO{n}$\tnote{$\star$} &
$\displaystyle\frac{2^{\omega+2}-14}{8^{\omega}-9{\cdot}4^\omega+26{\cdot}2^\omega-24}\MM(n)$\\[10pt]
\bottomrule
\end{tabular}
    \begin{tablenotes}
    \item[$\star$] If $|\F| \ge mn$, \cref{rk:ippluq} can be
        used to reduce the number of pointer registers to $\bigO{\log{n}}$.
    \end{tablenotes}
  \end{threeparttable}
\end{table}

\section{Fast in-place polynomial multiplication with
  accumulation}\label{sec:inpaccpol}
\Cref{alg:bilin} can also be used for polynomial multiplication.
An additional difficulty is that this does not completely fit the setting,
as a multiplication of two size-$n$ inputs will in general span a (double)
size-$2n$ output.
This is not an issue until one has to distribute separately the two
halves of this $2n$ values (or more generally to different parts of
different outputs).
In the following we show that this can anyway always be done
for polynomial multiplications.

\subsection{Double-size output in polynomial multiplication}
To illustrate the handling of double-sized output,
we consider, for instance, an in-place Karatsuba
polynomial multiplication. We start with
\begin{multline}
(Ya_1 + a_0)(Yb_1 +b_0)\\=a_0b_0+Y^2(a_1b_1)
    +Y(a_0b_0+a_1b_1-(a_0-a_1)(b_0-b_1)).
\label{eq:kara}
\end{multline}
From~\cref{eq:kara}, we can express the coefficient vector of the result
as $\mu\left(\alpha\begin{smatrix}a_0\\a_1\end{smatrix}\odot \beta\begin{smatrix}b_0\\b_1\end{smatrix}\right)$, where
\begin{equation}
\mu=\begin{bmatrix}
1&0&0\\
1&1&-1\\
0&1&0
\end{bmatrix},\quad\quad
\alpha=\begin{bmatrix}
1&0\\
0&1\\
1&-1
\end{bmatrix},\quad\quad
\beta=\begin{bmatrix}
1&0\\
0&1\\
1&-1
\end{bmatrix}.\label{eq:bilinkara}
\end{equation}

Then, with $Y=X^\delta$ and $a_i$, $b_i$, $c_i$ polynomials in $X$ (and $a_0$,
$b_0$, $c_0$ of degree less than $t$), this is
detailed, with accumulation, in~\cref{eq:karasplit}:
\begin{equation}\label{eq:karasplit}
\fbox{\scalebox{.975}[0.975]{\ensuremath{\begin{aligned}
A(Y)& = Ya_1 + a_0;\quad
B(Y) = Yb_1 + b_0;\\
C(Y)& = Y^3c_{11} + Y^2c_{10} + Yc_{01} + c_{00};\\
m_0& = a_0\cdot{b_0} = m_{01}Y+m_{00};\quad
m_1 = a_1\cdot{b_1} = m_{11}Y+m_{10};\\
m_2& = (a_0 - a_1)\cdot(b_0 - b_1)= m_{21}Y+m_{20};\\
t_{00} &= c_{00}+m_{00};\quad
t_{01}  = c_{01}+m_{01}+m_{00}+m_{10}-m_{20};\\
t_{10} &= c_{10}+m_{10}+m_{01}+m_{11}-m_{21};\quad
t_{11} = c_{11}+m_{11};\\
\text{\algorithmicthen}&\quad C+AB ={Y^3t_{11}+Y^2t_{10}+Yt_{01}+t_{00}}
\end{aligned}}}}
\end{equation}
To deal with the distributions of each half of the
products of~\cref{eq:karasplit},
each coefficient of $\mu$ in~\cref{eq:bilinkara} can be expanded into
$2{\times}2$ identity blocks, and the middle rows combined two by two,
as each tensor product actually spans
two sub-parts of the result; we obtain~\cref{eq:bilinkarasplit}:
\begin{equation}\label{eq:bilinkarasplit}\setlength\arraycolsep{3pt}
\ensuremath{\mu^{(2)}{=}
\begin{bmatrix} I_2&0_2&0_2\\0&0&0\\0&0&0\end{bmatrix}
{+}
\begin{bmatrix} 0&0&0\\I_2&I_2&-I_2\\ 0&0&0\end{bmatrix}
{+}
\begin{bmatrix} 0&0&0\\0&0&0\\0_2&I_2&0_2\end{bmatrix}
{=}
\begin{bmatrix}
1&0&0&0&0&0\\
1&1&1&0&-1&0\\
0&1&1&1&0&-1\\
0&0&0&1&0&0
\end{bmatrix}}
\end{equation}
Finally, \cref{eq:karasplit} then translates into an in-place algorithm
thanks to~\cref{alg:bilin,eq:bilinkara,eq:bilinkarasplit}.
We formalize this process in the following~\cref{ssec:consecutive}.

\subsection{In-place bilinear accumulation in consecutive blocks}\label{ssec:consecutive}
The first point of polynomial multiplications is that
products double the degree: This corresponds to a constraint that the
two blocks obtained after a multiplication have to remain together
when distributed.
In other words, this means that the matrix $\mat{\mu}^{(2)}$ needs to be
considered two consecutive columns by two consecutive columns.
This is always possible if the two columns are of full rank~$2$.
Indeed, consider a $2\times{2}$ invertible sub-matrix
$M=\begin{smatrix} v & w\\x & y\end{smatrix}$ of these two columns.
Then computing
$\begin{smatrix}c_i\\c_j\end{smatrix}\pe{}M\begin{smatrix}\rho_0\\\rho_1\end{smatrix}$
is equivalent to computing a $2\times{2}$ version of~\cref{eq:basemul}:
\begin{equation}\label{eq:twobytwo}
\left\lbrace\begin{bmatrix}c_i\\c_j\end{bmatrix} \fe M^{-1}; \quad
\begin{bmatrix}c_i\\c_j\end{bmatrix} \pe \begin{bmatrix} \rho_0\\\rho_1\end{bmatrix}; \quad
\begin{bmatrix}c_i\\c_j\end{bmatrix} \fe M\right\rbrace.
\end{equation}
The other rows of these two columns can be dealt with as before by
pre- and post-multiplying or dividing by a constant and pre- and
post-adding or subtracting the adequate $c_i$ and $c_j$.
Now to apply a matrix $M=\begin{smatrix}a&b\\c&d\end{smatrix}$ to a
vector of results $\begin{smatrix}\vec{u}\\\vec{v}\end{smatrix}$,
it is sufficient that one of its coefficients is invertible.
W.l.o.g suppose that its upper left element, $a$, is invertible.
Thus,
$\begin{smatrix}a&b\\c&d\end{smatrix}=\begin{smatrix}1&0\\ca^{-1}&1\end{smatrix}\begin{smatrix}a&b\\0&d-ca^{-1}b\end{smatrix}$.
Then the in-place evaluation of~\cref{eq:twobytwomul} performs this
application, using the two (known in advance) constants $\beta=ca^{-1}$ and $\alpha=d-ca^{-1}b$:
\begin{equation}\label{eq:twobytwomul}
  \fbox{\ensuremath{\left.\begin{aligned}
	  \vec{u}&\fe{a}\\[-5pt]
	  \vec{u}&\pe{b\cdot\vec{v}}\\[-5pt]
	  \vec{v}&\fe{\alpha}\\[-5pt]
	  \vec{v}&\pe{\beta\cdot\vec{u}}
   \end{aligned}\right\rbrace
   \quad
   \begin{array}{l}
	\text{computes in-place:}\\
	\begin{bmatrix}\vec{u}\\\vec{v}\end{bmatrix}
	\gets\begin{bmatrix}a&b\\c&d\end{bmatrix}
	\odot
	\begin{bmatrix}\vec{u}\\\vec{v}\end{bmatrix}
	=\begin{bmatrix}a\vec{u}+b\vec{v}\\c\vec{u}+d\vec{v}\end{bmatrix}\\
	\text{for}~\beta=ca^{-1}~\text{and}~\alpha=d-\beta{}b
      \end{array}
    }}
\end{equation}

\begin{remark}\label{rq:zerotopleft}
  In practice for $2\times{2}$ blocks, if $a$ is not
  invertible, permuting the rows is sufficient (since $c$ has then to be
  invertible for the matrix to be invertible): For
  $J=\begin{smatrix}0&1\\1&0\end{smatrix}$,
  if $\tilde{M}=\begin{smatrix}c&d\\0&b\end{smatrix}=J{\cdot}{M}$, then
  $M=J{\cdot}\tilde{M}$ and $M^{-1}=\tilde{M}^{-1}{\cdot}{J}$.

  This can be simplified further by setting
  $\gamma=cd^{-1}$, $\delta=-\gamma{}b$, if $d$ is also invertible,
  so that now the algorithm becomes:
  $\vec{v}\fe{d}$; $\vec{v}\pe{b\cdot\vec{u}}$;
  $\vec{u}\fe\delta$; $\vec{u}\pe{\gamma\cdot\vec{v}}$.

  Finally, if $d$ is not invertible, an additional step seems to be
  required, with for instance:
  $\vec{v}\fe{b}$; $\vec{u}\fe(-c)$;
  $\vec{u}\pe\vec{v}$; $\vec{v}\me\vec{u}$; $\vec{u}\pe\vec{v}$.
\end{remark}

We now have the tools for in-place polynomial algorithms.
We start, in~\cref{alg:doublebilin}, with a version
of~\cref{alg:bilin} for which the multiplications are accumulated into
two consecutive blocks (denoted \MUL-2D).

\begin{algorithm}[htb]
  \caption{In-place bilinear $2$ by $2$ formula.}\label{alg:doublebilin}
  \begin{algorithmic}[1]
    \REQUIRE $\vec{a}\in\F^m$, $\vec{b}\in\F^n$, $\vec{c}\in\F^s$;
    $\mat{\alpha}\in\F^{t{\times}m}$, $\mat{\beta}\in\F^{t{\times}n}$,
    $\mat{\mu^{(2)}}\in\F^{s{\times}(2t)}=\begin{smatrix}M_1&\cdots&M_t\end{smatrix}$,
      with no zero-rows in $\mat{\alpha}$, $\mat{\beta}$, $\mat{\mu^{(2)}}$,
      s.t. $(a_i\cdot{b_j})$ fits two result variables $c_k$, $c_l$
      and s.t. $M_i\in\F^{s{\times}2}$ is of full-rank $2$ for $i=1..t$.
    \READONLY$\mat{\alpha},\mat{\beta},\mat{\mu^{(2)}}$.
    \ENSURE $\vec{c}\pe\mat{\mu^{(2)}}\vec{m}$, for
    $\vec{m}=(\mat{\alpha}\vec{a})\odot(\mat{\beta}\vec{b})$
    \FOR{$\ell=1$ \To $t$}
    \STATE Let $i$ s.t. $\alpha_{\ell,i}\neq{0}$;
    $a_i\fe\alpha_{\ell,i}$;
    \ForDoEnd[lin:doublealpha]{$\lambda=1$ \To $m$, $\lambda\neq{i}$,
      $\alpha_{\ell,\lambda}\neq{0}$
    }{$a_i\pe\alpha_{\ell,\lambda}a_\lambda$}
    \STATE Let $j$ s.t. $\beta_{\ell,j}\neq{0}$;
    $b_j\fe\beta_{\ell,j}$;
    \ForDoEnd[lin:doublebeta]{$\lambda=1$ \To $n$, $\lambda\neq{j}$,
      $\beta_{\ell,\lambda}\neq{0}$}{$b_j\pe\beta_{\ell,\lambda}b_\lambda$}
    \STATE Let $k$, $f$
    s.t. $M=\begin{smatrix}\mu^{(2)}_{k,2\ell}&\mu^{(2)}_{k,2\ell+1}\\\mu^{(2)}_{f,2\ell}&\mu^{(2)}_{f,2\ell+1}\end{smatrix}$
    is invertible;
    \STATE\label{lin:invmul}$\begin{smatrix} c_k\\c_f \end{smatrix} \leftarrow M^{-1}
    \begin{smatrix} c_k\\c_f \end{smatrix}$
    \hfill\COMMENT{via~\cref{eq:twobytwomul,rq:zerotopleft}}
    \ForDoEnd[lin:divsube]{$\lambda=1$ \To $s$,
      $\lambda\not\in\{f,k\}$,
      $\mu^{(2)}_{\lambda,2\ell}\neq{0}$}{$c_\lambda\me\mu^{(2)}_{\lambda,2\ell}{c_k}$}
    \ForDoEnd[lin:divsubo]{$\lambda=1$ \To $s$,
      $\lambda\not\in\{f,k\}$,
      $\mu^{(2)}_{\lambda,2\ell+1}\neq{0}$}{$c_\lambda\me\mu^{(2)}_{\lambda,2\ell+1}{c_f}$}
    \STATE\label{lin:doubleproduct}$\begin{smatrix} c_k\\c_f
    \end{smatrix}\pe{a_i\cdot{b_j}}$\hfill\COMMENT{this is the
      accumulation of the product $\begin{smatrix} m_k\\m_f\end{smatrix}$}
    \ForDoEnd{$\lambda=1$ \To $s$, $\lambda\not\in\{f,k\}$,
      $\mu^{(2)}_{\lambda,2\ell+1}\neq{0}$}{$c_\lambda\pe\mu^{(2)}_{\lambda,2\ell+1}{c_f}$}
    \ForDoEnd{$\lambda=1$ \To $s$, $\lambda\not\in\{f,k\}$,
      $\mu^{(2)}_{\lambda,2\ell}\neq{0}$}{$c_\lambda\pe\mu^{(2)}_{\lambda,2\ell}{c_k}$}
    \STATE$\begin{smatrix} c_k\\c_f \end{smatrix} \leftarrow M\begin{smatrix} c_k\\c_f \end{smatrix}$
    \hfill\COMMENT{via~\cref{eq:twobytwomul,rq:zerotopleft}, undo~\ref{lin:invmul}}
    \ForDoEnd{$\lambda=1$ \To $n$, $\lambda\neq{j}$,
      $\beta_{\ell,\lambda}\neq{0}$}{$b_j\me\beta_{\ell,\lambda}b_\lambda$};~$b_j\de\beta_{\ell,j}$;
    \ForDoEnd{$\lambda=1$ \To $m$, $\lambda\neq{i}$,
      $\alpha_{\ell,\lambda}\neq{0}$}{$a_i\me\alpha_{\ell,\lambda}a_\lambda$};~$a_i\de\alpha_{\ell,i}$;
    \ENDFOR
  \end{algorithmic}
\end{algorithm}
A C++ implementation of~\cref{alg:doublebilin}
(\href{https://github.com/jgdumas/plinopt/blob/main/trilplacer.cpp}{\texttt{trilplacer -e}})
is also available in the
\plinopt~library.
\begin{theorem}\label{thm:doublebilin}
\Cref{alg:doublebilin} is correct, in-place, and requires
$t$ \MUL-2D,
$2(\#\alpha+\#\beta+\#\mu^{(2)}-t)$ \ADD and
$2(\sharp\alpha+\sharp\beta+\sharp\mu^{(2)}+2t)$ \SCA operations.
\end{theorem}
\begin{proof}
  Thanks to~\cref{eq:twobytwo,eq:twobytwomul,rq:zerotopleft},
  correctness is similar to that of~\cref{alg:bilin}
  in~\cref{thm:bilin}.
Then, \cref{eq:twobytwomul} requires $4$ \SCA
and $2$ \ADD operations and is called $2t$ times.
The rest is similar to~\cref{alg:bilin} and amounts to
$2t+2(\#\alpha-t+\#\beta-t+\#\mu^{(2)}-2t)+(2t)2$ \ADD and
$2(\sharp\alpha+\sharp\beta+\sharp\mu^{(2)}-2t)+(2t)4$ \SCA operations.
\end{proof}

There remains to use a double expansion of the post matrix
$\mu$ to simulate the double size of the
intermediate products (\MUL-2D), producing
$\mu^{(2)}$, as in~\cref{eq:bilinkarasplit} (that can then be used as
an input in~\cref{alg:doublebilin}).
This double expansion matrix
$\mu^{(2)}\in\F^{s{\times}(2t)}$ is simply obtained by
duplicating and interleaving all the entries of
$\mu\in\F^{(s-1){\times}t}$ as follows:
\begin{equation}\label{eq:doubleexpansion}
\forall{i}\in[1..(s-1)], \forall{j}\in[1..t],
\begin{cases}
\mu^{(2)}(1,2j)=0\\
\mu^{(2)}(i,2j-1)=\mu(i,j)\\
\mu^{(2)}(i+1,2j)=\mu(i,j)\\
\mu^{(2)}(s,2j-1)=0
\end{cases}
\end{equation}

We prove, in \cref{lem:fullrank},
that in fact any such double expansion of a representative matrix
is suitable for the in-place computation of~\cref{alg:doublebilin}.

\begin{lemma}\label{lem:fullrank}
  If $\mu$ does not contain any zero column, then each pair of columns
  of $\mu^{(2)}$, resulting from the expansion of a single column in
  $\mu$, as in~\cref{eq:doubleexpansion},
  contains an invertible lower triangular $2{\times}2$ sub-matrix.
\end{lemma}
\begin{proof}
The top most non-zero element of a column
is expanded as a $2{\times}2$ identity matrix whose second row is
merged with the first row of the next identity matrix:
$\begin{smatrix} a \\ b\end{smatrix}$ is expanded to
$\begin{smatrix} a &0 \\ b & a\\ * & b\end{smatrix}$.
\end{proof}

\subsection{In-place accumulating Karatsuba
  multiplication}
We now can come back to the Karatsuba case.

For instance with $m_{00}+Ym_{01}=a_0b_0=\rho_0+Y\rho_1$,
consider the upper left $2\times{2}$ block of $\mu^{(2)}$
in~\cref{eq:bilinkarasplit}, that is
$M=\begin{smatrix} 1&0\\1&1\end{smatrix}$, whose inverse is
$M^{-1}=\begin{smatrix} 1&0\\-1&1\end{smatrix}$.
One has first to precompute
$M^{-1}\begin{smatrix}c_{00}\\c_{01}\end{smatrix}$, that is nothing to
$c_{00}$ and $c_{01}\me{}c_{00}$ for the second coefficient.
Then, afterwards, the third row, for $c_{10}$, will just be $-m_{01}$:
for this just pre-subtract $c_{10}\me{}c_{01}$, and post-add
$c_{10}\pe{}c_{01}$ after the product actual computation.
This example is in lines \ref{lin:m0BEG} and \ref{lin:m0END}
of~\cref{alg:accinplmulkara} thereafter.
To complete~\cref{eq:karasplit}, the computation of $m_1$ is dealt
with in the same manner, while that
of $m_2$ is direct in the results.
Note that as both $t_{01}$ and $t_{10}$ receive together
$m_{01}+m_{10}$, some pre- and post-additions are simplified out
in~\cref{alg:accinplmulkara}.
The second point is to deal with unbalanced dimensions and degrees
for $Y=X^\delta$ and recursive calls. First split the largest
polynomial into two parts, so that two sub-products are performed: a
large balanced one, and, recursively, a smaller unbalanced one.
Second, for the balanced case, the idea is to ensure that three out of
four parts of the result, $t_{00}$, $t_{01}$ and $t_{10}$, have the
same size and that the last one $t_{11}$ is smaller. This ensures that
all accumulations can be performed in-place.
The in-place algorithm can be obtained automatically (up to some sign
swaps) via the combination of~\cref{eq:doubleexpansion,alg:doublebilin} with the \plinopt~library, running:
\begin{verbatim}
./bin/trilplacer -e data/1o1o2_3_Karatsuba_{L,R,P}.sms
\end{verbatim}
The resulting details can be found in~\cref{alg:accinplmulkara}.

\begin{algorithm}[!ht]
\caption{In-place accumulating Karatsuba polynomial multiplication.}\label{alg:accinplmulkara}
\begin{minipage}{\columnwidth}
\begin{algorithmic}[1]
\REQUIRE $A$, $B$, $C$ polynomials of degrees $m$, $n$, $m+n$
with $m\geq{n}$.
\ENSURE $C\pe{AB}$
\IF{$n\leq\threshold$}\hfill\COMMENT{constant-time if $\threshold\in\bigO{1}$}
\STATE Apply the quadratic in-place multiplication. \hfill\COMMENT{\cref{alg:classicmulpoly}}
\ELSIF{$m>n$}
\STATE Let $A(X)=A_0(X)+X^{n+1}A_1(X)$
\STATE $C_{0..2n}\pe{A_0B}$\hfill\COMMENT{recursive call}
\IF{$m\geq{2n}$}
\STATE $C_{(n+1)..(n+m)}\pe{A_1B}$\hfill\COMMENT{recursive call}
\ELSE
\STATE $C_{(n+1)..(n+m)}\pe{BA_1}$\hfill\COMMENT{recursive call}
\ENDIF
\ELSE\hfill\COMMENT{now $m=n$}
\STATE Let $\delta=\lceil(2n+1)/4\rceil$; \hfill\COMMENT{$\delta-1\geq{2n-3\delta}$ and thus $\delta>n-\delta$}
\STATE Let $A=a_0+X^{\delta}a_1$; $B=b_0+X^{\delta}b_1$;
\STATE Let $C=c_{00}+c_{01}X^{\delta}+c_{10}X^{2\delta}+c_{11}X^{3\delta}$;
\hfill\COMMENT{$d^\circ{c_{11}}=2n-3\delta$}
\STATE\label{lin:m0BEG}$c_{01}\me{c_{00}}$;\quad $c_{10}\me{c_{01}}$;
\STATE $\begin{bmatrix} c_{00} \\ c_{01}\end{bmatrix} \pe a_0\cdot{b_0} $
\hfill\COMMENT{recursive call for $m_0$}
\STATE $c_{11}\me{c_{10}{\scriptstyle[0..2n{-}3\delta]}}$;
\hfill\COMMENT{first $2n-3\delta+1$ coefficients of $c_{10}$}
\STATE $\begin{bmatrix} c_{01} \\ c_{10}\end{bmatrix} \pe a_1\cdot{b_1} $
\hfill\COMMENT{recursive call for $m_1$}
\STATE $c_{11}\pe{c_{10}{\scriptstyle[0..2n{-}3\delta]}}$;
\hfill\COMMENT{as $d^\circ m_{11} \leq 2n-3\delta$}
\STATE\label{lin:m0END}$c_{10}\pe{c_{01}}$;\quad $c_{01}\pe{c_{00}}$;
\STATE $a_{0}\me{a_{1}}$;\quad\quad $b_{0}\me{b_{1}}$;\hfill\COMMENT{$d^\circ{a_0}=\delta-1\geq{n-\delta}=d^\circ{a_1}$}
\STATE $\begin{bmatrix} c_{01} \\ c_{10}\end{bmatrix} \me a_0\cdot{b_0}$
    \hfill\COMMENT{recursive call\footnote{Variant that computes $C\me AB$, obtained by changing signs at each recursive call.} for $m_2$}
\STATE $b_{0}\pe{b_{1}}$;\quad\quad $a_{0}\pe{a_{1}}$;
\ENDIF
\end{algorithmic}
\end{minipage}
\end{algorithm}

\begin{proposition}
\cref{alg:accinplmulkara} is correct and requires
$\bigO{mn^{\log_2(3)-1}}$ operations.
\end{proposition}
\begin{proof}
With the above analysis, correctness comes from that
of~\cref{eq:doubleexpansion,alg:doublebilin} applied
to~\cref{eq:bilinkara}.
When $m=n$, with $3$ recursive calls and \bigO{n} extra operations,
the algorithm thus requires overall $\bigO{n^{\log_2(3)}}$ operations.
Otherwise, it requires $\left\lfloor\frac{m}{n}\right\rfloor$ equal
degree calls, then a recursive call with $n$ and $(m\bmod{n})$.
Let $u_1=m$, $u_2=n$, $u_3$, \dots, $u_k$ denote the
successive residues in the Euclidean algorithm on inputs $m$ and $n$
(where $u_k$ is the last non-zero residue).
Then, \cref{alg:accinplmulkara} requires
less than
$\bigO{\sum_{i=1}^{k-1}\lfloor\frac{u_i}{u_{i+1}}\rfloor{}u_{i+1}^{\log_2(3)}}
\leq \bigO{\sum_{i=1}^{k-1} u_iu_{i+1}^{\log_2(3)-1}}$
operations.
But, $u_{i+1}\leq{u_2}=n$ and if we let $s_i=u_i+u_{i+1}$,
$u_i\leq{s_i}$.
This means that the number of operations is bounded by
$\bigO{\sum_{i=1}^{k-1} s_i n^{\log_2(3)-1}}$.
From~\cite[Corollary~2.6]{Grenet:2020:euclide}, we have that
$s_i\leq s_1(2/3)^{i-1}$. Therefore, $\sum_{i=1}^{k-1} s_i
\leq s_1 \sum_{i\geq0}(2/3)^i = \bigO{m+n}$,
and the number of operations is $\bigO{mn^{\log_2(3)-1}}$.
\end{proof}

\begin{figure}[!ht]
\caption{In-place Karatsuba polynomial multiplication.}\label{fig:ipkara}
\includegraphics[width=.9\textwidth]{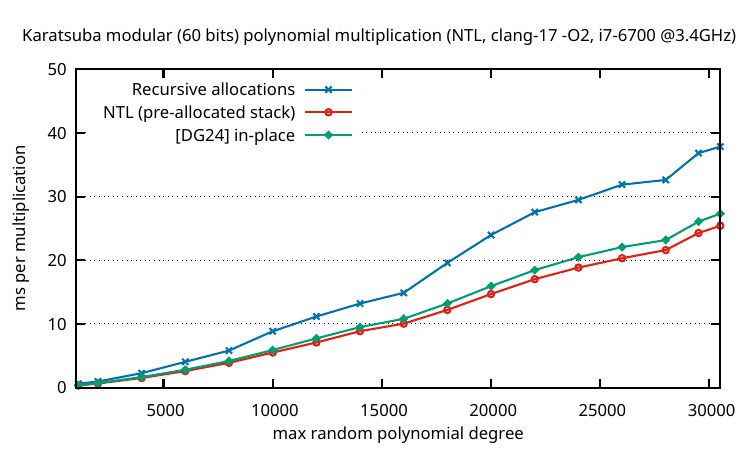}
\includegraphics[width=.9\textwidth]{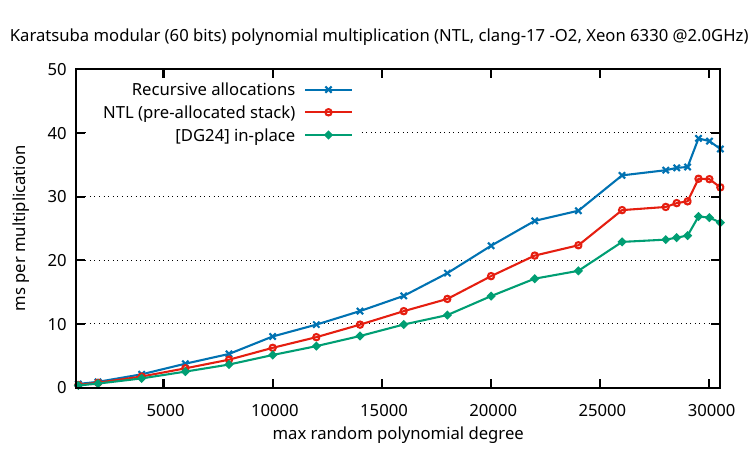}
\end{figure}

Note that all coefficients of $\mat{\alpha}$,
$\mat{\beta}$ and $\mu^{(2)}$ being $1$ or $-1$, \cref{alg:accinplmulkara}
does compute the accumulation $C\pe{AB}$ without constant
multiplications.
Also, the de-duplication enables some
natural reuse.
There is thus a cost of
$2(\#\alpha-t+\#\beta-t)=2(4-3+4-3)=4$ additions with
$a_0,a_1,b_0,b_1$.
Then $2(2(\#\mu-t)-1)=2(\#\mu^{(2)}-2t-1)=2(10-6-1)$ additions with
$c_{ij}$ (\emph{i.e.}, $10-6$ minus the one saved by factoring
$m_{01}+m_{10}$).
This is a total of $3$ recursive
accumulating calls and at most $10$ half-block additions.
For degree $n-1$ (size $n$) polynomials, this is between $5n-\tfrac12$ and
$5n-5$ additional additions, and with $2$ operations for an
accumulating base case, this gives
a dominant term between $11.75n^{\log_2(3)}$ and $9.5n^{\log_2(3)}$.
A careful Karatsuba implementation for both polynomials of {\em size} $n$ a
power of two requires instead $6.5n^{\log_2(3)}$
operations~\cite{Roche:2009:spacetime}.
Now in practice, the supplementary operations are in fact, at least
mostly, compensated by the gain in memory allocations or movements
as shown in~\cref{fig:ipkara}.
We indeed have implemented~\cref{alg:accinplmulkara} with the
\href{https://github.com/libntl/ntl}{NTL}
library,\footnote{\url{https://libntl.org}.}
and we compare it to two other Karatsuba  modular ($60$ bit prime)
implementations:
\begin{itemize}
\item a multiplication that recursively allocates temporary
  polynomials when needed;
\item the state-of-the-art library \href{https://github.com/libntl/ntl}{NTL}
  multiplication implementation that pre-allocates once a stack
  larger than the extra memory requirements at any point in the
  recursive algorithm.
\end{itemize}
The observed results are dependent on the architecture and the compiler.
We provide two graphs with two different architectures, both using the same
compiler clang-17 which appears to produce faster code. In both cases,
the implementation with on-the-fly allocations
is significantly slower than the others.
On one architecture, the in-place version is very close to the stacked
implementation, in fact never more than $10\%$ slower. And on the second
architecture, the stacked implementation is at least $20\%$ slower.

We also compare in~\cref{tab:kara} the procedure given
in~\cref{alg:accinplmulkara} (obtained via the automatic application
of~\cref{eq:doubleexpansion,alg:doublebilin}) with previous Karatsuba-like algorithms for
polynomial multiplications, designed to reduce their memory footprint
(see also \cite[Table~2.2]{Giorgi:2019:hdr}).

\begin{table}[!ht]\centering
\caption{Reduced-memory algorithms for Karatsuba polynomial
  multiplication}\label{tab:kara}
\begin{tabular}{lcccc}
\toprule
\multirow{2}{*}{Algorithm} & \multicolumn{2}{c}{Memory Reg.} & \multirow{2}{*}{Inputs} & \multirow{2}{*}{Accumulation} \\
& Algebraic& Pointer & &\\
\midrule
\cite{Maeder:1993:DISCO:kara} & $2n$ & $5\log{n}$ & {\color{teal} read-only} & {\xno}\\
\cite{Thome:2002:karatemp} & $n$ & $5\log{n}$ & {\color{teal} read-only} & {\xno}\\
\cite{Roche:2009:spacetime,Roche:2011:waterloo} & $0$ & $5\log{n}$ & {\color{teal} read-only} & {\xno}\\
\cite{Giorgi:2019:issac:reductions} &  \multicolumn{2}{c}{\textcolor{teal}{$\bigO{1}$}} & {\color{teal} read-only} & {\xno}\\
\cref{alg:accinplmulkara} & $0$ & $5\log n$ & mutable & {\xyes}\\
\bottomrule
\end{tabular}
\end{table}

\subsection{In-place accumulating Toom-Cook multiplications}\label{ssec:toom}

We have shown that any bilinear algorithm for polynomial
multiplication can be transformed into an
in-place version. This approach thus also works for any Toom-$k$
algorithm using $2k-1$ interpolations points instead of the three
points of Karatsuba (Toom-$2$).

For instance Toom-$3$ uses interpolations at
$0,1,-1,2,\infty$. Therefore, $\alpha$ and $\beta$ are the Vandermonde matrices
of these points for the $3$ parts of the input polynomials and
$\mu$ is the inverse of the Vandermonde matrix of these points for the
$5$ parts of the result, as shown in~\cref{eq:toom3} thereafter.

\begin{equation}\label{eq:toom3}\setlength\arraycolsep{3pt}
\mu=\begin{bmatrix}
1 & 0 & 0 & 0 & 0 \\
1 & 1 & 1 & 1 & 1 \\
1 & -1 & 1 & -1 & 1 \\
1 & 2 & 4 & 8 & 16 \\
0 & 0 & 0 & 0 & 1
\end{bmatrix}^{-1}\!\!\!\!=
\begin{bmatrix}
 1 & 0 & 0 & 0 & 0 \\
{\scriptscriptstyle{-}}\frac{1}{2} & 1 & {\scriptscriptstyle{-}}\frac{1}{3} &  {\scriptscriptstyle{-}}\frac{1}{6} & 2 \\
 -1 & \frac{1}{2} & \frac{1}{2} & 0 & -1 \\
\frac{1}{2} & {\scriptscriptstyle{-}}\frac{1}{2} & {\scriptscriptstyle{-}}\frac{1}{6} & \frac{1}{6} & -2 \\
 0 & 0 & 0 & 0 & 1
\end{bmatrix}\!\!;~\alpha=\beta=\begin{bmatrix}
1 & 0 & 0 \\
1 & 1 & 1 \\
1 & -1 & 1 \\
1 & 2 & 4 \\
0 & 0 & 1
\end{bmatrix}
\end{equation}

It is then possible to apply~\cref{eq:doubleexpansion} and obtain $\mu^{(2)}$:
\begin{equation}\label{eq:toom3mu2}
\begin{bmatrix}
1&0&0&0&0&0&0&0&0&0\\
-\frac{1}{2}&1&1&0&-\frac{1}{3}&0&-\frac{1}{6}&0&2&0\\
-1&-\frac{1}{2}&\frac{1}{2}&1&\frac{1}{2}&-\frac{1}{3}&0&-\frac{1}{6}&-1&2\\
\frac{1}{2}&-1&-\frac{1}{2}&\frac{1}{2}&-\frac{1}{6}&\frac{1}{2}&\frac{1}{6}&0&-2&-1\\
0&\frac{1}{2}&0&-\frac{1}{2}&0&-\frac{1}{6}&0&\frac{1}{6}&1&-2\\
0&0&0&0&0&0&0&0&0&1\\
\end{bmatrix}\end{equation}

Applying~\cref{thm:doublebilin} to $\alpha$ and $\beta$
in~\cref{eq:toom3}, and $\mu^{(2)}$ in~\cref{eq:toom3mu2} gives an
operation count of $2(11+11-2*5)+2(2(16-5))=68$ additions and
$2(2+2+2(11))=52$ scalar multiplications.

The \plinopt~implementation of~\cref{alg:doublebilin} on these
matrices is:
\begin{verbatim}
./bin/trilplacer -e data/2o2o4_5_Toom3_{L,R,P}.sms
\end{verbatim}

This automatically generates an in-place straight-line program (SLP)
using only $58$ additions, as it simplifies some intermediate post and
pre-additions.
Further direct simplifications can be found
(running \plinopt's \verb!./bin/optimizer! for the operations in-between
recursive calls), to produce an in-place SLP using then only
$43$ extra additions and $34$ extra scalings.

Further optimizations should be tried to further reduce the
complexity,
for instance, trying all possible orderings of the recursive calls and
incorporating the tricks presented
in~\cite{Bodrato:2007:WAIFI:toomcook,Bodrato:2007:toomcook,Brent:2011:Modern,Giorgi:2019:hdr}.

\subsection{In-place accumulating FFT-based multiplication}\label{ssec:fft}

When sufficiently large roots of unity exist, polynomial
multiplications can be computed fast in our in-place model via a
discrete Fourier transform and its inverse.
For simplicity,
we consider a ring $\D$ that contains a principal $N$th root of unity $\omega\in\D$
for some $N = 2^p$. (In particular, $2$ is a unit in $\D$.)

Let $F\in\D[X]$ of degree less than $N$.
The discrete Fourier transform of $F$ at $\omega$ is defined as
$\DFT_{N}(F,\omega) = (F(\omega^0), F(\omega^1), \dotsc, F(\omega^{N-1}))$.
The map is invertible, of inverse $\DFT^{-1}_{N}(\cdot,\omega) = \frac{1}{N} \DFT_{N}(\cdot,
\omega^{-1})$. Further, the DFT can be computed over-place, replacing the input by the
output~\cite{1965:CooleyTukey:MathComp:FFT}. Actually, for over-place
algorithms and their extensions to the \emph{truncated Fourier transform}, it is
more natural to work with the \emph{bit-reversed DFT} defined by
$\brDFT_{N}(F,\omega) = (F(\omega^{[0]_{p}}), F(\omega^{[1]_{p}}), \dotsc, F(\omega^{[N-1]_{p}}))$
where $[i]_{p}=\sum_{j=0}^{p-1} d_j 2^{p-j}$ is the length-$p$ bit reversal of $i = \sum_{j=0}^{p-1} d_j 2^j$,
$d_j\in\{0,1\}$.
Its inverse is $\brDFT_N^{-1}(\Lambda,\omega) = \frac{1}{N}\DFT_N((\Lambda_{[0]_p},\dots,\Lambda_{[N-1]_p}),\omega^{-1})$.

\begin{remark}
    The Fast Fourier Transform (FFT) algorithm has two main variants:
    \emph{decimation in time} (DIT) and \emph{decimation in frequency} (DIF).
    Both algorithms can be performed over-place, replacing the input by the
    output. Without applying any permutation to the entries of the input/output
    vector, the over-place DIF-FFT algorithm naturally computes
    $\brDFT_{N}(\cdot,\omega)$, while the over-place DIT-FFT algorithm on $\omega^{-1}$
    computes $N\cdot\brDFT_{N}^{-1}(\cdot, \omega)$~\cite[Exercise~49]{Bini:1994:polymatcomp}.
\end{remark}

We start with the in-place~\cref{alg:fft2pow} for the power of two case.

\begin{algorithm}[!ht]
\caption{In-place power of two accumulating multiplication.}\label{alg:fft2pow}
\begin{algorithmic}[1]
    \REQUIRE $\vec{a}$, $\vec{b}$ and $\vec{c}$ of respective lengths $n$, $n$ and $N = 2n$, containing the coefficients of $A$, $B$, $C\in\D[X]$ respectively; $\omega\in\D$ principal $N$th root of unity, with $N = 2^p$.
\ENSURE $\vec{c}$ contains the coefficients of $C+A\cdot B$.
\STATE $\brDFT_{2n}(\vec{c},\omega)$;
    \hfill\COMMENT{over-place}
\STATE\label{lin:DFTab}$\brDFT_n(\vec{a},\omega^2)$; $\brDFT_n(\vec{b},\omega^2)$
    \hfill\COMMENT{over-place}
\ForDoEnd{$i=0$ \To $n-1$}{$c_i\pe a_i\times b_i$}
\STATE $\brDFT^{-1}_n(\vec{a},\omega^2)$; $\brDFT^{-1}_n(\vec{b},\omega^2)$
    \hfill\COMMENT{undo \ref{lin:DFTab}}
\ForDoEnd[lin:mulw]{$i=0$ \To $n-1$}{$a_i\fe \omega^i$; $b_i\fe \omega^i$}
\STATE\label{lin:DFTab2}$\brDFT_n(\vec{a},\omega^2)$; $\brDFT_n(\vec{b},\omega^2)$
    \hfill\COMMENT{over-place}
\ForDoEnd{$i=0$ \To $n-1$}{$c_{i+n}\pe a_i\times b_i$}
\STATE $\brDFT^{-1}_n(\vec{a},\omega^2)$; $\brDFT^{-1}_n(\vec{b},\omega^2)$
    \hfill\COMMENT{undo \ref{lin:DFTab2}}
\ForDoEnd{$i=0$ \To $n-1$}{$a_i\de \omega^i$; $b_i\de \omega^i$}
    \hfill\COMMENT{undo \ref{lin:mulw}}
\STATE $\brDFT^{-1}_{2n}(\vec{c},\omega)$
\end{algorithmic}
\end{algorithm}

\begin{theorem}\label{thm:fft2pow}
  Using an over-place $\brDFT$ algorithm with
  complexity bounded by $\bigO{n\log n}$, \cref{alg:fft2pow} is
  correct, in-place and has complexity bounded by $\bigO{n\log n}$.
\end{theorem}
\begin{proof}
    \cref{alg:fft2pow} follows the pattern of the standard FFT-based
    multiplication
    algorithm. Our goal is to compute $\brDFT_{2n}(A,\omega)$, $\brDFT_{2n}(B,\omega)$ and
    $\brDFT_{2n}(C,\omega)$,
    then obtain $\brDFT_{2n}(C+AB,\omega)$ and finally $C+AB$ using an inverse
    $\brDFT$. Computations on $C$ and then $C+AB$ are performed over-place using
    any standard over-place $\brDFT$ algorithm. The difficulty happens for $A$ and
    $B$ that are stored in length-$n$ arrays. We use the following property of
    the bit reversed order: for $k < n$, $[k]_{p} = 2[k]_{p-1}$, and for
    $k \ge n$, $[k]_{p} = 2[k-n]_{p-1}+1$. Therefore, the first $n$
    coefficients of $\brDFT_{2n}(A,\omega)$ are $(A(\omega^{2[0]_{p-1}}), \dotsc
    A(\omega^{2[n-1]_{p-1}})) = \brDFT_n(A,\omega^2)$. Similarly, the next $n$
    coefficients are $\brDFT_n(A(\omega X), \omega^2)$. Therefore, one can compute
    $\brDFT_n(A,\omega^2)$ and $\brDFT_n(B,\omega^2)$ in $\vec{a}$ and $\vec{b}$ respectively,
    and update the first $n$ entries of~$\vec{c}$. Next we restore $\vec{a}$
    and $\vec{b}$ using $\brDFT^{-1}_n(\cdot,\omega^2)$. We compute $A(\omega X)$ and
    $B(\omega X)$ and again $\brDFT_n(A(\omega X),\omega^2)$ and $\brDFT_n(B(\omega X),\omega^2)$ to update the
    last $n$ entries of $\vec{c}$. Finally, we restore $\vec{a}$ and
    $\vec{b}$ and perform $\brDFT^{-1}$ on $\vec{c}$.
    The cost is dominated by the ten $\brDFT^{\pm1}$ computations.
\end{proof}

The standard (not in-place) algorithm uses two $\brDFT$ and one $\brDFT^{-1}$ in size $2n$.
A $\brDFT^{\pm1}$ in size $n$ requires $\frac{3}{2}n\log n+\bigO{n}$ ring operations if the required
powers of $\omega$ are precomputed~\cite{MCA2013}. Therefore, the standard algorithm uses
$9n\log n + \bigO{n}$ ring operations. By contrast, our in-place variant uses 2 size-$2n$ and 8 size-$n$ $\brDFT^{\pm1}$ and cannot precompute the powers of $\omega$. Each call to $\brDFT^{\pm1}$ in size
$n$ then requires $2n\log n +\bigO{n}$ ring operations, and the full algorithm requires $24n\log n+\bigO{n}$ ring
operations. The dominant term is therefore $\frac{8}{3}$ times as large in the in-place variant.

The case where the sizes are not powers of two is loosely similar, using as a routine
a truncated Fourier transform (TFT) rather than a
DFT~\cite{2004:vanderHoeven:ISSAC:TFT}. Let $\omega$ still be an $N$th root of
unity for some $N = 2^p$, and $n < N$.
The length-$n$ (bit-reversed) TFT of a polynomial $F\in\D[X]$ at $\omega$
is $\brTFT_n(F,\omega) = (F(\omega^{[0]_p}), \dotsc, F(\omega^{[n-1]_p}))$, that is
the $n$ first coefficients of $\brDFT_{N}(F,\omega)$.
As for the DFT, the (bit-reversed) TFT and its inverse
can be computed over-place~\cite{Harvey:2010:issactft, Roche:2011:waterloo,
2013:Arnold:ISSAC:TFT, Coxon:2022:JSC:inplaceTFT}.

Given inputs $A$ and $B\in\D[X]$ of respective lengths $m$ and $n$ and an output
$C\in\D[X]$ of length $m+n-1\leq N$, we aim to replace $C$ by $C+AB$. As in the power-of-two
case, we first replace $C$ by
$\brTFT_{m+n-1}(C,\omega)$ in  $\vec{c}$.
Then we progressively update $\vec{c}$ using small $\brTFT$'s on the inputs, using the following lemma.

\begin{lemma}[\cite{Harvey:2010:issactft,Roche:2011:waterloo}]\label{lem:roche}
    Let $F\in\D[X]$, $\ell$, $s\in\Z{>0}$
    where $2^\ell$ divides
    $s$, and $\omega$ be a $2^p$th principal root of unity. If $F_{s,\ell}(X) =
    F(\omega^{[s]_p}X)\bmod X^{2^\ell-1}$,
    \(\brDFT_{2^\ell}(F_{s,\ell},\omega^{2^{p-\ell}}) = (F(\omega^{[s]_p}),\dotsc,F(\omega^{[s+2^\ell-1]_p}))\).
\end{lemma}
\begin{proof}
    Let $\omega_\ell = \omega^{2^{p-\ell}}$. This is a principal $2^\ell$th root of unity
    since $\omega$ is a principal $2^p$th root of unity. In particular, for any $i <
    2^\ell$, $F_{s,\ell}(\omega_\ell^{[i]_\ell}) = F(\omega^{[s]_p}\omega_\ell^{[i]_\ell})$.
    Now, $\omega_\ell^{[i]_\ell} = \omega^{[i]_p}$ since $2^{p-\ell}[i]_\ell = [i]_p$.
    Furthermore, $[s]_p+[i]_p = [s+i]_p$ since $i < 2^\ell$ and $2^\ell$ divides
    $s$.
    Finally, $F_{s,\ell}(\omega_\ell^{[i]_\ell}) = F(\omega^{[s+i]_p})$.
\end{proof}

\begin{corollary}\label{cor:partTFT}
    Let $F\in\D[X]$ stored in an array $\vec{f}$ of length $n$,
    $\ell$, $k\in\Z_{>0}$ and $\omega$ be a $2^p$th principal root of unity, with $2^\ell\le
    n$ and $(k+1)2^\ell \le 2^p$. There exists an algorithm,
    $\partTFT_{k,\ell}(\vec{f},\omega)$, that replaces the first $2^\ell$ entries of
    $\vec{f}$ by $F(\omega^{[k\cdot 2^\ell]_p})$, \dots,
    $F(\omega^{[(k+1)\cdot2^\ell-1]_p})$, and an inverse algorithm
    $\partTFT^{-1}_{k,\ell}$ that restores $\vec{f}$ to its initial state. Both
    algorithms are in-place have complexity bounded by
    $\bigO{n+\ell\cdot2^\ell}$.
\end{corollary}
\begin{proof}
    Algorithm $\partTFT_{k,\ell}(\vec{f}, \omega)$ is the following:
    \begin{algorithmic}[1]
    \ForDoEnd{$i=0$ \To $n-1$}{$f_i\fe \omega^{i[k\cdot2^\ell]_p}$}
    \ForDoEnd{$i=2^\ell$ \To $n-1$}{$f_{i-2^\ell}\pe f_i$}
    \STATE $\brDFT_{2^\ell}(\vec{f}_{0..2^\ell-1},\omega^{2^{p-\ell}})$
    \end{algorithmic}
    Its correctness is ensured by~\cref{lem:roche}, while,
    $\partTFT^{-1}_{k,\ell}(\vec{f},\omega)$, its inverse algorithm, does the converse:
    \begin{algorithmic}[1]
    \STATE $\brDFT^{-1}_{2^\ell}(\vec{f}_{0..2^\ell-1},\omega^{2^{p-\ell}})$
    \ForDoEnd{$i=2^\ell$ \To $n-1$}{$f_{i-2^\ell}\me f_i$}
    \ForDoEnd{$i=0$ \To $n-1$}{$f_i\de \omega^{i[k\cdot2^\ell]_p}$}
    \end{algorithmic}
    In both algorithms, the call to $\brDFT^{\pm1}$ has cost
    $\bigO{\ell\cdot{2^\ell}}$, and the two other steps have cost~$\bigO{n}$.
\end{proof}

To implement the previously sketched strategy, we assume that $m\le n$ for
simplicity. We let $\ell$, $t$ be such that $2^\ell\le m<2^{\ell+1}$ and
$2^{\ell+t}\le n<2^{\ell+t+1}$. Using $\partTFT^{\pm 1}$, we are able to compute
$(A(\omega^{[k\cdot 2^\ell]_p}), \dotsc, A(\omega^{[(k+1)\cdot 2^{\ell}-1]_p}))$ for any
$k$ and restore $A$ afterwards.
Similarly, it is possible to compute
$(B(\omega^{[k\cdot 2^{\ell+t}]_p}), \dotsc, B(\omega^{[(k+1)\cdot 2^{\ell+t}-1]_p}))$ and restore $B$.

\begin{algorithm}[!ht]
\caption{In-place fast accumulating polynomial multiplication.}\label{alg:fftaccu}
\begin{algorithmic}[1]
\REQUIRE $\vec{a}$, $\vec{b}$ and $\vec{c}$ of length $m$, $n$ and $m+n-1$,
$m\le n$, containing the coefficients of $A$, $B$, $C\in\D[X]$ respectively;
$\omega\in\D$ principal $2^p$th root of unity with $2^{p-1} < m+n-1 < 2^p$
\ENSURE $\vec{c}$ contains the coefficients of $C+A\cdot B$.
\STATE $\brTFT_{m+n-1}(\vec{c},\omega)$;
    \hfill\COMMENT{over-place}
\STATE Let $r = m+n-1$
\WHILE{$r \ge 0$}
\STATE\label{lin:boundslt}Let $\ell = \lfloor\log_2\min\{r, m\}\rfloor$, $t = \lfloor\log_2\min\{r, n\}\rfloor-\ell$;
    $k = m+n-1-r$
    \STATE\label{lin:TFTb}$\partTFT_{k,\ell+t}(\vec{b},\omega)$
        \hfill\COMMENT{over-place: $B(\omega^{[k\cdot2^{\ell+t}]_p]}), \dotsc, B(\omega^{[(k+1)\cdot 2^{\ell+t}-1]_p})$}
	\FOR{$s=0$ \To $2^t-1$}
	\STATE\label{lin:TFTa}$\partTFT_{s+k\cdot2^t,\ell}(\vec{a},\omega)$
	\Statex\hfill\COMMENT{over-place: $A(\omega^{[(k\cdot2^t+s)2^\ell]_p]}), \dotsc, A(\omega^{[(k\cdot 2^t+s+1)2^\ell-1]_p})$}
	\ForDoEnd{$i=0$ \To $2^\ell-1$}{$c_{i+(k\cdot 2^t+s)2^\ell} \pe  a_i b_{i+s\cdot 2^\ell}$}
	\STATE $\partTFT^{-1}_{s+k\cdot2^t,\ell}(\vec{a},\omega)$
	\hfill\COMMENT{undo~\ref{lin:TFTa} over-place}
	\ENDFOR
    \STATE $\partTFT^{-1}_{k,\ell+t}(\vec{b},\omega)$
    \hfill\COMMENT{undo~\ref{lin:TFTb} over-place}
    \STATE Let $r = r - 2^{\ell+t}$
\ENDWHILE
\STATE $\brTFT^{-1}_{m+n-1}(\vec{c},\omega)$
\end{algorithmic}
\end{algorithm}

\begin{theorem}
    \Cref{alg:fftaccu} is correct and in-place.  If the algorithm
    $\brDFT$ used inside $\partTFT$ has complexity $\bigO{n\log n}$, then the running
    time of~\cref{alg:fftaccu} is $\bigO{n\log n}$.
\end{theorem}

\begin{proof}
    The fact that the algorithm is in-place comes
    from~\cref{cor:partTFT}.
    The only slight difficulty is to produce, fast
    and in-place, the relevant roots of unity. This is actually dealt with in
    the original over-place TFT algorithm~\cite{Harvey:2010:issactft} and can be
    done the same way here.

    To assess its correctness, first note that the values
    of~\cref{lin:boundslt}
    are computed so that $2^\ell\le r,m$ and
    $2^{\ell+t}\le r,n$. One iteration of the while loop updates the entries
    $c_k$ to $c_{k+2^{\ell+t}-1}$ where $k = m+n-1-r$. To this end,
    we first compute $B(\omega^{[k\cdot2^{\ell+t}]_p]})$ to $B(\omega^{[(k+1)\cdot
    2^{\ell+t}-1]_p})$ in $\vec{b}$ using $\partTFT$. Then, since $\vec{a}$ may
    be too small to store $2^{\ell+t}$ values, we compute the corresponding
    evaluations of $A$ by groups of $2^\ell$, using a smaller $\partTFT$. After
    each computation in $\vec{a}$, we update the corresponding entries in
    $\vec{c}$ and restore $\vec{a}$. Finally, at the end of the iteration,
    entries $k$ to $k+2^{\ell+t}-1$ of $\vec{c}$ have been updated and $\vec{b}$
    can be restored. This proves the correctness of the algorithm.

    To bound its complexity,
    we first bound the number of iterations of the
    while loop. We identify two phases, first iterations where $r \ge n$ and
    then iterations with $r < n$. The first phase has at most $3$ iterations since
    $2^{\ell+t}>\frac{n}{2}$ entries of $\vec{c}$ are updated per iteration.
    The second phase starts with $r < n$ and each iteration updates $2^{\ell+t}>\frac{r}{2}$ entries.
    That is, $r$ is halved and this
    second phase has at most $\log_2{n}$ iterations.
    The cost of an iteration is dominated by the calls to $\partTFT^{\pm 1}$. The cost of
    a call to $\partTFT^{\pm 1}_{k,\ell}$ with a size-$m$ input is the sum of a
    linear term $\bigO{m}$ and a non-linear term $\bigO{\ell\cdot 2^\ell}$.
    At each iteration, there are two calls to $\partTFT^{\pm 1}$ on $\vec{b}$ and
    $2^{t+1}$ calls to $\partTFT^{\pm 1}$ on $\vec{a}$. The linear terms sum to
    $\bigO{n+m\cdot 2^t} = \bigO{n}$ since $m\cdot
    2^t < 2^{\ell+1+t} \le 2n$. Over the $\log_2{n}$ iterations, the global
    cost due to these linear terms is $\bigO{n\log n}$.
    The cost due to the non-linear terms in one iteration is
    $\bigO{(\ell+t)\cdot{2^{\ell+t}}}$. In the first iterations,
    $2^{\ell+t} \le n$ and these costs sum to $\bigO{n\log n}$.
    In the next iterations, $2^{\ell+t} \le r < n$. Since $r$ is
    halved at each iteration, the non-linear costs in these
    iterations sum to
    $\bigOdisplay{\sum_i\frac{n}{2^i}\log\frac{n}{2^i}}=\bigO{n\log{n}}$.
\end{proof}

Then, \cref{alg:fftaccu} is compared with previous FFT-based
algorithms for polynomial multiplications designed to reduce their
memory footprint in~\cref{tab:fft} (see also
\cite[Table~2.2]{Giorgi:2019:hdr}).
Note that no call stack is needed for computing the FFT, therefore
these algorithms only require $\bigO{1}$ pointer registers.

\begin{table}[!ht]\centering
\caption{Reduced-memory algorithms for FFT polynomial
  multiplication}\label{tab:fft}
\begin{tabular}{lccc}
\toprule
Algorithm & Algebraic Reg. & Inputs & Accumulation\\
\midrule
\cite{1965:CooleyTukey:MathComp:FFT} & $2n$ & {\color{teal} read-only} & {\xno}\\
\cite{Roche:2009:spacetime} & $\bigO{2^{\lceil\log_2 n\rceil}-n}$ & {\color{teal} read-only} & {\xno}\\
\cite{Harvey:2010:issactft} & \textcolor{teal}{$\bigO{1}$} & {\color{teal} read-only} & {\xno}\\
\cref{alg:fftaccu} & \textcolor{teal}{$\bigO{1}$} & mutable & {\xyes}\\
\bottomrule
\end{tabular}
\end{table}

\section{Fast in-place convolution with
  accumulation}\label{sec:fconv}

In this section, we reduce in-place accumulating convolutions to
in-place multiplication with accumulation. This allows us in the next
section to describe in-place algorithms for structured matrix operations,
and to ultimately obtain an in-place fast polynomial remainder
in the subsequent section.

From algorithms for polynomial multiplications at a lower level, one
can devise an algorithm for the generalized accumulated convolution
$C\pe{AB}\mod{(X^n-f)}$.
To compute such convolutions, one can reduce the computations to full
in-place products. The previous section proves that these computations
have cost $O(\M(n))$.
Then, the initial idea is to unroll a first recursive iteration and then to
call any in-place accumulated polynomial multiplication on halves.

We describe several algorithms. In \cref{ssec:evendim}, we deal with
the case where $n$ is even. \Cref{alg:accinplfconvol} uses a Karatsuba-like
iteration in the case $f\notin\{0,1\}$ while \cref{alg:accinplconvol}
deals with the case $f = 1$. \Cref{ssec:odddim} treats the case of an
odd $n$ when $f$ is invertible, and \cref{ssec:shortprod} the remaining
cases of a short product, that is when $f = 0$. \Cref{ssec:fconv} summarizes
these algorithms and provides the required proofs.

\subsection{Even dimension}\label{ssec:evendim}

Suppose indeed first that $n$ is even. Let $t=n/2$,
and
$A(X)=a_0(X)+X^ta_1(X)$, $B(X)=b_0(X)+X^tb_1(X)$,
$C(X)=c_{0}(X)+X^tc_{1}(X)$, all of degree $n-1=2t-1$
(so that  all of $a_0$, $a_1$, $b_0$, $b_1$, $c_0$ and $c_1$ are of
degree at most $t-1$).
Then let $\tau_0+X^t\tau_1=a_0b_1+a_1b_0$.
Since $X^{2t}=X^n\equiv{f}\mod{X^n-f}$, we have that:
$C + AB\mod{(X^n-f)}=C+a_0b_0+X^t\tau_0+f\cdot\tau_1+f\cdot{a_1b_1}$.
This can be computed with $4$ full accumulated sequential products,
each of degree no more than $2t-2=n-2$, and by exchanging the lower
and upper parts when accumulating $a_0b_1X^t$ and $a_0b_1X^t$:
this comes from the fact that
$(\tau_0+X^t\tau_1)X^t\equiv{X^t\tau_0+f\cdot\tau_1}\mod{X^n-f}$.

\emph{\`A la} Karatsuba, a more efficient version could use only $3$ full
polynomial products: Let instead
$m_0=m_{00}+m_{01}X^t=a_0b_0$,
$m_1=m_{10}+m_{11}X^t=(a_0+a_1)(b_0+b_1)$
and
$m_2=m_{20}+m_{21}X^t=a_1b_1$
be these $3$ full products, to be
computed and accumulated in-place.
As $X^{2t}=X^n\equiv{f}\mod{X^n}$, then $C+{AB}\mod{(X^n-f)}$ is also:
\begin{multline}\label{eq:karaevenconvol}
c_0+m_{00}+fm_{20}+f(m_{11}-m_{01}-m_{21})\\
+\Bigl (c_1+m_{01}+fm_{21}+(m_{10}-m_{00}-m_{20})\Bigr)X^t.
\end{multline}
\Cref{eq:karaevenconvol} is an in-place accumulating linear
computation:
it computes $\vec{c}\pe\mat{\mu}\vec{m}$
where $\vec{m} = (\mat{\alpha}\vec{a})\odot(\mat{\beta}\vec{b})$,
for $\mat{\alpha} = \mat{\beta} = \begin{smatrix} 1&0\\1&1\\0&1\end{smatrix}\in\F^{3\times2}$
and
\(\mat{\mu} =\left[
\begin{smallmatrix}1&-f\\-1&1\end{smallmatrix}\middle|
\begin{smallmatrix}0&f\\1&0\end{smallmatrix}\middle|
\begin{smallmatrix}f&-f\\-1&f\end{smallmatrix}
\right]\in\F^{2\times6}\).
If~$f\notin\{0,1\}$ and $\mat{\mu} = \left[M_0\middle|M_1\middle| M_2\right]$,
\(M_0^{-1}=(1-f)^{-1}\begin{smatrix}1&f\\1&1\end{smatrix}\),
\(M_1^{-1}=\begin{smatrix}0&1\\f^{-1}&0\end{smatrix}\),
and
\(M_2^{-1}=(f^2-f)^{-1}\begin{smatrix}f&f\\1&f\end{smatrix}\).
From this, \cref{eq:doubleexpansion,alg:doublebilin} derives an
accumulating in-place algorithm, using $2\times{2}$ invertible blocks,
that computes
$\begin{smatrix}c_i\\c_j\end{smatrix}\pe{}M\begin{smatrix}\rho_0\\\rho_1\end{smatrix}$,
via the equivalent algorithm:
$\begin{smatrix}c_i\\c_j\end{smatrix}\fe M^{-1}$; $\begin{smatrix}c_i\\c_j\end{smatrix} \pe\begin{smatrix}\rho_0\\\rho_1\end{smatrix}$; $\begin{smatrix}c_i\\c_j\end{smatrix}\fe M$.
After some simplifications described below, we obtain \Cref{alg:accinplfconvol}.

\begin{algorithm}[!ht]
\caption{In-place even degree accumulating $f$-convolution.}\label{alg:accinplfconvol}
\begin{algorithmic}[1]
\REQUIRE $A(X)$, $B(X)$, $C(X)$ polynomials of odd degree $n-1$; $f\in\F\setminus\{0,1\}$.
\ENSURE $C\pe{AB}\mod{(X^n-f)}$
\IF{$n\leq\threshold$}\hfill\COMMENT{constant-time if $\threshold\in\bigO{1}$}
\STATE Apply the quadratic in-place polynomial multiplication.
\ELSE
\STATE Let $t=n/2$;
\hfill\COMMENT{$n$ is even},
\STATE Let $A=a_0+X^ta_1$; $B=b_0+X^tb_1$; $C=c_{0}+X^tc_{1}$; \hfill\COMMENT{degrees $<n/2$}
\STATE $c_{1}\pe{c_{0}}$;
\STATE $c_{1}\de{(1-f)}$;\hfill\COMMENT{simplification \ref{trick:inv}}
\STATE $c_{0}\pe{f\cdot{c_{1}}}$;
\STATE $\begin{bmatrix} c_{0} \\ c_{1}\end{bmatrix} \pe a_0\cdot{b_0} $;
\hfill\COMMENT{acc. full prod. $m_0$~with
  $a_0b_0=\begin{bmatrix}m_{00}\\m_{01}\end{bmatrix}$}
\STATE $c_0\de{f}$;\hfill\COMMENT{simplifications \ref{trick:m2im0}
  and  \ref{trick:neg}}
\STATE $\begin{bmatrix} c_{1} \\ c_{0}\end{bmatrix} \me a_1\cdot{b_1} $;
\hfill\COMMENT{acc. full prod. $m_2$~with
  $a_1b_1=\begin{bmatrix}m_{20}\\m_{21}\end{bmatrix}$}
\STATE $c_{0}\me{c_{1}}$;\hfill\COMMENT{this is $c_0/f+m_{00}/f-m_{01}+m_{20}-m_{21}$}
\STATE $c_{1}\fe{(1-f)}$;\hfill\COMMENT{simplification \ref{trick:mm1im2}}
\STATE $c_{1}\me{f\cdot{c_{0}}}$;\hfill\COMMENT{this is $c_1-m_{00}+m_{01}-m_{20}+fm_{21}$}
\STATE $a_{0}\pe{a_{1}}$;
\STATE $b_{0}\pe{b_{1}}$;
\STATE $\begin{bmatrix} c_{1} \\ c_{0}\end{bmatrix} \pe a_0\cdot{b_0} $;
\hfill\COMMENT{acc. full prod. $m_1$~with
  $(a_0+a_1)(b_0+b_1)=\begin{bmatrix}m_{10}\\m_{11}\end{bmatrix}$}
\STATE $b_{0}\me{b_{1}}$;
\STATE $a_{0}\me{a_{1}}$;
\STATE $c_{0}\fe{f}$;\hfill\COMMENT{simplification \ref{trick:m1j}}
\ENDIF
\end{algorithmic}
\end{algorithm}

The algorithm is obtained from the output
of~\cref{eq:doubleexpansion,alg:doublebilin} after applying the following
simplifications:
\begin{enumerate}[label=(\roman*)]
\item\label{trick:inv} As $1+f(1-f)^{-1}=(1-f)^{-1}$, then
  \(M_0^{-1}\begin{smatrix}c_0\\c_1\end{smatrix}=(1-f)^{-1}\begin{smatrix}1&f\\1&1\end{smatrix}\begin{smatrix}c_0\\c_1\end{smatrix}\)
  can be sequentially computed via $c_1\pe{c_0}$; $c_1\de(1-f)$;
  $c_0\pe{fc_1}$.
\item\label{trick:m2im0} As
\(
(M_2^{-1}
\cdot
M_0)
=
\begin{smatrix}0&-1\\-f^{-1}&0\end{smatrix}
=
-\begin{smatrix}1&0\\0&f^{-1}\end{smatrix}\begin{smatrix}0&1\\1&0\end{smatrix}
\), then computing $M_2$ right after $M_0$
allows simplifying the intermediate in-place operations into a
variable swap, negations and multiplication by $f^{-1}$.
\item\label{trick:neg} That negation can be delayed, since
  \((-M\cdot{c})\pe\rho\) is equivalent to \(-((M\cdot{c})\me\rho)\).
\item\label{trick:mm1im2}
Similarly,
\(
-(M_1^{-1}
\cdot
M_2)
=
\begin{smatrix}1&-f\\-1&1\end{smatrix}
\). Up to some swaps, this is simplified as
\(
\begin{smatrix}1&0\\-f&1\end{smatrix}{\!\cdot\!}
\begin{smatrix}1&0\\0&1-f\end{smatrix}{\!\cdot\!}
\begin{smatrix}1&-1\\0&1\end{smatrix}
\), equivalent to the
sequential computation:
$c_0\me{c_1}$; $c_1\fe(1-f)$; $c_1\me{fc_0}$.
\item\label{trick:m1j}
  Finally,
  $M_1=\begin{smatrix}f&0\\0&1\end{smatrix}\begin{smatrix}0&1\\1&0\end{smatrix}$
  and thus this combination is again just a swap of variables and a
  multiplication by $f$.
\end{enumerate}
Overall, we obtain~\cref{alg:accinplfconvol} that works for even $n$
with $f\notin\{0,1\}$.
We then need to design variant algorithms for the other cases.
\Cref{alg:accinplconvol} deals with the case where $f=1$
(as $M_0$ and $M_2$ are not invertible when $f=1$).
We present here a version with $4$ full products, but a more
efficient version with $3$ products only,
like~\cref{alg:accinplfconvol}, could also be derived.

\begin{algorithm}[!ht]
\caption{In-place even degree accumulating $1$-convolution.}\label{alg:accinplconvol}
\begin{algorithmic}[1]
\REQUIRE $A(X)$, $B(X)$, $C(X)$ polynomials of odd degree $n-1$;
\ENSURE $C\pe{AB}\mod{(X^n-1)}$
\IF{$n\leq\threshold$}\hfill\COMMENT{constant-time if $\threshold\in\bigO{1}$}
\STATE Apply the quadratic in-place polynomial multiplication.
\ELSE
\STATE Let $t=n/2$;
\hfill\COMMENT{$n$ is even, let $Y=X^t$, so that $Y^2\equiv{1}$},
\STATE Let $A=a_0+X^ta_1$; $B=b_0+X^tb_1$; $C=c_{0}+c_{1}X^t$;
\STATE $\begin{bmatrix} c_{0} \\ c_{1}\end{bmatrix} \pe a_0\cdot{b_0} $
\hfill\COMMENT{acc. full prod. of degree $2t-2\leq{n-1}$}
\STATE $\begin{bmatrix}c_{0}\\c_{1}\end{bmatrix}\pe{a_1\cdot{b_1}}$
\hfill\COMMENT{acc. full prod. since $Y^2\equiv{1}$}
\STATE $\begin{bmatrix} c_{1} \\ c_{0}\end{bmatrix} \pe a_0\cdot{b_1} $
\hfill\COMMENT{acc. full prod. since $(u+vY)Y\equiv{v+uY}$}
\STATE $\begin{bmatrix} c_{1} \\ c_{0}\end{bmatrix} \pe a_1\cdot{b_0} $
\hfill\COMMENT{acc. full prod. since $(u+vY)Y\equiv{v+uY}$}
\ENDIF
\end{algorithmic}
\end{algorithm}

\subsection{Odd dimension}\label{ssec:odddim}

\Cref{alg:accinplfconvoleven} deals with the odd-$n$ case where
$f$ is invertible. For the sake of simplicity, we also only present the
version with $4$ full products. The additional
difficulty here is that the degrees of the lower and upper parts are
different. Therefore,~\cref{alg:accinplfconvoleven} ensures that there
is always the correct space to accumulate.

\begin{algorithm}[!ht]
\caption{In-place odd degree accumulating $f$-convolution.}\label{alg:accinplfconvoleven}
\begin{algorithmic}[1]
\REQUIRE $A(X)$, $B(X)$, $C(X)$ polynomials of even degree $n-1$; $f\in\F^*$.
\ENSURE $C\pe{AB}\mod{(X^n-f)}$
\IF{$n\leq\threshold$}\hfill\COMMENT{constant-time if $\threshold\in\bigO{1}$}
\STATE Apply the quadratic in-place polynomial multiplication.
\ELSE
\STATE Let $t=(n+1)/2$;
\hfill\COMMENT{$n$ is odd, so that $X^{2t}\equiv{fX}$},
\STATE Let $A=a_0+X^ta_1$; $B=b_0+X^tb_1$; $C=c_{0}+c_{1}X^t$;
\Statex\hfill\COMMENT{$a_1$, $b_1$, $c_1$ of degree $n-1-t=t-2$}
\STATE $\begin{bmatrix} c_{0} \\ c_{1}\end{bmatrix} \pe a_0\cdot{b_0}$
\hfill\COMMENT{acc. full prod. of degree $2t-2\leq{n-1}$}
\STATE $\begin{bmatrix} c_{1..(t-1)} \\ c_{1}\end{bmatrix} \pe a_1\cdot{b_1}$
\hfill\COMMENT{acc. full prod. of degree $(2t-4)+1\leq{n-1}$}
\STATE $c_{0..(t-2)}\de{f}$;
\STATE $\begin{bmatrix} c_{t..(2t-2)} \\ c_{0..(t-2)}\end{bmatrix} \pe a_0\cdot{b_1} $
\hfill\COMMENT{acc. full prod. since $\left\lbrace\begin{smallmatrix} 2t-3<n-1\\(u+vX^{t-1})X^t\equiv{fv+uX^t}\end{smallmatrix}\right.$}
\STATE $\begin{bmatrix} c_{t..(2t-2)} \\ c_{0..(t-2)}\end{bmatrix} \pe a_1\cdot{b_0} $
\hfill\COMMENT{acc. full prod. since $\left\lbrace\begin{smallmatrix}2t-3<n-1\\(u+vX^{t-1})X^t\equiv{fv+uX^t}\end{smallmatrix}\right.$}
\STATE $c_{0..(t-2)}\fe{f}$;
\ENDIF
\end{algorithmic}
\end{algorithm}

\subsection{Short product}\label{ssec:shortprod}

Finally, we also have to deal with the case $f=0$, that is with the
in-place short product $C\pe{AB}\mod{X^n}$.
A first idea would be to again split the input polynomials in two
parts,
$A(Y)= X^{n/2}a_1 + a_0$
and
$B(Y) = X^{n/2}b_1 + b_0$,
then compute $C(Y)\pe (a_0b_0)+(a_1b_0+a_0b_1 \mod X^{n/2})$.
This would be one full in-place polynomial multiplication with
accumulation, for $a_0b_0$,
and two recursive calls of half degree, for $a_1b_0$ and $a_0b_1$.
Unfortunately, these two recursive calls when splitting in two would
induce some extra logarithmic factors in the complexity when the full
multiplication is not $\M(m)=\Theta(m^{1+\epsilon})$ for some
$\epsilon>0$.

We have obtained different alternative solutions,
either with less recursive calls, or splitting the input polynomials
in more parts, and some of them with restrictions on the field
characteristic.
We first present one variant in detail
in~\cref{ssec:s3l4r1}.
We then give only the \textsc{hm} representations of more variants,
and compare their respective
complexity bounds in~\cref{ssec:shortprods,ssec:spefficiency}.

\begin{remark}\label{rq:oldsp}
The algorithms described below have a better complexity than
\cite[Alg. 4]{jgd:2024:inplacerem}. Another approach, suitable
over fields with at least three elements, is to compute the
short product as a sum of two convolutions modulo $X^n-f$ for
two non-zero distinct values $f$,
\emph{cf.}~\cite[Remark~1]{jgd:2024:inplacerem}.
This approach is also slightly less efficient than the ones
presented here.
\end{remark}

\subsubsection{Splitting in three, with a single recursive call}\label{ssec:s3l4r1}
We here split the input and output polynomials in $3$~blocks of size
close to~$n/3$. A first possibility is given
in~\cref{eq:karaoneinfive}, with overlapping intermediate polynomials~$t_i$:
\begin{equation}\label{eq:karaoneinfive}
\fbox{\scalebox{.85}[0.85]{
\ensuremath{\begin{aligned}
A(Y)& = Y^2a_2 + Ya_1 + a_0;
B(Y) = Y^2b_2 + Yb_1 + b_0;
C(Y) = Y^2c_2 + Yc_1 + c_0;\\
m_0 &= a_0(b_0{-}b_2);
m_1= a_0b_2;
m_2= (a_0{+}a_1)b_1;
m_3= a_1(b_0{-}b_1);
m_4= (a_1{+}a_2)b_0;\\
t_0&=c_0 + m_0 + m_1;\quad
t_1=c_1 + m_2 + m_3;\quad
t_2=c_2 + m_1 - m_3 + m_4;\\
\text{\textbf{then}}&~C+AB\mod{Y^3}\equiv{Y^2t_2+Yt_1+t_0\mod{Y^3}}
\end{aligned}}}}
\end{equation}

The \textsc{hm} representation corresponding to the complete
operations of~\cref{eq:karaoneinfive} (\emph{i.e.}, $Y^2t_2+Yt_1+t_0$ without
discarding the degrees larger than $Y^3$) is given in~\cref{eq:s3l4r1}:
\begin{equation}\label{eq:s3l4r1}
\mat{\mu}=\begin{bmatrix}
1&1&0&0&0\\
0&0&1&1&0\\
0&1&0&-1&1\\
\end{bmatrix};\quad
\mat{\alpha}=\begin{bmatrix}
1&0&0\\
1&0&0\\
1&1&0\\
0&1&0\\
0&1&1\\
\end{bmatrix};\quad
\mat{\beta}=\begin{bmatrix}
1&0&-1\\
0&0&1\\
0&1&0\\
1&-1&0\\
1&0&0\\
\end{bmatrix}
\end{equation}

Splitting the five products $m_i=m_{i0}+m_{i1}Y$, as
in~\cref{eq:bilinkarasplit}, we obtain the $2\times{2}$ expansion
of~$\mat{\mu}$ in~\cref{eq:s3l4r1} as a $4\times{10}$ matrix
$\mat{\mu}^{(2)}$.
The technique to get a short product algorithm from this, is to not
compute the high degree coefficients: this is just discarding the last
row of $\mat{\mu}^{(2)}$, to obtain the $3\times{10}$ matrix given
in~\cref{eq:exponeinfive}:
\begin{equation}\label{eq:exponeinfive}
\renewcommand\arraystretch{.75}\setlength\arraycolsep{4pt}
\begin{bmatrix} c_0\\c_1\\c_2\end{bmatrix}
\pe
\left[\begin{array}{cc|cc|cc|cc|cc}
1&0 & 1&0 & 0&0 & 0&0 & 0&0\\
0&1 & 0&1 & 1&0 &1&0 & 0 &0\\
0&0 & 1&0 & 0&1 &-1&1 & 1 &0
\end{array}\right]\cdot
\begin{bmatrix}m_{00}\\m_{01}\\\vdots\\m_{41}\end{bmatrix}.
\end{equation}

Now the four first pairs of columns of~\cref{eq:exponeinfive} can be
dealt with the technique of~\cref{alg:doublebilin}
and the last pair is $m_{40}\mod{X^{n/3}}$, thus
performed via a single recursive call of degree $n/3$.
Note that one can directly read off
the matrix $\mat{\mu}$ in~\cref{eq:s3l4r1} that there is a single recursive call:
the number of recursive calls is indeed the
number of columns having a single non-zero in the last row (after
expansion and discarding, this column becomes a pair of non-full rank
columns in~\cref{eq:exponeinfive}).
\begin{remark}\label{rk:onereccall}
Note that since $m_0$ is used only once,
one could modify~\cref{eq:karaoneinfive} to get a seemingly
simplified version as follows:
replace
$m_0 = a_0(b_0{-}b_2)$; $m_1= a_0b_2$; $\ldots$
$t_0=c_0 + m_0 + m_1$
by instead
$m'_0 = a_0b_0$; $m_1= a_0b_2$; $\ldots$
$t_0=c_0 + m'_0$.
But this would now require a second recursive call, as the high
degree part of $m_1$ can not be computed nor stored anymore.
\end{remark}

For the first four calls to full-products, it is sufficient to extract
the first
four columns of $\mu$ and the first four rows of $\alpha$ and $\beta$
in~\cref{eq:s3l4r1}:
\[
\begin{smatrix}c_0\\c_1\\c_2\end{smatrix}
\pe
\mu^{(2)}_{1..3,1..8}
\cdot
\left(\mat{\alpha}_{1..4,*}\begin{smatrix}a_0&a_1&a_2\end{smatrix}
\odot
\mat{\beta}_{1..4,*}\begin{smatrix}b_0&b_1&b_2\end{smatrix}\right)
\]

\Cref{alg:doublebilin} can then be called on these sub-matrices via
running the \plinopt~library with:
\begin{verbatim}
./bin/trilplacer -e data/2o2o2_4_partSP_{L,R,P}.sms
\end{verbatim}
It is then sufficient to extract only the computations $c_0,c_1,c_2$
and remove that of $c_3$ (the highest
degrees of the result are not used when computing
modulo $X^n$ -- or $Y^3=(X^{n/3})^3$ in~\Cref{eq:karaoneinfive}).
This part thus needs only $10$ additions and $4$ calls to in-place
accumulating products of degrees $n/3$ to compute the first four
columns of~\cref{eq:exponeinfive}. This is shown in
lines~\ref{line:begfirstfourcol} to~\ref{line:endfirstfourcol}
of~\cref{alg:accinplshortoneinfive}.

Then, the recursive call is the low degree part of the last column of
$\mu$ and the last row of $\alpha$ and $\beta$:
$c_2\pe{\lbrace(a_1+a_2)\cdot{b_0}\rbrace_{\text{low}}}$. Finally one
deals with degrees above $3\lceil{n/3}\rceil$ directly and, overall,
we now obtain~\cref{alg:accinplshortoneinfive}.

\begin{algorithm}[htp]
\caption{In-place accumulating short product with $1$ recursive call.}\label{alg:accinplshortoneinfive}
\begin{algorithmic}[1]
\REQUIRE $A(X)$, $B(X)$, $C(X)$ polynomials of degree $<n$ in $\F[X]$;
\ENSURE $C\pe{AB}\mod{X^n}$.
\IF{$n\leq\threshold$}\hfill\COMMENT{constant-time if $\threshold\in\bigO{1}$}
\STATE Apply the quadratic in-place polynomial multiplication.
\ELSE
\STATE Let $t=\lfloor{n/3}\rfloor$;
\STATE Let $A\mod{X^{3t}}=a_0+X^ta_1+X^{2t}a_2$; $B\mod{X^{3t}}=b_0+X^tb_1+X^{2t}b_2$;
$C\mod{X^{3t}}=c_{0}+c_{1}X^t+X^{2t}c_{2}$;
\Statex\COMMENT{First four columns of~\Cref{eq:exponeinfive}
via~\cref{alg:doublebilin}:}
\STATE $b_{0}\me{b_2}$;\label{line:begfirstfourcol}
\STATE $\begin{bmatrix}c_{0}\\c_{1}\end{bmatrix}\pe{a_{0}\cdot{b_0}}$;
\hfill\COMMENT{acc. full prod.}
\STATE $b_{0}\pe{b_{2}}$; $c_{2}\me{c_{0}}$;
\STATE $\begin{bmatrix}c_{0}\\c_{1}\end{bmatrix}\pe{a_{0}\cdot{b_{2}}}$;
\hfill\COMMENT{acc. full prod.}
\STATE $a_{1}\pe{a_{0}}$;
\STATE $\begin{bmatrix}c_{1}\\c_{2}\end{bmatrix}\pe{a_{1}\cdot{b_{1}}}$;
\hfill\COMMENT{acc. full prod.}
\STATE $a_{1}\me{a_{0}}$; $b_{0}\me{b_{1}}$; $c_{2}\pe{c_{1}}$;
\STATE $\begin{bmatrix}c_{1}\\c_{2}\end{bmatrix}\pe{a_{1}\cdot{b_{0}}}$;
\hfill\COMMENT{acc. full prod.}
\STATE $b_{0}\pe{b_{1}}$; $c_{2}\me{c_{1}}$; $c_{2}\pe{c_{0}}$;\label{line:endfirstfourcol}
\Statex\COMMENT{Fifth pair of columns of~\Cref{eq:exponeinfive}:}
\STATE $a_1\pe{a_2}$;
\STATE $c_2\pe{a_1}\cdot{b_0}\mod{X^t}$ \hfill\COMMENT{recursive call: $c_2\pe{m_4}\mod{X^t}$}
\STATE $a_1\me{a_2}$;\label{line:restoreA}
\Statex\COMMENT{Directly computing highest degrees:}
\IF{$n\geq{3t+1}$}\hfill\COMMENT{scalar~accumulations}
\ForDoEnd{$i=0$ \To $3t$}{$c_{3t}\pe{A_i}\cdot{B_{3t-i}}$}
\ENDIF
\IF{$n={3t+2}$}\hfill\COMMENT{scalar accumulations}
\ForDoEnd{$i=0$ \To $3t+1$}{$c_{3t+1}\pe{A_i}\cdot{B_{3t+1-i}}$}
\ENDIF
\ENDIF
\end{algorithmic}
\end{algorithm}

\begin{remark}\label{rk:callstack}
Even though~\cref{alg:accinplshortoneinfive} only makes one recursive
call, this is not a tail recursive call. Therefore, a call stack is
\emph{a priori} needed. Yet, it is easily seen that this call stack
can be removed. Indeed, \cref{alg:accinplshortoneinfive} could be
split in two parts: the first one is a tail recursive algorithm made
of all instructions up to the recursive call; the second one is not
recursive, and it only restores $A$ (\cref{line:restoreA}) and
computes the coefficients $c_{3t}$ and $c_{3t+1}$ if needed. This way,
no call stack is required since the only recursive algorithm is tail
recursive.
\end{remark}

\begin{theorem} Using an in-place polynomial multiplication with
  complexity bounded by \M(n),
  \cref{alg:accinplshortoneinfive}
  is correct, in-place and has complexity bounded by
$\frac{4}{3^{1+\epsilon}-1}\M(n)+\bigO{n}$.
\end{theorem}
\begin{proof}
\cref{alg:accinplshortoneinfive} uses $4$
polynomial multiplications with polynomials of degree $n/3$, then $1$
recursive call of a short product of degree $n/3$, and $\bigO{n}$
extra computations.
Solving the recursion for this cost, gives an overall complexity bound
of~$\frac{4}{3^{1+\epsilon}-1}\M(n)+\bigO{n}$ if
$\M(n)=\Theta(n^{1+\epsilon})$ for $0<\epsilon\leq{1}$.
If instead $\M(n)=\softoh{n}$, one then gets $2\M(n)+\bigO{n}$ which
is also $\frac{4}{3^{1+\epsilon}-1}\M(n)$ with $\epsilon=0$.
\end{proof}

\subsubsection{Short product varying the number of recursive
  calls}\label{ssec:shortprods}
By setting up a polynomial system with the coefficients of the three
matrices in an \textsc{hm} representation as indeterminates,
one can solve for matrices that produce a short product.
With this we performed a search on the obtained solutions for
various splittings (in $2$ or $3$) and associated tensor rank
(respectively $3$ and $5$) and various recursive calls.

From this we obtained that:
\begin{itemize}
\item Splitting in $2$, there is no algorithm for the short product
  with no recursive calls (and $3$ full multiplications) working in
  even characteristic;
\item Splitting in $3$, there is no algorithm for the short product
  with strictly more than $2$ recursive calls (and tensor rank $5$);
\item Splitting in $3$, there is no algorithm for the short product
  with no recursive calls (and $5$ full multiplications) working in
  even characteristic.
\end{itemize}

Then for the other situations we have found several possibilities and
present some of them next, showing only their initial \textsc{hm}
representation.

\paragraph{Splitting in 3 with 2 recursive calls and 3 full multiplications}
We found~\cref{eq:s3l3r2}:
\begin{equation}\label{eq:s3l3r2}
\mu = \begin{bmatrix}
1&0&0&0&0\\
-1&-1&1&0&0\\
0&1&0&1&1\\
\end{bmatrix},
\quad\quad
\alpha=\begin{bmatrix}
1&0&0\\
0&1&0\\
1&1&0\\
1&0&0\\
0&0&1\\
\end{bmatrix},
\quad\quad
\beta=\begin{bmatrix}
1&0&0\\
0&1&0\\
1&1&0\\
0&0&1\\
1&0&0\\
\end{bmatrix}.
\end{equation}

On the one hand,~\cref{eq:s3l3r2} uses $3$ polynomial
multiplications with polynomials of degree $n/3$, then $2$ recursive
calls of short products of degree $n/3$, and $\bigO{n}$ extra
computations. This
gives an overall complexity bound
of~$\frac{3}{3^{1+\epsilon}-2}\M(n)+\bigO{n}$.
For instance with a quadratic full polynomial multiplication
($\epsilon=1$) the in-place short product is computed at a cost lower
than half that of the full one (more precisely, the ratio is $3/7$).

In terms of $\epsilon$, the cut-off point
between~\cref{eq:s3l3r2,alg:accinplshortoneinfive}
is then at $1+\epsilon=\log_3(5)\approx{1.465}$ (the exponent of
Toom-$3$ algorithms).

\paragraph{Split in 3 with no recursive calls, in odd characteristic}
\begin{equation}\label{eq:s3l5r0}
\mu = \begin{bmatrix}
1&0&0&1&0\\
0&\frac{1}{2}&\frac{1}{2}&0&1\\
0&0&1&1&1\\
\end{bmatrix},
\quad\quad
\alpha = \begin{bmatrix}
2&1&-1\\
-1&1&0\\
1&1&0\\
-1&-1&1\\
0&-1&0\\
\end{bmatrix},
\quad\quad
\beta = \begin{bmatrix}
1&0&0\\
1&-2&1\\
1&0&1\\
1&0&0\\
0&-1&1\\
\end{bmatrix}.
\end{equation}

\paragraph{Split in 2 with 1 recursive call}
\begin{equation}\label{eq:s2l2r1}
\mu = \begin{bmatrix}
0&1&-1\\
1&1&0\\
\end{bmatrix},
\quad\quad
\alpha = \begin{bmatrix}
1&0\\
0&1\\
-1&1\\
\end{bmatrix},
\quad\quad
\beta = \begin{bmatrix}
0&1\\
1&0\\
1&0\\
\end{bmatrix}.
\end{equation}

\paragraph{Split in 2 with no recursive calls, in odd characteristic
  (and 1 recursive call in even characteristic)}
\begin{equation}\label{eq:s2l3r0}
\mu = \begin{bmatrix}
-2&1&1\\
1&-1&0\\
\end{bmatrix},
\quad\quad
\alpha = \begin{bmatrix}
1&2\\
0&1\\
1&1\\
\end{bmatrix},
\quad\quad
\beta = \begin{bmatrix}
0&1\\
-1&2\\
1&2\\
\end{bmatrix}.
\end{equation}

\subsubsection{Short product efficiency}\label{ssec:spefficiency}
We now give in~\cref{tab:spefficiency} the dominant term of the
complexity bounds for the different short product algorithms
of~\cref{ssec:shortprods}.
For this we just present the ratios when used with an underlying full
polynomial multiplication of complexity
$\M(n)=\Theta(n^{1+\epsilon})$, for $0<\epsilon\leq{1}$.
Now, when $\M(n)=\softoh{n}$, letting $\epsilon=0$, also provides the
correct ratio.

\begin{table}[ht]\centering
  \caption{Short product complexity ratio with
    $\M(n)=\Theta(n^{1+\epsilon})$.}\label{tab:spefficiency}
\begin{tabular}{cccccc}
\toprule
Alg.
& $\frac{n}{2}$, $0$ rec.
& $\frac{n}{2}$, $1$ rec.
& $\frac{n}{3}$, $0$ rec.
& $\frac{n}{3}$, $1$ rec.
& $\frac{n}{3}$, $2$ rec.
\\
Equation & (\ref{eq:s2l3r0})& (\ref{eq:s2l2r1})& (\ref{eq:s3l5r0})&
(\ref{eq:s3l4r1}) & (\ref{eq:s3l3r2}) \\
Char. & odd & any & odd & any & any\\
Extra Mem. reg. & $\bigO{1}$  & $\bigO{1}$ & $\bigO{1}$ & $\bigO{1}$& $\bigO{1}$\\
Extra Ptr. reg. & $\bigO{1}$  & $\bigO{1}$ & $\bigO{1}$ & $\bigO{1}$& $\bigO{\log{n}}$\\
\midrule
Ratio & $\ddfrac{3}{2^{1+\epsilon}}$ & $\ddfrac{2}{2^{1+\epsilon}-1}$ & $\ddfrac{5}{3^{1+\epsilon}}$ & $\ddfrac{4}{3^{1+\epsilon}-1}$ & $\ddfrac{3}{3^{1+\epsilon}-2}$\\
\midrule
$1+\epsilon=2$ &$\frac{3}{4}$ &$\frac{2}{3}$ & $\frac{5}{9}$ & $\frac{1}{2}$ & $\frac{3}{7}$ \\
Karatsuba & ${1}$ & ${1}$& ${0.88}$& ${0.85}$& ${0.81}$ \\
Toom-$3$ & ${1.09}$ & ${1.14}$& ${1}$& ${1}$& ${1}$ \\
\multicolumn{6}{c}{$\vdots$} \\
$\M(n)=\softoh{n}$ &$\ddfrac{3}{2}$ &${2}$ & $\frac{5}{3}$ & ${2}$ & ${3}$ \\
\bottomrule
\end{tabular}
\end{table}

First, as noted in~\cref{rq:oldsp}, \cref{alg:accinplshortoneinfive}
(from~\cref{eq:s3l4r1}) is always better than
the in-place algorithms we gave in
\cite[Alg. 4]{jgd:2024:inplacerem}
and
\cite[Remark 1]{jgd:2024:inplacerem},
with ratio
$\frac{4}{3^{1+\epsilon}-1}$
compared to, respectively,
$\frac{5}{3^{1+\epsilon}-2}$
and
$2$.

Then, as long as the number of recursive calls is not larger than $1$,
the short product algorithms are as in-place as the underlying full
polynomial multiplication, thanks to~\cref{rk:callstack}.
For instance when used with~\cref{alg:fftaccu},
the algorithms obtained from
\cref{eq:s2l3r0,eq:s2l2r1,eq:s3l5r0,eq:s3l4r1}
are fully in-place ($\bigO{1}$ memory \emph{and} pointer registers);
while \cref{eq:s3l3r2} would require $\bigO{1}$ memory registers but
$\bigO{\log(n)}$ pointer registers.

As expected, on the one hand, we also see that when the underlying
polynomial multiplication is closer to quasi-linear, then performing
the least possible number of recursive call is more interesting.
On the other hand, when the underlying polynomial multiplication is
closer to quadratic, then performing as many recursive calls as
possible is better.

Finally, there is most probably room for improvements if one considers
splitting the initial polynomials in more than three blocks.

\subsection{Fast in-place convolution with accumulation}\label{ssec:fconv}
Finally,
\cref{alg:accinplfconvol,alg:accinplconvol,alg:accinplfconvoleven,alg:accinplshortoneinfive}
all together provide a complete solution for the fast in-place
convolution with accumulation, as given in~\cref{alg:accinplallconvol}.

\begin{algorithm}[!ht]
\caption{In-place convolution with accumulation.}\label{alg:accinplallconvol}
\begin{algorithmic}[1]
\REQUIRE $A(X)$, $B(X)$, $C(X)$ polynomials of degree $<n$; $f\in\F$.
\ENSURE $C\pe{AB}\mod{(X^n-f)}$
\IfThen{$f=0$}{Apply~\cref{alg:accinplshortoneinfive}}
\ElseIf{$n$ is odd}{Apply~\cref{alg:accinplfconvoleven}}\hfill\COMMENT{$f\neq{0}$}
\ElseIf{$f=1$}{Apply~\cref{alg:accinplconvol}}\hfill\COMMENT{$n$
  is even}
\ElseEnd{Apply~\cref{alg:accinplfconvol}.}\hfill\COMMENT{$f\not\in\{0,1\}$}
\end{algorithmic}
\end{algorithm}

\begin{theorem}\label{thm:convol}
  Using an in-place polynomial multiplication with
  complexity bounded by \M(n), \cref{alg:accinplallconvol} is
  correct, in-place and has complexity bounded by $\bigO{\M(n)}$.
\end{theorem}
\begin{proof}
The correctness of~\cref{alg:accinplfconvol} comes from that of
\cref{eq:karaevenconvol}.
The correctness of~\cref{alg:accinplconvol,alg:accinplfconvoleven} is
direct from the degrees of the sub-polynomials (as given in the
comments, line by line).
The correctness
of~\cref{alg:accinplshortoneinfive} comes
from that of \cref{eq:karaoneinfive}.
Also, all four algorithms are in-place as they use only in-place atomic
operations or in-place polynomial multiplication.

The complexities of
\cref{alg:accinplfconvol,alg:accinplconvol,alg:accinplfconvoleven} satisfy
$T(n) \leq 4\M(n/2)+\bigO{n}$ since they use
only a linear number of operations plus up to four calls to polynomial
multiplications of half-degrees. Since $\M(n)/n$ is non-decreasing,
$T(n) = \bigO{\M(n)}$.
Finally, \cref{alg:accinplshortoneinfive}
use a linear number of atomic operations, three or four {degree-$n/3$}
polynomial multiplications
and only two or one recursive calls with degree-$n/3$ polynomials.
Thus, their complexity bounds satisfy
$T(n)\leq{2T(n/3)+3\M(n/3)+\bigO{n}}$
or
$T(n)\leq{1T(n/3)+4\M(n/3)+\bigO{n}}$, whence
in both cases $T(n)=\bigO{\M(n)}$ since $\M(n)/n$ is non-decreasing.
\end{proof}

\section{Circulant and Toeplitz matrix operations with accumulation}
\label{sec:toeplitz}
Toeplitz matrix-vector multiplication can be reduced to circulant
matrix-vector multiplication, via an embedding into a double-size
circulant matrix. But this is not immediately in-place, since doubling
the size requires a double space.
We see in the following how we can instead double the operations while
keeping the same dimension.
We start by the usual definitions, also extending circulant matrices
to $f$-circulant matrices, following, e.g.,
\cite[Theorem~2.6.4]{Pan:2001:SMP}.

\begin{definition}
For $\vec{a}\in\F^m$, $\Circulant{}(\vec{a})$ is the \emph{circulant matrix}
represented by $\vec{a}$, that is, the
$m\times{m}$ matrix $(C_{ij})$, such that $C_{1j}=a_j$ and the
$(i+1)$st row is the cyclic right shift by $1$ of the $i$th row.
\end{definition}

\begin{definition}
For $f\in\F$ and $\vec{a}\in\F^m$, the (lower) \emph{$f$-circulant matrix}
represented by $\vec{a}$, $\Circulant{f}(\vec{a})$, is the
$m\times{m}$ matrix $(\Gamma_{ij})$, such that:
\[\text{for}~C=\Circulant{}(\vec{a}),
\begin{cases} \Gamma_{ij}=C_{ij} & \text{if}~i\leq{j}\text{,} \\
  \Gamma_{ij} = f\cdot{C_{ij}} & \text{otherwise.}
\end{cases}\]
\end{definition}

\subsection{Fast in-place accumulating ($f$-)circulant matrix vector product}\label{ssec:fcirculant}

It is well known that circulant matrices are diagonalized by a
discrete Fourier transform, and hence can be manipulated via fast
Fourier transforms as
$\Circulant{1}(\vec{a})=F_m^{-1}\text{diag}(F_m\vec{a})F_m$,
for $F_m$ a DFT-matrix,
see, e.g.,~\cite[\S~4.7.7]{Golub:1996:MatrixC}.
This gives an alternative way to compute circulant matrix-vector
multiplication in-place (using and restoring afterwards both the
matrix and the vector). Indeed, the (even truncated) Fourier transform
and its inverse can be computed
in-place~\cite{Roche:2009:spacetime,Harvey:2010:issactft,Coxon:2022:JSC:inplaceTFT}.
This gives us an in-place algorithm to compute the \emph{accumulation}
$\vec{c}\pe\Circulant{1}(\vec{a})\cdot\vec{b}$ as:
\begin{algorithmic}[1]
\STATE $\vec{a}\leftarrow{F_m\vec{a}}$, $\vec{b}\leftarrow{F_m\vec{b}}$
  and $\vec{c}\leftarrow{F_m\vec{c}}$;
\STATE $\vec{c}\pe\text{diag}(\vec{a})\cdot\vec{b}$;
\STATE $\vec{c}\leftarrow{F_m^{-1}\vec{c}}$;
  $\vec{b}\leftarrow{F_m^{-1}\vec{b}}$ and $\vec{a}\leftarrow{F_m^{-1}\vec{a}}$.
\end{algorithmic}

Now, this diagonalization enables us to compute fast in-place accumulated
circulant matrix-vector only when primitive roots of sufficiently
large order exist.
In the following we present more generic fast in-place algorithms that
work over any field.

The algebra of $f$-circulant matrices is in fact isomorphic to the
algebra of polynomials modulo
$X^n-f$~\cite[Theorem~2.6.1]{Pan:2001:SMP}.
This means that the product of an $f$-circulant matrix by a vector is
obtained by the convolution of this vector by the vector representing
the $f$-circulant matrix, and thus via~\cref{alg:accinplallconvol}.
Note that, as recalled above,
the case $f=1$ can also be
computed using a discrete Fourier transform, but only
when primitive roots of sufficiently large order exist.
By contrast, \cref{alg:accinplallconvol} has
no restriction on $f$ and is a reduction to any accumulated in-place
polynomial multiplication (including DFT ones, as, e.g.,
\cref{alg:fftaccu}).

\subsection{Fast in-place accumulating Toeplitz matrix-vector product}\label{ssec:toeplitz}
\begin{definition}
For $\vec{a}\in\F^{2m-1}$, $\Toeplitz{}(\vec{a})$ is the (square) \emph{Toeplitz matrix}
represented by $\vec{a}$, that is, the
$m\times{m}$ matrix $(T_{ij})$, such that
$T_{ij}=a_{m+j-i}$.
Similarly, for $\vec{a}\in\F^{m+n-1}$, we denote by
$\Toeplitz{m,n}(\vec{a})$ the $m\times{n}$ rectangular Toeplitz matrix
defined by its first column, $\vec{a}_{1..m}$ bottom to top, and its
first row, $\vec{a}_{m..(m+n-1)}$, left to right.
\end{definition}

The matrix-vector product for rectangular Toeplitz matrices
and the middle product of polynomials are the same task, see, e.g.,
\cite[\S~3.1]{Giorgi:2019:issac:reductions}.
We also immediately see that with these notations, we have for instance
$\Circulant{1}(\vec{a})=\Circulant{}(\vec{a})$,
$\Circulant{0}(\vec{a})=\frac{1}{2}(\Circulant{1}(\vec{a})+\Circulant{-1}(\vec{a}))$,
or also $\Circulant{f}(\vec{a})=\Toeplitz{}([f{\cdot}\vec{a}_{2..m},\vec{a}])$,
where $[\vec{u},\vec{v}]$ denotes the vector obtained by concatenation
of $\vec{u}$ and $\vec{v}$.

Fast algorithms for $f$-circulant matrices then provide algorithms,
by reduction, for accumulation with triangular
and square Toeplitz matrices first, as sums of $f$-circulant
in~\cref{alg:uppertriangtoeplitz,alg:squaretoeplitz},
and then for any Toeplitz matrix, again as sums of triangular Toeplitz
matrices in~\cref{alg:toeplitz}.

\begin{algorithm}[!ht]
  \caption{In-place accumulating Upp. Triang. Toeplitz m-v. mult.}
  \label{alg:uppertriangtoeplitz}
  \begin{algorithmic}[1]
    \REQUIRE $\vec{a},\vec{b},\vec{c}\in\F^m$.
    \ENSURE $\vec{c}\pe\Toeplitz{}([\vec{0},\vec{a}])\cdot\vec{b}$.
    \STATE $\vec{c}\pe\Circulant{0}(\vec{a})\cdot\vec{b}$.\hfill\COMMENT{\cref{alg:accinplallconvol}}
  \end{algorithmic}
\end{algorithm}

\begin{remark}\label{rk:trspalg}
Similarly to \cref{alg:accinplallconvol}, we can design an algorithm for accumulating lower triangular Toeplitz matrix-vector product. This algorithm is automatically obtained from \cref{alg:accinplallconvol} by reversing the indices in the matrix and vectors. More generally, given an algorithm that computes a matrix-vector product $\vec{c} = A\cdot\vec{b}$, the \emph{reversed} algorithm obtained by reversing all indices computes the reversed vector $\cev{c} = \Transpose{A}\cdot\cev{b}$.

In particular, we here present $f$-circulant matrices where the coefficient
acts on the lower left part of the matrix (excluding the diagonal). Algorithms
can be reversed to deal with the
other type of $f$-circulant matrices, where the coefficient would on
the upper right part of the matrix (excluding the diagonal).
\end{remark}

\begin{algorithm}[!ht]
  \caption{In-place accumulating square Toeplitz m-v. mult.}
  \label{alg:squaretoeplitz}
  \begin{algorithmic}[1]
    \REQUIRE $\vec{a_1}\in\F^{m}$, $\vec{a_2},\vec{b},\vec{c}\in\F^{m+1}$,
    \ENSURE
    $\vec{c}\pe\Toeplitz{}([\vec{a_1},\vec{a_2}])\cdot\vec{b}$.
    \STATE Let $\vec{b_1}=\vec{b}_{1..m}$ and $\vec{c_2}=\vec{c}_{2..m+1}$;
    \STATE
      $\cev{c_2}\pe\Circulant{0}(\cev{a_1})\cdot\cev{b_1}$;\hfill\COMMENT{\cref{alg:accinplallconvol} (reversed)}
    \STATE $\vec{c}\pe\Circulant{0}(\vec{a_2})\cdot\vec{b}$;\hfill\COMMENT{\cref{alg:accinplallconvol}}

  \end{algorithmic}
\end{algorithm}

\begin{lemma}\label{lem:upsqtoeplitz}
 \Cref{alg:uppertriangtoeplitz,alg:squaretoeplitz}
  are correct and have complexity bounded by $\bigO{\M(n)}$.
 \end{lemma}
\begin{proof}
  The complexity bound comes from that of~\cref{alg:accinplallconvol}.
  Correctness is obtained directly looking at the values of the
  matrices.
  First, for \cref{alg:uppertriangtoeplitz}, we have that
  $\Toeplitz{}([\vec{0},\vec{a}])=\Circulant{0}(\vec{a})$.
    Second, for \cref{alg:squaretoeplitz},
  $\Toeplitz{}([\vec{a_1},\vec{a_2}])=\Circulant{0}(\vec{a_2})+  \begin{smatrix}\Transpose{\vec{0}}&0\\\Transpose{\Circulant{0}(\cev{a_1})}&\vec{0}\end{smatrix}$.
    The computation $\Transpose{\Circulant{0}(\cev{a_1})}\cdot \vec{b_1}$ is the reversed algorithm of
    $\Circulant{0}{\cev{a_1}}\cdot\cev{b_1}$ as explained in \cref{rk:trspalg}.
\end{proof}

From this, we give in~\cref{alg:toeplitz} an in-place rectangular
Toeplitz matrix-vector multiplication.

\begin{algorithm}[!ht]
  \caption{In-place accumulating rectangular Toeplitz matrix-vector multiplication.}
  \label{alg:toeplitz}
  \begin{algorithmic}[1]
    \REQUIRE $\vec{a}\in\F^{m+n-1}$, $\vec{b}\in\F^n$, $\vec{c}\in\F^m$,
    \ENSURE $\vec{c}\pe\Toeplitz{m,n}(\vec{a})\cdot\vec{b}$.
    \IF{$m=n$}
    \STATE $\vec{c}\pe\Toeplitz{}(\vec{a})\cdot\vec{b}$;\hfill\COMMENT{\cref{alg:squaretoeplitz}}
    \ELSIF{$m>n$}
    \STATE Let $c_1=\vec{c}_{1..n}$ and $c_2=\vec{c}_{(n+1)..m}$;
    \STATE
    $c_1\pe\Toeplitz{}(\vec{a}_{(m-n+1)..(m+n-1)})\cdot\vec{b}$;\hfill\COMMENT{\cref{alg:squaretoeplitz}}
    \STATE
    $c_2\pe\Toeplitz{m-n,n}(\vec{a}_{1..(m-1)})\cdot\vec{b}$;\hfill\COMMENT{recursive call}
    \ELSE
    \STATE Let $b_1=\vec{b}_{1..m}$ and $b_2=\vec{b}_{(m+1)..n}$;
    \STATE
    $c\pe\Toeplitz{}(\vec{a}_{1..(2m-1)})\cdot{b_1}$;\hfill\COMMENT{\cref{alg:squaretoeplitz}}
    \STATE
    $c\pe\Toeplitz{m,n-m}(\vec{a}_{(m+1)..(m+n-1)})\cdot{b_2}$;\hfill\COMMENT{recursive call}
    \ENDIF
  \end{algorithmic}
\end{algorithm}
\begin{proposition}\label{prop:toeplitz}
  \Cref{alg:toeplitz} is correct. Its complexity is bounded by
  $\bigOdisplay{\frac{\max\{m,n\}}{\min\{m,n\}}\M(\min\{m,n\})}$ operations.
\end{proposition}
\begin{proof} If $m>n$, there are $\lfloor{m/n}\rfloor$
  calls to~\cref{alg:squaretoeplitz} in size $n$, requiring
  $\bigO{(m/n)\M(n)}\leq\bigO{\M(m)}$ operations.
  The remaining recursive call is then negligible.
  This is similar when~$m<n$.
\end{proof}

\begin{remark}\label{rk:generalizedtoeplitz}
\Cref{alg:uppertriangtoeplitz,alg:squaretoeplitz,alg:toeplitz} and their reversed versions can be combined to compute in-place the accumulation $\vec{c} \pe T\cdot\vec b$ for any $\vec{c}\in\F^m$, $\vec{b}\in\F^n$, and band Toeplitz matrix $T = \Toeplitz{m,n}([\vec{0}_k,\vec{a},\vec{0}_{t}])$ where $\vec{a}\in\F^\ell$ and $t = m+n-k-\ell-1$. The cost is $\bigOdisplay{\frac{\max(m,n)}{\min(m,n)}\M(\min(m,n))}$.

The idea is to tile the matrix $T$ with upper triangular, lower triangular, and rectangular Toeplitz matrices. Note that \cref{alg:toeplitz} cannot be used directly on $T$ since it requires the vectors $\vec{a_1}$ and $\vec{a_2}$ to be writable, while the zeroes of $T$ are not stored.
\end{remark}

\subsection{Fast over-place Toeplitz matrix operations}\label{ssec:overtoeplitz}

These in-place accumulated Toeplitz matrix-vector multiplications
allow us to obtain both over-place triangular Toeplitz
multiplication and system
solve, given in~\cref{alg:overtriangtoeplitz,alg:solvetriangtoeplitz}.
If $\M(m)=\Theta(m^{1+\epsilon})$ for some
$\epsilon>0$, then the obtained algorithm have the same asymptotic
cost as they not-in-place counterparts.
Otherwise, an extra logarithmic factor now appears in the complexity
bounds.

\begin{algorithm}[!ht]
  \caption{Over-place triang. Toeplitz m-v. mult.}
  \label{alg:overtriangtoeplitz}
  \begin{algorithmic}[1]
    \REQUIRE $\vec{a},\vec{b}\in\F^m$, s.t. $a_1\in\F^{*}$.
    \ENSURE
    $\vec{b}\leftarrow\Toeplitz{}([\vec{a},\vec{0}])\cdot\vec{b}$.
    \IF{$m\leq\threshold$}\hfill\COMMENT{constant-time if $\threshold\in\bigO{1}$}
    \STATE Apply  the quadratic in-place triang. m-v. mult.
    \hfill\COMMENT{\cref{alg:classical}}
    \ELSE
    \STATE Let $k=\lceil{m/2}\rceil$, $b_1=\vec{b}_{1..k}$ and $b_2=\vec{b}_{(k+1)..m}$;
    \STATE ${b}_2\leftarrow\Toeplitz{}([\vec{a}_{(k+1)..m},\vec{0}])\cdot{b}_2$;
    \hfill\COMMENT{recursive call}
    \STATE ${b}_2\pe\Toeplitz{m-k,k}([a_1,\ldots,a_{m-1}])\cdot{b}_1$;
    \hfill\COMMENT{\cref{alg:toeplitz}}
    \STATE ${b}_1\leftarrow\Toeplitz{}([a_{(m-k+1)..m},\vec{0}])\cdot{b}_1$;
    \hfill\COMMENT{recursive call}
    \ENDIF
  \end{algorithmic}
\end{algorithm}

\begin{proposition}\label{prop:overtriangtoeplitz}
\Cref{alg:overtriangtoeplitz} is correct and its complexity his
bounded by $\bigO{\M(m)\log(m)}$ operations,
or $\bigO{\M(m)}$ if $\M(m)=\Theta(m^{1+\epsilon})$ for some $\epsilon>0$.
\end{proposition}
\begin{proof}
For the correctness, let $T=\Toeplitz{}([\vec{a},\vec{0}])$ and consider
it as blocks
$T_1=\Toeplitz{}([a_{(m-k+1)..m},\vec{0}])$,
$T_2=\Toeplitz{}([\vec{a}_{(k+1)..m},\vec{0}])$
 and
$G=\Toeplitz{m-k,k}([a_1,\ldots,a_{m-1}])$. Then
$T=\begin{smatrix}T_1&0\\G&T_2\end{smatrix}$.
Thus $T\vec{b}=\begin{smatrix}T_1{b}_1\\G{b}_1+T_2{b}_2 \end{smatrix}$.
Let $\bar{b}_1=T_1{b}_1$,
$\hat{b}_2=T_2{b}_2$ and
$\bar{b}_2=G{b}_1+T_2{b}_2$.
Then
$\bar{b}_2=\hat{b}_2+G{b}_1$ and the algorithm is correct.
Now for the complexity bound,
the cost function is
$T(m)\leq{2T(m/2)+\bigO{\M(m)}}$,
that is $\bigO{\M(m)\log(m)}$,
or $\bigO{\M(m)}$ if $\M(m)=\Theta(m^{1+\epsilon})$ for some $\epsilon>0$.
\end{proof}

\begin{algorithm}[!ht]
  \caption{Over-place triang. Toeplitz system solve.}
  \label{alg:solvetriangtoeplitz}
  \begin{algorithmic}[1]
    \REQUIRE $\vec{a},\vec{b}\in\F^m$, s.t. $a_1\in\F^{*}$.
    \ENSURE
    $\vec{b}\leftarrow\Toeplitz{}([\vec{0},\vec{a}])^{-1}\cdot\vec{b}$.
    \IF{$m\leq\threshold$}\hfill\COMMENT{constant-time if $\threshold\in\bigO{1}$}
    \STATE Apply the quadratic in-place triang. syst.
    solve.
    \hfill\COMMENT{\cref{alg:classical}}
    \ELSE
    \STATE Let $k=\lceil{m/2}\rceil$, $b_1=\vec{b}_{1..k}$ and $b_2=\vec{b}_{(k+1)..m}$;
    \STATE ${b}_2\leftarrow\Toeplitz{}([\vec{0},\vec{a}_{1..(m-k)}])^{-1}\cdot{b}_2$;
    \hfill\COMMENT{recursive call}
    \STATE ${b}_1\me\Toeplitz{k,m-k}(\vec{a}_{2..m})\cdot{b}_2$;
    \hfill\COMMENT{\cref{alg:toeplitz}}
    \STATE ${b}_1\leftarrow\Toeplitz{}([\vec{0},\vec{a}_{1..k}])^{-1}\cdot{b}_1$;
    \hfill\COMMENT{recursive call}
    \ENDIF
  \end{algorithmic}
\end{algorithm}

\begin{proposition}\label{prop:solvetriangtoeplitz}
\Cref{alg:solvetriangtoeplitz} is correct
 and has its complexity bounded by $\bigO{\M(m)\log(m)}$ operations,
or $\bigO{\M(m)}$ if $\M(m)=\Theta(m^{1+\epsilon})$ for some $\epsilon>0$.
\end{proposition}
\begin{proof}
First, let $T=\Toeplitz{}([\vec{0},\vec{a}])$ and consider
it as blocks
$T_1=\Toeplitz{}([\vec{0},\vec{a}_{1..k}])$,
$T_2=\Toeplitz{}([\vec{0},\vec{a}_{1..(m-k)}])$ and
$G=\Toeplitz{k,m-k}(\vec{a}_{2..m})$. Then
$T=\begin{smatrix}T_1&G\\0&T_2\end{smatrix}$.
Now define $H$,
s.t. $T^{-1}=\begin{smatrix}T_1^{-1}&H\\0&T_2^{-1}\end{smatrix}$.
Then $H$ satisfies $T_1^{-1}G+HT_2=0$.
Also, we have
$T^{-1}\vec{b}=\Transpose{\begin{smatrix}T_1^{-1}{b}_1+H{b}_2&T_2^{-1}{b}_2\end{smatrix}}$.
Let $\bar{b}_2=T_2^{-1}{b}_2$ and
$\bar{b}_1=T_1^{-1}{b}_1+H{b}_2$.
Then
$\bar{b}_1=T_1^{-1}{b}_1+HT_2\bar{b}_2=T_1^{-1}{b}_1-T_1^{-1}G\bar{b}_2=T_1^{-1}\left({b}_1-G\bar{b}_2\right)$
and this shows that the algorithm is correct.
Now for the complexity bound,
the cost function is
$T(m)\leq{2T(m/2)+\bigO{\M(m)}}$,
that is $T(m) = \bigO{\M(m)\log(m)}$,
or $\bigO{\M(m)}$ if $\M(m)=\Theta(m^{1+\epsilon})$ for some $\epsilon>0$.
\end{proof}

Note that by means of \cref{rk:trspalg}, we also have over-place algorithms for
\emph{upper} triangular Toeplitz matrix-vector multiplication and for
\emph{lower} triangular Toeplitz system solving.

\subsection{Fast in-place accumulating Toeplitz-like matrix-vector product}\label{sec:disprank}

Structured matrices such as Toeplitz and circulant matrices can be generalized
and unified with the notion of displacement rank~\cite{Pan:2001:SMP}. In
particular, a matrix $A\in\F^{m\times n}$ is said Toeplitz-like of displacement
rank $\alpha$ if the matrix $\nabla_Z(A) = A-Z_mA\Transpose{Z_n}$ has rank
$\alpha$, where $Z_m\in\F^{m\times m}$ is defined by $(Z_m)_{i,i-1} = 1$ and
$(Z_m)_{i,j} = 0$ if $j\neq i-1$. In such a case, $A$ admits \emph{generators}
$G\in\F^{m\times\alpha}$ and $H\in\F^{n\times\alpha}$ such that $\nabla_Z(A) =
G\Transpose{H}$.

We describe here an algorithm that takes as inputs the generators $G$, $H$ of a
Toeplitz-like matrix $A$ and two vectors $\vec b$, $\vec c$, and computes the
accumulation $\vec c\pe A\vec b$ in place. The tool for this is the so-called
\emph{$\Sigma LU$-formula}: $\nabla_Z(A) = G\Transpose H$ if and only if \[A =
\sum_{i=1}^{\alpha} L(\vec{g_i})\cdot U(\vec{h_i})\] where $\vec{g_i}$ (resp.
$\vec{h_i}$) is the $i$th column of $G$ (resp. $H$), and $L(\vec g) =
\Toeplitz{m,\ell}([\cev g,\vec 0])$ and $U(\vec h) = \Toeplitz{\ell,n}([\vec
0,\vec h])$ for $\ell = \min(m,n)$.

\begin{algorithm}[!ht]
  \caption{In-place accumulating Toeplitz-like m-v. mult.}
  \label{alg:toeplitzlike}
  \begin{algorithmic}[1]
    \REQUIRE Generators $G\in\F^{m\times\alpha}$ and $H\in\F^{n\times\alpha}$ of a Toeplitz-like matrix $A\in\F^{m\times n}$, $\vec{b}\in\F^n$, $\vec{c}\in\F^m$,
    \ENSURE $\vec{c}\pe A\cdot\vec{b}$.
      \FOR{$i=1$ \To $\alpha$}
      \STATE Let $\vec g_i$, $\vec h_i$ be the $i$th columns of $G$ and $H$
      \STATE Let $k$ be the index of the first non-zero entry of $\vec h_i$
      \STATE Let $\ell = \min(m,n-k+1)$
      \STATE Let $L = \Toeplitz{m,\ell}([\cev g,\vec 0])$, $U_1 = \Toeplitz{\ell,\ell}([\vec 0,(\vec h_i)_{k,k+\ell-1}])$ and $\vec b_1 = \vec b_{k..k+\ell-1}$
      \STATE Let $U_2 = \Toeplitz{\ell,n-k+1-\ell}([\vec 0, (\vec h_i)_{k+\ell,n}])$ and $\vec b_2=\vec b_{k+\ell..n}$
        \hfill\COMMENT{only if $\ell = m$}
      \STATE $\vec b_1 \gets U_1\cdot\vec b_1$ \hfill\COMMENT{\cref{alg:overtriangtoeplitz} (reversed)}\label{lin:U1}
      \STATE $\vec b_1 \pe U_2\cdot \vec b_2$ \hfill\COMMENT{\cref{alg:toeplitz}, only if $\ell = m$}\label{lin:U2}
      \STATE $\vec c \pe L\cdot\vec b$ \hfill\COMMENT{\cref{rk:generalizedtoeplitz}}
      \STATE $\vec b_1 \me U_2\cdot \vec b_2$ \hfill\COMMENT{undo \cref{lin:U2}, \cref{alg:toeplitz}}
      \STATE $\vec b_1 \gets U_1^{-1}\cdot\vec b_1$ \hfill\COMMENT{undo \cref{lin:U1}, \cref{alg:solvetriangtoeplitz}}
      \ENDFOR
  \end{algorithmic}
\end{algorithm}

\begin{proposition}\label{prop:toeplitzlike}
\Cref{alg:toeplitzlike} is in-place, correct and requires
\[\bigOdisplay{\left(\frac{\max(m,n)}{\min(m,n)}+\log\min(m,n)\right)\M(\min(m,n))}\]
operations, or $\bigO{\frac{\max(m,n)}{\min(m,n)}\M(\min(m,n))}$ if $\M(n)
= \Theta(n^{1+\epsilon})$ for some $\epsilon > 0$.
\end{proposition}
\begin{proof}
To prove the correctness of the algorithm, we first consider the $\Sigma
LU$-formula, rewritten as $A = \sum_{i=1}^\alpha L_i U_i$ where $L_i =
\Toeplitz{m,\ell}([\cev g_i,\vec 0])$ and $U_i = \Toeplitz{\ell,n}([\vec
0,\vec h_i])$. The goal is to compute $A\cdot \vec b$ as $\sum_i L_i
U_i\vec b$ by first computing $U_i\vec b$ into $\vec b$, then accumulating
$L_i\vec b$ into $\vec c$ and finally restoring $\vec b$.  If the first
non-zero entry of $\vec h_i$ has index $k$, the $k-1$ left columns of $U_i$
are made of zeroes. Therefore, $L_i U_i\vec b = LU\vec b_{k..n}$ where $L =
\Toeplitz{m,\ell}([\cev g_i,\vec 0])$ and $U = \Toeplitz{\ell,n-k+1}([\vec
0, (\vec h_i)_{k..n}])$.  Now if $\ell = m$, the matrix $U$ is rectangular.
We split it into $[U_1 | U_2]$ where $U_1$ is an upper triangular square
Toeplitz matrix. Then $U\cdot\vec b_{k..n} = U_1\vec b_1 + U_2\vec b_2$.
This can be computed in $\vec b_1$, and $\vec b_1$ can be then restored.

For the complexity, we consider that no $\vec h_i$ starts with a zero, and let
$\ell = \min(m,n)$. The over-place triangular Toeplitz matrix computations
(\cref{alg:overtriangtoeplitz,alg:solvetriangtoeplitz}) have cost
$\bigO{\M(\ell)\log(\ell)}$ or $\bigO{\M(\ell)}$ if $\M(\ell) =
\Theta(\ell^{1+\epsilon})$ for some $\epsilon>0$. If $\ell = m$, operations
    with $U_2$ and $L$ have total cost $\bigO{\frac{n}{m}\M(m)}$. If $\ell = n$, $U_2$
is empty and the cost of $\vec c \pe L\cdot\vec b_1$ is
$\bigO{\frac{m}{n}\M(n)}$. Altogether, we obtain the announced cost.
\end{proof}

\section{Fast in-place modular remainder}\label{sec:reminplace}
We consider now the fast in-place (resp. over-place) computation of the
Euclidean polynomial modular remainder $R = A\bmod B$
(resp. $A=A\bmod B$)
with $A$ and $B$ of respective degrees $m+n$ and $n$.
Standard algorithms for the remainder require
$\bigOdisplay{\frac{m}{n}\M(n)}$ arithmetic operations and, apart from
that of $A$ and $B$, at least $\bigO{m}$ extra memory~\cite{Giorgi:2020:issac:inplace}
to store the whole quotient $Q$
such that $A=BQ+R$ with $\deg{R}<\deg{B}$.
We first show how to avoid the storage of the whole quotient (as was hinted
in~\cite{Giorgi:2020:issac:inplace}),
and propose an algorithm still using $\bigOdisplay{\frac{m}{n}\M(n)}$
arithmetic operations but only $n$ extra space.

Second, we combine this with the techniques
of~\cref{sec:fconv,sec:toeplitz} and use the input space of $A$ or $B$
for intermediate computations in order to derive in-place and
over-place algorithms for the modular remainder using at most
$\bigOdisplay{\M(m)\log(m)}$ arithmetic operations,
or $\bigO{\M(m)}$ if $\M(m) = \Theta(m^{1+\epsilon})$ for some
$\epsilon>0$.
In practice, this means that the asymptotic arithmetic cost of
not-in-place algorithms is preserved by our in-place versions, except
when the not-in-place base case is in the FFT regime, for which we
have an extra $\log{n}$ factor.

Our first step is to interpret the results of \cref{ssec:toeplitz,ssec:overtoeplitz} in terms of polynomial
and power series operations.

\begin{remark}
A linear-algebraic derivation of the results of this section
is presented in the conference paper~\cite{jgd:2024:inplacerem}.
\end{remark}

\subsection{Toeplitz computations as polynomial operations}

In \cref{ssec:toeplitz,ssec:overtoeplitz}, several in-place and over-place algorithms for
Toeplitz matrix-vector operations have been described. We give here their
interpretation in terms of polynomial or power series operations, in order
to use them in our in place remainder algorithms.

We identify a polynomial $A\in\F[X]$ with its vector of coefficients
$\vec{a}\in\F^n$ where $\deg(A) = n-1$. Then $A = \sum_{i=0}^{n-1} a_i X^i$.
(Note that here vectors are indexed from $0$ to $n-1$.)

Let
$\vec a$, $\vec b$, $\vec c\in\F^n$ and $A$, $B$, $C\in\F[X]$ be their corresponding
polynomials of degree $<n$.
For the lower triangular Toeplitz matrix $L = \Toeplitz{}([\cev a,\vec 0])$,
the matrix-vector product $\vec c\gets L\cdot\vec b$
corresponds to the \emph{short product} $C\gets A\cdot B\bmod X^n$.
This is the natural product for truncated power series known at precision $n$.

We shall need the \emph{reversed} operation, sometimes called a \emph{high short product},
as well as its inverse.
For a polynomial
$A$, let $\cev{A} = X^{\deg(A)}A(1/X)$ be its reversed polynomial.

\begin{definition}
Let
$\vec a$, $\vec b$, $\vec c\in\F^n$ and $A$, $B$, $C\in\F[X]$ their corresponding
    polynomials of degree $<n$, and assume that $a_{n-1}$ is non-zero. Let
    $U_{\vec a} = \Toeplitz{}([\vec 0,\cev a])$ be an upper triangular Toeplitz matrix.
\begin{itemize}
\item The \emph{reversed short product} of $A$ and $B$, denoted $A\revmul B$, is the polynomial
    $C$ of degree $<n$ defined by one of the following equivalent equations: $C = A\cdot B\bquo X^{n-1}$, $\cev C = \cev A\cdot \cev B\bmod X^n$, or $\vec c = U_{\vec a}\cdot\vec b$.
\item The \emph{reversed power series division} of $A$ and $B$, denoted $A\revdiv B$, is the polynomial
    $C$ of degree $<n$ defined by one of the following equivalent equations: $\cev C = \cev A/\cev B\bmod X^n$, or $\vec c = U_{\vec{a}}^{-1}\cdot\vec b$.
\end{itemize}
\end{definition}

Let now $\vec a\in\F^{m+n-1}$, $\vec b\in\F^n$ and $\vec c\in\F^m$ and let
$T = \Toeplitz{}([\cev a])$ be a rectangular Toeplitz matrix. The matrix-vector
product $\vec c\gets T\cdot\vec b$ corresponds to the \emph{middle product}
$C \gets (A\cdot B)\bmod X^{m+n-1} \bquo X^n$. In particular, when $m = n$ this
is the standard middle product of polynomials~\cite{Hanrot:2004:MiddleProd}.

\begin{proposition}\label{prop:toeplitzpoly}
Let $A$, $B$, $C\in\F[X]$ of degrees $<n$. Then
\begin{enumerate}[label=(\roman*)]
    \item\label{item:revmul} \Cref{alg:uppertriangtoeplitz} computes $C\pe B\revmul A$ in-place
        in $\bigO{\M(n)}$ operations;
    \item\label{item:shortprod} \Cref{alg:overtriangtoeplitz} computes $B \gets B\cdot A\bmod X^n$ over-place in $\bigO{\M(n)\log n}$ operations, or $\M(n)$ if $\M(n) = \Theta(n^{1+\epsilon})$ for some $\epsilon > 0$;
    \item\label{item:revdiv} \Cref{alg:solvetriangtoeplitz} computes $B \gets B\revdiv A$ over-place in $\bigO{\M(n)\log n}$ operations, or $\M(n)$ if $\M(n) = \Theta(n^{1+\epsilon})$ for some $\epsilon > 0$.

    \item\label{item:midprod} If $\deg(A) = m+n-2$, $\deg(B) = n-1$ and $\deg(C) = m-1$, \cref{alg:squaretoeplitz,alg:toeplitz} compute
    $C\pe (A\cdot B)\bmod X^{m+n-1}\bquo X^n$ in $\bigO{\frac{\mu}{\nu}\M(\nu)}$ operations where $\mu=\max(m,n)$
    and $\nu = \min(m,n)$.
\end{enumerate}
\end{proposition}
\begin{remark}\label{rk:toeplitzpoly}
By \cref{rk:trspalg}, the reversed version of \cref{alg:overtriangtoeplitz}
computes $B\gets B\revmul A$ in place, in the same complexity.  By
\cref{rk:generalizedtoeplitz}, \cref{alg:overtriangtoeplitz} also works in
place in the same complexity if $\deg(A) < n-1$.
\end{remark}

\subsection{Small-space remainder by overwriting the quotient}

The starting point of our derivation is the standard long division algorithm
applied block by block, when $A$ is large compared to $B$. Let $n = \deg(B)$
and write $A = \sum_{i=0}^{k-1} A_i\cdot X^{ni}$ where each $A_i$ has degree
at most $n-1$. The long division algorithm recalled in \cref{alg:longdiv} computes
the Euclidean division of $A$ by $B$ as follows.

\begin{algorithm}[!ht]
  \caption{Long division of polynomials.}\label{alg:longdiv}
  \begin{algorithmic}[1]
    \REQUIRE $A$, $B$, $Q$, $R$ in $\F[X]$ of respective degrees $m+n$,
    $n$, $m$ and $n-1$.
    \READONLY $A$, $B$.
    \ENSURE $Q = A\bquo B$ and $R=A\bmod B$.
    \STATE Let $k = \lfloor\frac{m}{n}\rfloor$ and write $A = \sum_{i=0}^{k-1} A_i X^{ni}$
        \hfill\COMMENT{polynomials of deg. $n-1$}
    \STATE $R \gets A_{k-1}$
    \STATE $Q \gets 0$
    \FOR{$i=k-2$ \DownTo $0$}
        \STATE $R \gets RX^n+A_i$ \label{line:addAi}
        \STATE $T \gets R\bquo B$ \label{line:computeT}
        \STATE $R \me BT$ \label{line:updateR}
        \STATE $Q \gets X^nQ+T$
    \ENDFOR
  \end{algorithmic}
\end{algorithm}

The first remark is that to compute the remainder, one does not
need to retain $Q$ but only $T$. The second aspect is to dive into the
quotient computation $R\bquo B$. The fast algorithm consists in a power
series division. More precisely,
at \cref{line:computeT},
$\cev{T} = \cev{R}/\cev{B}\bmod X^n$
where the division is a (truncated) power series division (\emph{cf.} for
instance~\cite[Chapter~9]{MCA2013}). Since the computation is at precision $n$,
only the first $n$ coefficients of $\cev{R}$ are needed. That is,
the update at \cref{line:addAi} can be postponed and merged with
\cref{line:updateR}, and
\cref{line:computeT} becomes a (reversed) power series division $T = R\revdiv B$.
\cref{line:updateR} becomes $R = RX^n+A_i-BT$ and since the result of this computation
has degree $<n$, the computation can be performed modulo $X^n$. It becomes
$R = A_i - (BT\bmod X^n)$. Altogether, we obtain \cref{alg:remainder}.

\begin{algorithm}[!ht]
  \caption{Overwritten-quotient Euclidean remainder.}\label{alg:remainder}
  \begin{algorithmic}[1]
    \REQUIRE $A$, $B$, $R$ in $\F[X]$ of respective degrees $m+n$, $n$ and $n-1$.
    \READONLY $A$, $B$.
    \ENSURE $R={A}\bmod{B}$.
    \IF{$m< 0$}
      \STATE $R\gets A$;
    \ELSE
    \STATE\label{lin:padd}Let $k = \lfloor\frac{m}{n}\rfloor$ and
      write $A = \sum_{i=0}^{k-1} A_i X^{ni}$
        \hfill\COMMENT{polynomials of deg. $n-1$}
    \STATE Let $B^* = B\bquo X$ and $B_* = B\bmod X^n$
    \STATE $R \gets A_{k-1}$
    \FOR{$i=k-2$ \DownTo $0$}
	\STATE\label{line:quotient} $T \gets R \revdiv B^*$; \hfill\COMMENT{rev. power series division}
        \STATE\label{line:remain}  $R \gets -B_*T\bmod X^n$; \hfill\COMMENT{short product}
	\STATE\label{line:substract} $R \pe A_i$;
    \ENDFOR
    \ENDIF
  \end{algorithmic}
\end{algorithm}

\begin{theorem}\label{thm:quotientfree}
\Cref{alg:remainder} is correct and requires $\bigOdisplay{\frac{m}{n}\M(n)}$
arithmetic operations and $n$ extra memory space.
\end{theorem}
\begin{proof}
Correctness follows from the previous discussion. For the complexity bounds,
(reversed) power series division can be computed in place in $\bigO{\M(n)}$
ring operations if the dividend can be erased~\cite[Corollary~2.6]{Giorgi:2020:issac:inplace}.
Here $R$ can be safely erased. Then the short product can also be computed in-place
in $O(\M(n))$ ring operations~\cite[Theorem~5.1]{Giorgi:2019:issac:reductions}.
Altogether, the time complexity is $O(k \M(n)) = O(\frac{m}{n}\M(n))$.
The only required extra space is the degree-$(n-1)$ polynomial $T$.
\end{proof}

\begin{remark}
For a polynomial $B$, denote by $\revdiv_B$ the operator defined by $\revdiv_B(A) = A\revdiv B$, and $\bmul_B$ the operator defined by $\bmul_B(A) = A\cdot B\bmod X^{\deg(B)}$. Then \cref{alg:remainder} gives rise to a formula for the remainder, namely
\[A\bmod B = \sum_{i=0}^{k-1} \left(\bmul_{B_*}\circ\revdiv_{B^*}\right)^i(A_i)\]
where the exponent notation denotes the $i$th iterate of a function. This formula is the exact counterpart of the linear-algebraic formula derived in~\cite[Eq.~(11)]{jgd:2024:inplacerem}.
\end{remark}

\subsection{In-place remainders via Toeplitz techniques}\label{ssec:toeprem}
We now derive algorithms that use only $\bigO{1}$ extra memory
space in the in-place model of~\cref{ssec:inplace}: Modifying the
inputs is possible if and only if all inputs are restored to their
initial state after the completion of the algorithm.
This allows us to store some intermediate results, over-writing the
inputs, provided that we can afterwards recompute the initial inputs in
their entirety. Further, this enables recursive calls, as intermediate
values are used but restored along the recursive descent.
The general idea is then to use variants of \cref{alg:remainder} based
on results from~\cref{sec:fconv,sec:toeplitz}.

\newcommand{\IPER}{\ensuremath{\text{IPER}}\xspace}
\newcommand{\OPER}{\ensuremath{\text{OPER}}\xspace}
\newcommand{\APER}{\ensuremath{\text{APER}}\xspace}

We present three variants of in-place polynomial remaindering:
\begin{itemize}
\item \IPER: given $A$ and $B$, it computes in-place $R = A\bmod B$
    using only the output space for $R$ and that of the modulus $B$
    (\emph{i.e.}, $A$ is read-only and $B$ is restored to its initial state after completion);
\item \OPER: given $A$ and $B$, it computes both $Q = A\bquo B$ and $R = A\bmod B$
    (such that $A =BQ+R$), replacing $A$ by $\langle Q,R\rangle$
    (with $B$ restored after completion);
\item \APER: given $A$, $B$ and $R$, it computes $R\pe{A}\bmod{B}$, accumulating
    the remainder into $A$
    (with both $A$ and $B$ restored after completion).
\end{itemize}
We present \IPER in~\cref{alg:inplaceremainder}:
this variant
replaces only~\cref{line:quotient,line:remain,line:substract}
of~\cref{alg:remainder} by their over-place variants,
\cref{alg:solvetriangtoeplitz,alg:overtriangtoeplitz}, that modify and
restore parts of $B$.
\begin{algorithm}[!ht]
  \caption{\IPER$(R,A,B)$: In-place Polynomial Euclidean Remainder.}\label{alg:inplaceremainder}
  \begin{algorithmic}[1]
    \REQUIRE $A$, $B$, $R$ in $\F[X]$ of respective degrees $m+n$, $n$ and $n-1$.
    \READONLY $A$.
    \ENSURE $R=A\bmod B$.
    \IF{$m<0$}
    \STATE $R\gets A$;
    \ELSE
    \STATE Let $k = \lfloor\frac{m}{n}\rfloor$ and write
      $A = \sum_{i=0}^{k-1} A_i X^{ni}$ \hfill\COMMENT{polynomials of deg. $n-1$}
    \STATE Let $B^* = B\bquo X$ and $B_* = B\bmod X^n$
    \STATE $R\gets A_{k-1}$;
    \FOR{$i=k-2$ \DownTo $0$}
	\STATE\label{line:IPquotient} $R \gets R\revdiv B^*$;
        \hfill\COMMENT{\cref{prop:toeplitzpoly}~\ref{item:revdiv}}
        \STATE\label{line:IPremain}  $R \gets -B_*R\bmod X^n$;
        \hfill\COMMENT{\cref{prop:toeplitzpoly}~\ref{item:shortprod}}
	\STATE\label{line:IPsubstract} $R\pe A_i$;
    \ENDFOR
    \ENDIF
  \end{algorithmic}
\end{algorithm}

\begin{theorem}\label{thm:remainder}
  \Cref{alg:inplaceremainder} is correct, in-place and has complexity
  bounded by $\bigO{\frac{m}{n}\M(n)\log(n)}$ operations,
  or $\bigO{\frac{m}{n}\M(n)}$ if $\M(n) = \Theta(n^{1+\epsilon})$ for some $\epsilon>0$.
\end{theorem}
\begin{proof} \Cref{alg:inplaceremainder} calls $\bigO{k} = \bigO{\frac{m}{n}}$ times
  \cref{alg:overtriangtoeplitz,alg:solvetriangtoeplitz}. And each call requires
  $\bigO{\M(n)\log(n)}$ operations,
  or $\bigO{\M(n)}$ if $\M(n) = \Theta(n^{1+\epsilon})$ for some $\epsilon>0$,
  by~\cref{prop:overtriangtoeplitz,prop:solvetriangtoeplitz}.
\end{proof}

\subsection{Over-place and accumulating remainders}\label{sec:operaper}
We give two variants of~\cref{alg:inplaceremainder}. In \OPER
(\cref{alg:overplaceremainder}), $A$ is replaced by $[A\bquo B,A\bmod B]$ while
$B$ is ultimately restored. Algorithm~\APER (\cref{alg:accinplremainder}) computes
$R\pe A\bmod B$.

\begin{algorithm}[!ht]
  \caption{$\OPER(A,B)$: Over-place Polynomial Euclidean Quotient and Remainder.}\label{alg:overplaceremainder}
  \begin{algorithmic}[1]
    \REQUIRE $A$, $B$ in $\F[X]$ of respective degrees $m+n$ and $n$.
      \ENSURE $A = [A\bquo B,A\bmod{B}]$
    of degrees $m$ and at most $n-1$.
    \IF{$m\ge 0$}
    \STATE Let $k = \lfloor\frac{m}{n}\rfloor$ and write
      $A = \sum_{i=0}^{k-1} A_i X^{ni}$ \hfill\COMMENT{polynomials of deg. $n-1$}
    \STATE Let $B^* = B\bquo X$ and $B_* = B\bmod X^n$
      \STATE Let  $s = m\bmod n$ be the degree of $A_{k-1}$ and $B_1 = B\bquo X^{n-s-1}$
      \STATE\label{line:smallinv}$A_{k-1} \gets A_{k-1}\revdiv B_1$
        \hfill\COMMENT{\cref{prop:toeplitzpoly}~\ref{item:revdiv}}
      \STATE\label{line:skinnyG}$A_{k-2} \me B_*A_{k-1}\bmod X^n$
      \hfill\COMMENT{\cref{rk:toeplitzpoly}}
    \FOR{$i=k-2$ \DownTo $1$}
	\STATE $A_i\gets A_i\revdiv B^*$
        \hfill\COMMENT{\cref{prop:toeplitzpoly}~\ref{item:revdiv}}
        \STATE $A_{i-1}\me B_*A_i\bmod X^n$
        \hfill\COMMENT{\cref{prop:toeplitzpoly}~\ref{item:shortprod}}
    \ENDFOR
    \ENDIF
  \end{algorithmic}
\end{algorithm}

The idea of \OPER is to compute the remainder progressively in the blocks of $A$, making
use of the \emph{over-place} algorithm for reversed power series division and the in-place
accumulated short product.
A nice property of \OPER~is that it is reversible since each operation is reversible. The
operation $A_{i-1} \me B_*A_i\bmod X^n$ is undone by $A_{i-1}\pe B_*A_i\bmod X^n$ and
$A_i\gets A_i\revdiv B^*$ by $A_i\gets A_i\revmul B^*$. Performing the reverses in reversed
order recovers~$A$.
We denote by $\OPER^{-1}$ this recovery.

With this, \APER~is now just the application of \OPER~and $\OPER^{-1}$
interleaved with an update of the remainder. This is shown
in~\cref{alg:accinplremainder}.
\begin{algorithm}[!ht]
  \caption{$\APER(R,A,B)$: Accumulated in-place Polynomial Euclidean Remainder.}\label{alg:accinplremainder}
  \begin{algorithmic}[1]
    \REQUIRE $R$, $A$, $B$ in $\F[X]$ of resp. degrees $n-1$, $m+n$ and $n$.
    \ENSURE $R\pe{A}\bmod{B}$ of degree at most $n-1$.
    \IfThenEnd{$m\ge 0$}{$\OPER(A,B)$;\label{lin:oper}}\hfill\COMMENT{\cref{alg:overplaceremainder}}
    \STATE $R\pe A\bmod X^n$; \hfill\COMMENT{contains the remainder}
    \IfThenEnd{$m\ge 0$}{$\OPER^{-1}(A,B)$;}\hfill\COMMENT{undo~\cref{lin:oper}}
  \end{algorithmic}
\end{algorithm}

\begin{theorem}\label{thm:oper:aper}
    \Cref{alg:overplaceremainder,alg:accinplremainder} are correct,
    in-place, and their complexity is bounded by $\bigO{\frac{m}{n}\M(n)\log(n)}$
    operations, or $\bigO{\M(n)}$ if $\M(n) = \Theta(n^{1+\epsilon})$
    for some $\epsilon >0$.
\end{theorem}

\subsection{Fast in-place modular multiplication}\label{sec:axpyin}

The goal is to compute $R\pe AC\bmod B$ where $\deg(R)$, $\deg(A)$, $\deg(C) < n$ and $\deg(B) = n$. 
This corresponds to multiplication in polynomial extensions of finite fields for instance.
Let $D = AC$ of degree $2n-2$. Write $D = D_1X^n + D_0$ where $\deg(D_1) = n-2$ and $\deg(D_0) < n$.
Then $D\bmod B = D_0 + (X^nD_1\bmod B)$. To compute $X^nD_1\bmod B$, we first compute the degree-$(n-2)$
quotient as $Q = D_1\revdiv B^*$ where $B^* = B\bquo X^2$. Then we get $X^nD_1\bmod B$ as $(X^nD_1-Q\cdot B)\bmod X^n = -(Q\cdot B)\bmod X^n$.
Since $D$ is known only as $A\cdot C$, we first compute $D_1$ into $C$ as $D_1 = C^*\revmul A^*$ where $A^*=A\bquo X$ and $C^* = C\bquo X$ and undo the computation afterwards.

\newcommand{\AXPYIN}{\ensuremath{\text{AXPYIN}}\xspace}
\begin{algorithm}[htb]
  \caption{$\AXPYIN(R,A,C,B)$: Accumulated in-place modular multiplication.}\label{alg:modmulin}
  \begin{algorithmic}[1]
    \REQUIRE $R$, $A$, $C$, $B$ in $\F[X]$ of
      resp. degrees $n-1$, $n-1$, $n-1$ and $n$
    \ENSURE $R\pe{AC}\bmod{B}$ of degree at most $n-1$.
    \STATE Let $B_* = B\bmod X^n$ and $B^* = B\bquo X^2$
    \STATE Let $A^* = A\bmod X$ and $C^* = C\bmod X$
    \STATE\label{line:revmul}$C^* \gets C^*\revmul A^*$
      \hfill\COMMENT{compute $D_1$ into $C^*$ by \cref{prop:toeplitzpoly}~\ref{item:revmul}}
    \STATE\label{line:revdiv}$C^* \gets C^*\revdiv B^*$
      \hfill\COMMENT{compute $Q$ into $C^*$ by \cref{prop:toeplitzpoly}~\ref{item:revdiv}}
    \STATE $R \me  B_*\cdot C^*\bmod X^n$
      \hfill\COMMENT{subtract $BQ\bmod X^n$ using \cref{alg:accinplshortoneinfive}}
    \STATE $C^* \gets C^*\revmul B^*$
      \hfill\COMMENT{undo \cref{line:revdiv}}
    \STATE $C^* \gets C^*\revdiv A^*$
      \hfill\COMMENT{undo \cref{line:revmul}} 
    \STATE $R \pe (A\cdot C)\bmod X^n$
      \hfill\COMMENT{add $D_0$ using \cref{alg:accinplshortoneinfive}}
 \end{algorithmic}
\end{algorithm}

\begin{proposition}\label{prop:modmulin}
  \Cref{alg:modmulin} is correct and performs 
  $\bigOdisplay{\M((n)\log(n)}$ operations,
  or $\bigOdisplay{\M(n)}$ only
  if $\M(n)=\Theta(n^{1+\epsilon})$ for some $\epsilon>0$.
\end{proposition}
\begin{proof}
    The correctness is justified by the paragraph preceding the algorithm. The complexity is given by \cref{prop:toeplitzpoly}.
\end{proof}

Combining this algorithm with \OPER let us perform any accumulating modular multiplication.

\begin{algorithm}[htb]
    \caption{Accumulated in-place modular multiplication.}\label{alg:fullmodmulin}
    \begin{algorithmic}[1]
    \REQUIRE $R$, $A$, $C$, $B$ in $\F[X]$ of
      resp. degrees $n-1$, $\ell$, $m$ and $n$
    \ENSURE $R\pe{AC}\bmod{B}$ of degree at most $n-1$.
        \STATE $\OPER(A,B)$ \hfill\COMMENT{$A\bmod B$ in $A_0$ via \cref{alg:overplaceremainder}, only if $\ell\ge n$}
        \STATE $\OPER(C,B)$ \hfill\COMMENT{$C\bmod B$ in $C_0$ via \cref{alg:overplaceremainder}, only if $m\ge n$}
        \STATE $\AXPYIN(R,A_0,C_0,B)$ \hfill\COMMENT{\cref{alg:modmulin}}
        \STATE $\OPER^{-1}(C,B)$ \hfill\COMMENT{recover $C$, only if $\ell\ge n$}
        \STATE $\OPER^{-1}(A,B)$ \hfill\COMMENT{recover $A$, only if $m\ge n$}
 \end{algorithmic}
\end{algorithm}

\begin{proposition}\label{prop:fullmodmulin}
    \Cref{alg:fullmodmulin} is correct and performs $\bigO{\frac{\ell+m}{n} \M(n)\log n)}$ operations, or only $\bigO{\frac{\ell+m}{n}\M(n)}$ if $\M(n) = \Theta(n^{1+\epsilon})$ for some $\epsilon > 0$.
\end{proposition}

\begin{proof}
    This is the combination of \cref{thm:oper:aper} and \cref{prop:modmulin}.
\end{proof}

\section{Conclusion}

We here provide a generic technique mapping any bilinear formula
(and more generally any linear accumulation)
into an in-place algorithm.
This allows us for instance to provide the first accumulating in-place
Strassen-like matrix multiplication algorithm. This algorithm compares
favorably in practice with the standard not-in-place variants (\cref{fig:fflas}).
We also extend the result to many in-place and over-place linear
algebra routines, some of which are actually not bilinear.

We apply the same technique to provide fast in-place accumulating polynomial
multiplication algorithms. Our implementation of an in-place variant of
Karatsuba polynomial multiplication has very close performance to that
of the state-of-the-art
library NTL (\cref{fig:ipkara}). From these polynomial multiplication
algorithms, we provide a series of reductions to get fast in-place
accumulating algorithms for generalized convolutions, short product
and power series division and remainder. We also get over-place
variants, and in particular describe a fast algorithm that replaces the
dividend by the quotient and remainder in the Euclidean division.
These results are obtained through their equivalent representations
as $f$-circulant or Toeplitz matrix-vector products, or system solving.
The web of reductions is depicted on \cref{fig:reductions}.

\begin{figure}[htp]
    \caption{Main polynomial reductions in the paper.}\label{fig:reductions}
    \crefname{algorithm}{Alg.}{Alg.}
    \crefname{proposition}{Prop.}{Prop.}
    \crefname{remark}{Rem.}{Rem.}
\noindent
\begin{tikzpicture}[%
        common/.style={
            thick,rounded corners,%
            draw=#1,fill=gray!20,%
            align=center,
            text width=20em
        },%
        cadre/.style={rectangle,common=#1},%
        double/.style={rectangle split,rectangle split parts=2,common=#1},%
        rightcol/.append style={text width=10.5em,xshift=.5cm},%
        node distance=.75cm,
        every edge/.style={green,very thick,->,draw}%
    ]
    \node[cadre=green] (prod) at (0,0)  {$C\pe A\cdot B$\\\cref{sec:inpaccpol}};
    \node[double=green,anchor=north west,rightcol](conv) at (prod.north east){%
            $C\pe A\cdot B\bmod X^n-f$\\\cref{alg:accinplfconvol,alg:accinplconvol,alg:accinplfconvoleven} %
            \nodepart{two} 
            Acc. $f$-circulant m-v. prod. \hfill\cref{ssec:fcirculant}%
    };
    \node[double=green,below=of prod] (sp) {%
        $C\pe A\cdot B\bmod X^n$ and $C\pe B\revmul A$\\\cref{alg:accinplshortoneinfive,prop:toeplitzpoly}~\ref{item:revmul}%
        \nodepart{two}
        Acc. triangular Toeplitz m-v. prod.\\\cref{alg:uppertriangtoeplitz,rk:trspalg}%
    };
    \node[double=green,below=of sp]  (mp) {
        $C\pe (A\cdot B)\bmod X^{m+n-1}\bquo X^n$\\\cref{prop:toeplitzpoly}~\ref{item:midprod}%
        \nodepart{two}
        Acc. Toeplitz m-v. product\\\cref{alg:squaretoeplitz,alg:toeplitz}%
    };
    \node[double=red,below=of mp]  (over) {
        $B\gets BA\bmod X^n$, $B\gets B\revmul A$ and $B\gets B\revdiv A$\hfill \cref{prop:toeplitzpoly} 
        and \cref{rk:toeplitzpoly}%
        \nodepart{two}
        Over-place triang. Toepl. m-v. prod. and\\syst. solv.\hfill \cref{alg:overtriangtoeplitz,alg:solvetriangtoeplitz}%
    };
    \node[cadre=red,rightcol,anchor=north west] (toeplike) at (over.north east) {Acc. Toeplitz-like m-v. prod.\cref{alg:toeplitzlike}};
    \node[cadre=red,rightcol,anchor=south west] (mod) at (over.south east){$R\gets A\bmod B$\\\cref{alg:remainder} (\IPER)};
    \node[cadre=red,below=of over] (overmod) {$(A,B) \longleftrightarrow (Q|R,B)$\\\cref{alg:overplaceremainder} (\OPER)};
    \node[cadre=red,right=0cm of overmod,rightcol] (modacc) {$R\pe A\bmod B$\\\cref{alg:accinplremainder} (\APER)};
    \node[cadre=red,below=of overmod] (modmul) {$R\pe AC\bmod B$\\\cref{alg:modmulin} (\AXPYIN)};
    \draw (prod) edge (conv.text west)
          (prod) edge (sp)
          (sp) edge (mp)
          (mp) edge[->,red] node[right] {$\log n$} (over)
          (over.text east) edge (toeplike)
          (over.two east) edge (mod)
          (over) edge (overmod)
          (overmod) edge (modmul)
          (overmod) edge (modacc)
    ;
    \node[cadre=green,text width=,minimum width=.5cm, minimum height=.5cm,anchor=west,below left=.75cm and -1cm of modmul] (carregreen) {};
    \node[cadre=red,text width=,minimum width=.5cm, minimum height=.5cm,anchor=west,below=of carregreen] (carrered) {};
    \node[right=0pt of carregreen,align=left] (legend) {Time: $O(\M(n))$\\Space: $O(1)$};
    \node[right=0pt of carrered,align=left] {Time: $O(\M(n)\log n)$ or $\M(n)$ if $\M(n) = \Omega(n^{1+\epsilon})$ for some $\epsilon > 0$\\Space: $O(1)$ algebraic registers, $O(\log n)$ pointers};
    \node[below left=.75cm and 0cm of carrered] (g) {};
    \node[below=.5em of g] (r) {};
    \draw[very thick, green,->] (g) -- +(.75,0);
    \draw[very thick, red,->] (r) -- +(.75,0);
    \node[right=of g] {Space- and time-preserving reduction};
    \node[right=of r] {Reduction with a call stack and an extra logarithmic factor in the time};

\end{tikzpicture}
\end{figure}

We have here the first fast in-place, over-place and
accumulating algorithms computing only the remainder of the polynomial
Euclidean division.
There, one open problem remains: to remove the extra logarithmic
factor in the complexity that appears in this case, when $\M(n) $ is not
$\Omega(n^{1+\epsilon})$ for some $\epsilon>0$.

Finally, \Cref{sec:axpyin} also shows a direct application of our
techniques for the multiplication in a polynomial extension of a
finite field.

Another possible improvement would be to study the behavior of these
algorithms with floating point numbers. In this case our ``undoing'' approach
might not restore the exact original states. It would then be
necessary to study not only the accuracy of the obtained programs but
also the effect on their inputs.

\clearpage
\printbibliography

\end{document}